\documentclass[11pt]{article}
\pdfoutput=1

\newcommand{\Fhat}{{\hat F}}
\newcommand{\Ghat}{{\hat G}}
\newcommand{\phat}{{\hat p}}

\newcommand{\Shat}{{\hat S}}

\newcommand{\Yhat}{{\hat Y}}
\newcommand{\xbar}{{\bar x}}
\newcommand{\Xbar}{{\bar X}}
\newcommand{\ybar}{{\bar y}}

\newcommand{\betahat}{{\hat\beta}}
\newcommand{\deltahat}{{\hat\delta}}
\newcommand{\thetahat}{{\hat\theta}}

\newcommand{\sigmahat}{{\hat\sigma}}

\newcommand{\iidB}{{i.i.d.~}}
\newcommand{\tSub}{t_{\alpha/2,n-1}}
\newcommand{\zSub}{z_{\alpha/2}}

\newcommand{\Var}{\mathrm{Var}}

\newcommand{\SE}{\mathrm{SE}}
\newcommand{\SD}{\mathrm{SD}}

\newcommand{\boxx}[2]
{
\begin{table*}[htbp!]
\begin{center}
\begin{tabular}{|p{4.6in}|} \hline
\vspace{.1mm} {\bf #1} \\
#2 \\[2.5mm] \hline
\end{tabular}
\end{center}
\end{table*}
}

\usepackage{graphicx}
\graphicspath{{.}{./figures/}%
{/Users/rocket/bootstrap/chapter/latex/}%
{/Users/rocket/bootstrap/Chihara/svn1/stat_inference/trunk/Images/}}

\usepackage{hyperref}
\usepackage{myAlgorithmic}
\usepackage{colortbl}
\usepackage{listdefs}

\usepackage{natbib} % Use with rss.bst

\newcommand{\citeP}[1]{\citep{#1}}   % use with natbib.sty & chicago.bst
\newcommand{\comment}[1]{}  % Use this to comment out text, maybe many lines
\newcommand{\figurewidth}{5in}
\newcommand{\mycaption}{} % redefined below

\newcommand{\figureplace}{tbp!} % tbp or htbp or tbph
\newcommand{\tableplace}{tbp!} % tbp or htbp or tbph or !tbp
\begin{document}

\title{What Teachers Should Know about the Bootstrap: Resampling in the Undergraduate Statistics Curriculum}
\author{Tim Hesterberg  \\ Google \\
\texttt{timhesterberg@gmail.com} }

\maketitle

% \newpage
%
% \mbox{}
% \vspace*{2in}
% \begin{center}
% \textbf{Author's Footnote:}
% \end{center}
% Some text about me.

%\newpage
%\begin{center}
%\textbf{Abstract}
%\end{center}
\section*{Abstract}

I have three goals in this article:
(1) To show the enormous potential
of bootstrapping and permutation tests to help students
understand statistical concepts including sampling distributions,
standard errors, bias, confidence intervals, null distributions,
and $P$-values.
(2) To dig deeper, understand why these methods work and when they
don't,
% compare them to formula approximations,
things to watch out for, and how to deal with these issues
when teaching.
(3) To change statistical practice---by
comparing these methods to common $t$ tests and intervals,
we see how inaccurate the latter are;
we confirm this with asymptotics. $n \ge 30$ isn't
enough---think $n \ge 5000$.
Resampling provides diagnostics, and more accurate alternatives.
Sadly, the common bootstrap percentile interval badly under-covers in small
samples; there are better alternatives.
% Along the way we'll see examples of the methods
% in practice. %, and compare them to common formula approximations.
The tone is informal, with a few stories and jokes.

\vspace*{.3in}

\noindent\textsc{Keywords}: {Teaching, bootstrap, permutation test, randomization test}

% I'll remove the TOC for submitting this to TAS, but keep it for
% the white paper and for use now.
\newpage
\tableofcontents

%\newpage

\section{Overview}
\label{section:overview}

I focus in this article on how to use relatively simple
bootstrap methods and permutation tests to help students understand
statistical concepts, and what instructors should know about these
methods. I have Stat 101 and Mathematical Statistics in mind, though
the methods can be used elsewhere in the curriculum.
For more background on the bootstrap and a broader array of applications,
see \citeP{efro93,davi97}.

Undergraduate textbooks that consistently use resampling as tools in
their own right and to motivate classical methods are beginning to appear,
including \cite{lock2013statistics} for Introductory Statistics
and \cite{chih11} for Mathematical Statistics.
Other texts incorporate at least some resampling.
% Diez book "Intro Stat with Randomization and Simulation"
% http://www.openintro.org/stat/textbook.php
% does not qualify for mention. It starts with randomization tests. However,
% bootstrapping is tacked on, and not understood.
% But I could mention it in my website.

Section~\ref{section:begin} is an introduction to one- and two-sample
bootstraps and two-sample permutation tests,
and how to use them to help students understand
sampling distributions, standard errors, bias, confidence intervals,
hypothesis tests, and $P$-values.
We discuss the idea behind the bootstrap, why it works,
and principles that guide our application.

In Section~\ref{section:variationBootstrapDistributions}
we take a visual approach toward understanding when the bootstrap
works and when it doesn't. We compare the effect on bootstrap
distributions of two sources of variation---the original sample, and
bootstrap sampling.

In Section~\ref{section:biasSkewness} we look at three
things that affect inferences---bias, skewness, and transformations---and
something that can cause odd results for bootstrapping, whether a
statistic is functional.
This section also discusses how inaccurate classical $t$ procedures
are when the population is skewed.
I have a broader goal beyond better pedagogy---to change
statistical practice.
Resampling provides diagnostics, and alternatives.

This leads to Section~\ref{section:confidenceIntervals},
on confidence intervals;
beginning with a visual approach to how confidence intervals
should handle bias and skewness, then a description of different
confidence intervals procedures and their merits,
and finishing with a discussion
of accuracy, using simulation and asymptotics.

In Section~\ref{section:samplingMethods} we consider
sampling methods for different situations, in particular regression,
and ways to sample to avoid certain problems.

We return to permutation tests in Section~\ref{section:permutationTests},
to look beyond the two-sample test to other applications
where these tests do or do not apply,
and finish with a short discussion of bootstrap tests.

Section~\ref{section:summary} summarizes key issues.

Teachers are encouraged to use the examples in this article in their
own classes. I'll include a few bad jokes; you're welcome to those too.
Examples and figures are created in {\em R} \citeP{r14}, using the
{\em resample} package \citeP{hest14}.
I'll put datasets and scripts at
\url{http://www.timhesterberg.net/bootstrap}.
% TODO(put them there)

I suggest that all readers begin by skimming the paper,
reading the boxes and Figures~\ref{figure:coverage2}
and~\ref{figure:coverage3}, before returning here for a full pass.

There are sections you may wish to read out of order.
If you have experience with resampling you may want to read the summary first,
Section~\ref{section:summary}.
To focus on permutation tests read Section~\ref{section:permutationTests}
after Section~\ref{section:pedagogicalValuePermutationTests}.
To see a broader range of bootstrap sampling methods earlier,
read Section~\ref{section:samplingMethods} after
Section~\ref{section:ideaBehindBootstrapping}.
And you may skip the Notation section, and refer to it as needed later.

\subsection{Notation}
\label{section:notation}

This section is for reference;
the notation is explained when it comes up.

We write $F$ for a population, with corresponding parameter $\theta$;
in specific applications we may have e.g. $\theta = \mu$ or
$\theta = \mu_1 - \mu_2$; the corresponding sample estimates are
$\thetahat$, $\xbar$, or $\xbar_1-\xbar_2$.

$\Fhat$ is an estimate for $F$.
Often $\Fhat$ is the empirical distribution $\Fhat_n$, with
probability $1/n$ on each observation in the original sample.
When drawing samples from $\Fhat$,
the corresponding estimates are
$\thetahat^*$, $\xbar^*$, or $\xbar_1^*-\xbar_2^*$.

$s^2 = (n-1)^{-1}\sum (x_i-\xbar)^2$ is the usual sample variance,
and $\sigmahat^2 = n^{-1}\sum (x_i-\xbar)^2 = (n-1)s^2/n$ is
the variance of $\Fhat_n$.

When we say ``sampling distribution'', we mean the sampling
distribution for $\thetahat$ or $\Xbar$ when sampling from $F$,
unless otherwise noted.

$r$ is the number of resamples in a bootstrap or permutation
distribution. The mean of the bootstrap distribution
is $\overline{\thetahat^*}$ or $\overline{\xbar^*}$,
and the standard deviation of the bootstrap distribution
(the bootstrap standard error)
is $s_B = \sqrt{(r-1)^{-1}\sum_{i=1}^r (\thetahat^*_i - \overline{\thetahat^*})^2}$
or $s_B = \sqrt{(r-1)^{-1}\sum_{i=1}^r (\xbar^*_i - \overline{\xbar^*})^2}$.

The $t$ interval with bootstrap standard error is
$\thetahat \pm \tSub s_B$.

$G$ represents a theoretical bootstrap or permutation distribution,
and $\Ghat$ is the approximation by sampling; the $\alpha$ quantile of this
distribution is $q_\alpha = \Ghat^{-1}(\alpha)$.

The bootstrap percentile interval is
$(q_{\alpha/2}, q_{1-\alpha/2})$, where $q$ are quantiles of $\thetahat^*$.
The expanded percentile interval is
$(q_{\alpha'/2}, q_{1-\alpha'/2})$,
where $\alpha'/2 = \Phi(-\sqrt{n/(n-1)} \tSub)$.
The reverse percentile interval is
$(2\thetahat - q_{1-\alpha/2}, 2\thetahat - q_{\alpha/2})$.

The bootstrap~$t$ interval is
$(\thetahat - q_{1-\alpha/2}\Shat, \thetahat - q_{\alpha/2}\Shat)$
where $q$ are quantiles for $(\thetahat^* - \thetahat)/\Shat^*$
and $\Shat$ is a standard error for $\thetahat$.

Johnson's (skewness-adjusted) $t$~statistic is
$t_1 = t + \kappa\  (2 t^2 + 1)$
where $\kappa = \mbox{skewness}/(6\sqrt{n})$.
The skewness-adjusted $t$~interval is
$\xbar + (\kappa\  (1 + 2 t_{\alpha/2}^2) \pm t_{\alpha/2}) (s/ \sqrt{n})$.

\section{Introduction to the Bootstrap and Permutation Tests}
\label{section:begin}

We'll begin with an example to illustrate
the bootstrap and permutation tests procedures,
discuss pedagogical advantages of these procedures,
and the idea behind bootstrapping.

% Barrett Rogers
Student B.~R.~was
annoyed by TV commercials. He suspected that
there were more commercials in the ``basic'' TV channels, the
ones that come with a cable TV subscription, than in the ``extended''
channels you pay extra for. To check this, he collected the data shown in
Table~\ref{table:tvCommercials}.

\begin{table}[\tableplace]
 \centering
\begin{tabular}{llllll}\hline
Basic    & 6.95 & 10.013 & 10.62 & 10.15 & 8.583 \\
         & 7.62 & 8.233 & 10.35 & 11.016 & 8.516 \\
Extended & 3.383 & 7.8 & 9.416 & 4.66 & 5.36\\
         & 7.63 & 4.95 & 8.013 & 7.8 & 9.58\\ \hline
\end{tabular}
\renewcommand{\mycaption}{Minutes of commercials per half-hour of TV.}
\caption[\mycaption]{\label{table:tvCommercials}
{\em \mycaption}}
\end{table}

He measured an average of
9.21 minutes of commercials per half hour in the basic channels,
vs only 6.87 minutes in the extended channels.
This seems to support his hypothesis.
But there is not much data---perhaps the difference was just random.
The poor guy could only stand to watch 20 random half hours of TV.
Actually, he didn't even do that---he got his girlfriend to watch half of it.
(Are you as appalled by the deluge of commercials as I am?
This is per half-hour!)

\subsection{Permutation Test}
How easy would it be for a difference of 2.34 minutes to occur
just by chance?
To answer this, we suppose there really is no difference between the
two groups, that ``basic'' and ``extended'' are just labels.
So what would happen if we assign labels randomly?
How often would a difference like 2.34 occur?

\begin{figure}[\figureplace]
% Scripts/TV.R
\centerline{\includegraphics[width=\figurewidth]{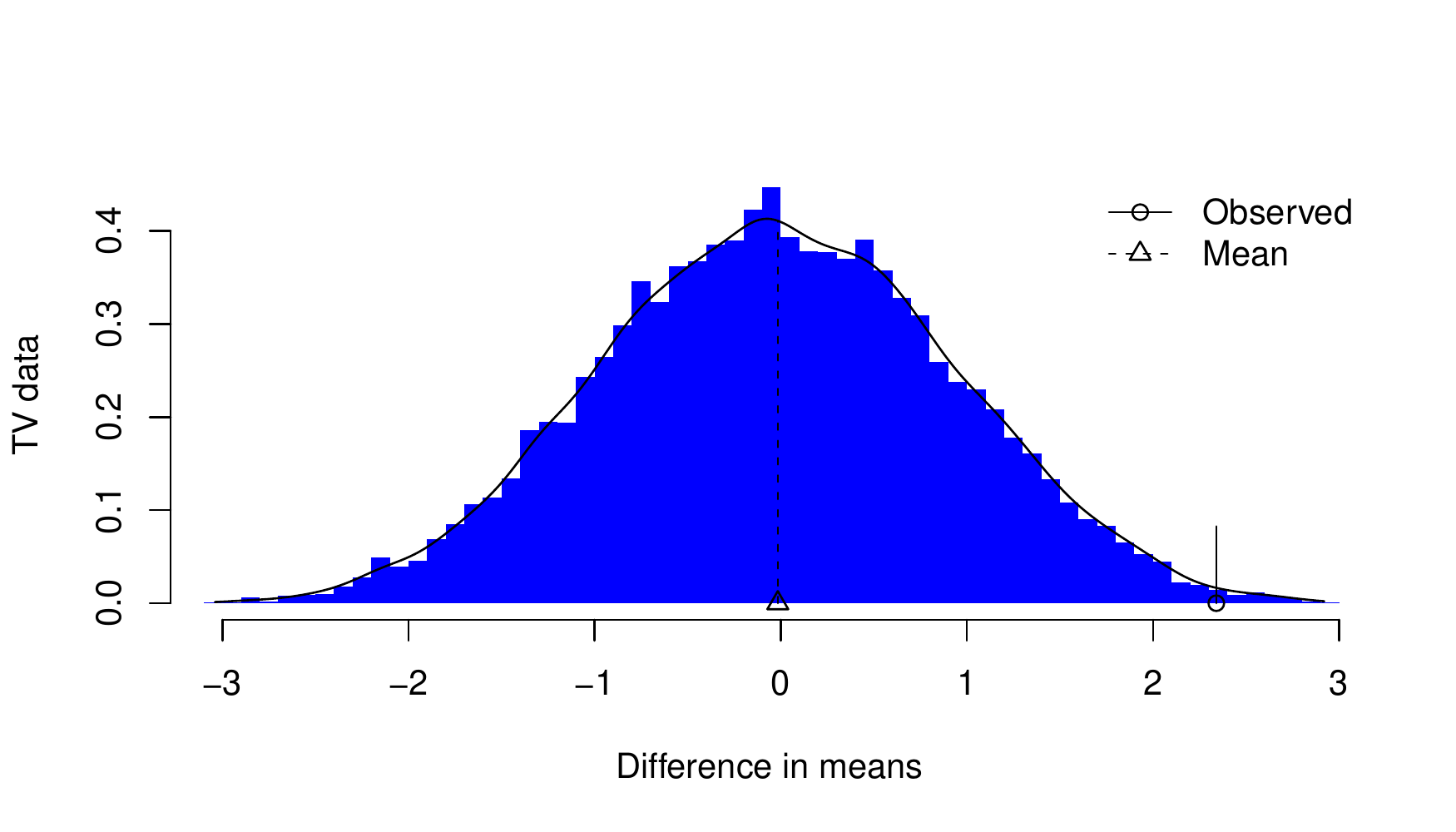}}
\renewcommand{\mycaption}{Permutation distribution for the difference in means between
basic and extended channels.}
\caption[\mycaption]{\label{figure:permTV}
{\em \mycaption}
The observed difference of $2.34$ is shown; a fraction $0.005$ of the
distribution is to the right of that value ($\ge 2.34$).
}
% The closing brace would normally go here, but that causes this and
% subsequent figures to be placed lower
\end{figure}

We'll pool all twenty observations,
randomly pick 10 of them to label ``basic'' and label the rest ``extended'',
and compute the difference in means between the two groups.
We'll repeat that many times, say ten thousand,
to get the
{\it permutation distribution} shown in
Figure~\ref{figure:permTV}.
The observed statistic $2.34$ is also shown; the fraction of the distribution
to the right of that value ($\ge 2.34$)
is the probability that random labeling would
give a difference that large.
In this case, the probability, the $P$-value, is $0.005$;
it would be rare for a difference this large to occur by chance.
Hence we conclude there is a real difference between the groups.

\boxx{Two-Sample Permutation Test}
{
\begin{algorithmic}
\STATE Pool the $n_1+n_2$ values
\REPEATX{9999 times}
%\REPEAT
\STATE Draw a resample of size $n_1$ without replacement.
\STATE Use the remaining $n_2$ observations for the other sample.
\STATE Calculate the difference in means, or another statistic that
compares samples.
\ENDREPEATX
%\UNTIL{we have enough samples}
\STATE Plot a histogram of the random statistic values; show the observed statistic.
\STATE Calculate the $P$-value as the fraction of times the random statistics
exceed or equal the observed statistic (add 1 to numerator and denominator);
multiply by 2 for a two-sided test.
\end{algorithmic}%
}

We defer some details
until Section~\ref{section:permutationDetails}, including why we
add 1 to numerator and denominator, and why we calculate a two-sided
$P$-value this way.

\subsection{Pedagogical Value}
\label{section:pedagogicalValuePermutationTests}

This procedure provides nice visual representation for
what are otherwise abstract concepts---a null distribution, and
a $P$-value.
Students can use the same tools they previously used for looking
at data, like histograms, to inspect the null distribution.

And it makes the convoluted logic of hypothesis testing quite natural.
(Suppose the null hypothesis is true, how
often we would get a statistic this large or larger?)
Students can learn that ``statistical significance'' means
``this result would rarely occur just by chance''.

It has the advantage that students can work directly with the statistic
of interest---the difference in means---rather than switching to some
other statistic like a $t$~statistic.
% It completely avoids the question of whether to pool variances;
% they are automatically pooled, which is appropriate for hypothesis tests.

It generalizes nicely to other statistics. We could work with the
difference in medians, for example, or a difference in trimmed means,
without needing new formulas.

% This lets us teach more consistently, and break free of the usual
% Stat 101 curriculum that covers both the mean and median in
% early data analysis chapters, then forgets all about the median when
% it comes time for hypothesis tests and confidence intervals.

\boxx{Pedagogical Value of Two-Sample Permutation Test}
{
\begin{squeezeitemize}
\item Make abstract concepts concrete---null distribution, $P$-value.
\item Use familiar tools, like histograms.
\item Work with the statistic of interest, e.g.\ difference of means.
\item Generalizes to other statistics, don't need new formulas.
  \item Can check answers obtained using formulas.
\end{squeezeitemize}%
}

\subsection{One-Sample Bootstrap}
\label{section:oneSampleBootstrap}

In addition to using the permutation test to see whether there is a
difference, we can
also use resampling, in particular the bootstrap, to quantify
the random variability in the two sample estimates, and in the
estimated difference. We'll start with one sample at a time.

In the bootstrap, we draw $n$ observations
with replacement from the original data
to create a {\em bootstrap sample} or
{\em resample}, and calculate the mean for this resample.
We repeat that many times, say 10000.
The bootstrap means comprise the {\em bootstrap distribution}.

The bootstrap distribution is a {\em sampling distribution},
for $\thetahat^*$ (with sampling from the empirical distribution);
we'll talk more below about how it
relates to the sampling distribution of $\thetahat$ (sampling from
the population $F$).
(In the sequel, when we say ``sampling distribution'' we mean
the latter, not the bootstrap distribution, unless noted.)

\begin{figure}[\figureplace]    % scripts/TV.R
\centerline{\includegraphics[width=\figurewidth]{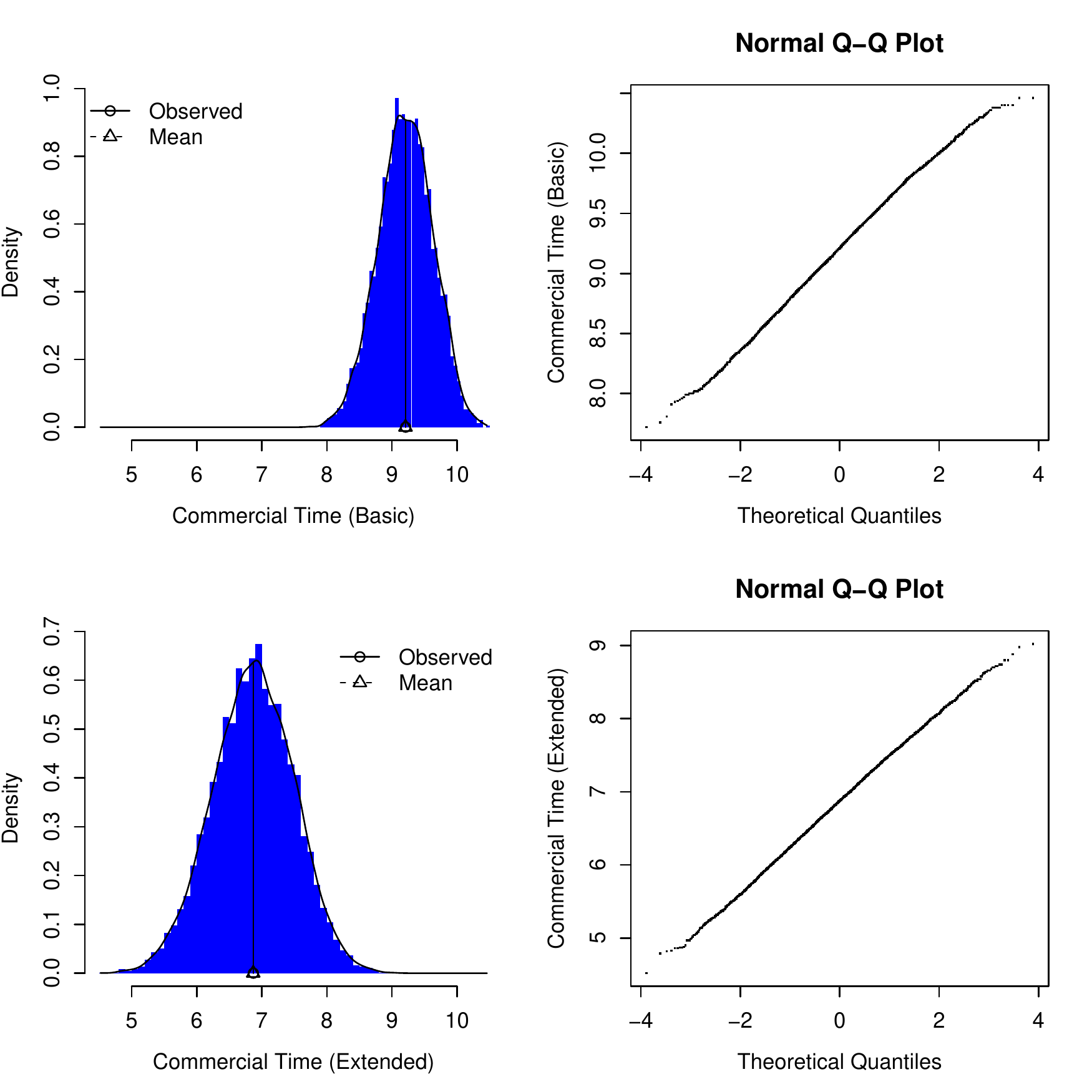}}
\renewcommand{\mycaption}{Bootstrap distributions for TV data.}
\caption[\mycaption]{\label{figure:bootTV1}
{\em \mycaption}
Bootstrap distributions for the mean of the basic channels (top)
and extended channels (bottom). The observed values, and means of the
bootstrap distributions, are shown.
These are sampling distributions for $\xbar^*_1$ and $\xbar^*_2$.
}    \end{figure}

Figure~\ref{figure:bootTV1} shows the bootstrap distributions for
the Basic and Extended data.
For each distribution, we look at the center, spread, and shape:
\begin{description}
\item[center:] Each distribution is centered approximately at the
observed statistic; this indicates that the sample mean is approximately
unbiased for the population mean.
We discuss bias in Section~\ref{section:bias}.
\item[spread:] The spread of each distribution estimates how much
the sample mean varies due to random sampling.
The {\em bootstrap standard error} is the sample standard deviation of
the bootstrap distribution,
%$s_B = \sqrt{(r-1)^{-1}\sum_{i=1}^r (\xbar^*_i - \overline{\xbar^*})^2} = 0.42$.
%
\item[shape:] Each distribution is approximately normally distributed.
\end{description}
A quick-and-dirty confidence interval, the {\em bootstrap percentile
confidence interval}, is the range of the middle 95\% of
the bootstrap distribution;
this is $(8.38, 9.99)$ for the Basic channels and
$(5.61, 8.06)$ for the Extended channels.
(Caveat---percentile intervals are too short in samples this small,
see Sections~\ref{section:smallSamplePicture} and~\ref{section:stat101},
and Figures~\ref{figure:coverage2}--\ref{figure:coverage4}).

Here are the summaries of the bootstrap distributions for basic
and extended channels
% produce by the {\em resample} package in R:
\begin{verbatim}
Summary Statistics:
      Observed        SE     Mean      Bias
Basic     9.21 0.4159658 9.207614 -0.002386

         Observed        SE     Mean      Bias
Extended     6.87 0.6217893 6.868101 -0.001899
\end{verbatim}
% For these data, each bootstrap distribution is centered at approximately
% the corresponding $\xbar$.
The spread for Extended is larger, due to
the larger standard deviation in the original data.
Here, and elsewhere unless noted, we use $10^4$ resamples for the bootstrap
or $10^4-1$ for permutation tests.

\boxx{One-Sample Bootstrap}
{
\begin{algorithmic}
\REPEATX{$r=10000$ times}
\STATE Draw a sample of size $n$ with replacement from the original data
(a {\em bootstrap sample} or {\em resample}).
\STATE Compute the sample mean (or other statistic) for the resample.
\ENDREPEATX
\STATE The 10000 bootstrap statistics comprise the {\em bootstrap distribution}.
\STATE Plot the bootstrap distribution.
\STATE The {\em bootstrap standard error} is the standard deviation of the bootstrap distribution,
$s_B = \sqrt{\sum (\thetahat_i^*-\overline{\thetahat^*})^2 / (r-1)}$.
\STATE The {\em bootstrap percentile confidence interval} is the range of the middle
95\% of the bootstrap distribution.
\STATE The {\em bootstrap bias estimate} is mean of the bootstrap distribution, minus the observed statistic, $\overline{\thetahat^*} - \thetahat$.
\end{algorithmic}%
}

\subsection{Two-Sample Bootstrap}

\begin{figure}[\figureplace]    % scripts/TV.R
\centerline{\includegraphics[width=\figurewidth]{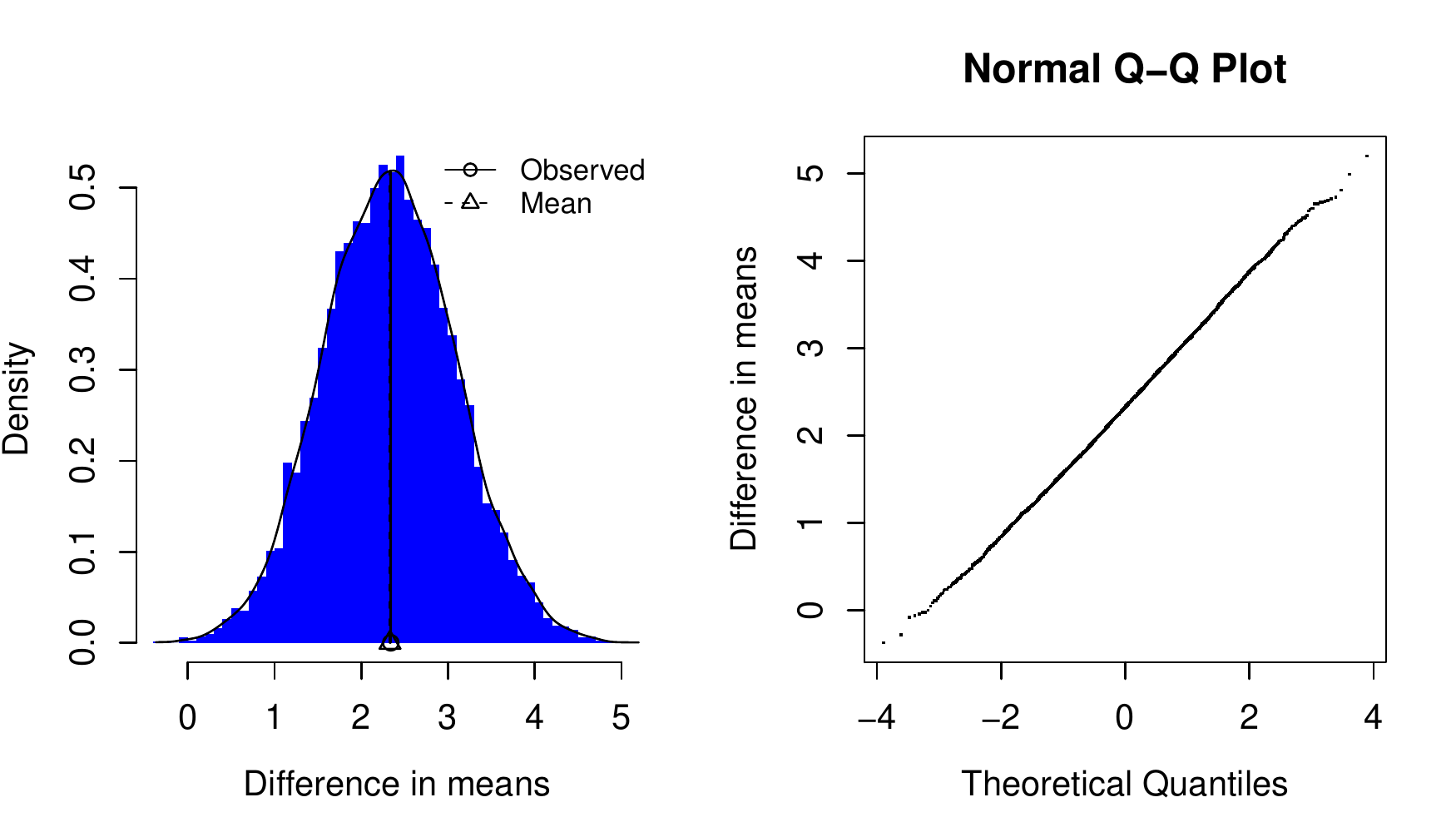}}
\renewcommand{\mycaption}{Two-sample bootstrap for TV commercials data.}
\caption[\mycaption]{\label{figure:bootTV2}
{\em \mycaption}
Bootstrap distribution for the difference of means between
extended and basic channels.
This is the sampling distribution of $\xbar^*_1-\xbar^*_2$.
}    \end{figure}

For a two-sample bootstrap, we independently draw bootstrap samples from
each sample,
and compute the statistic that compares the samples.
For the TV commercials data, we draw a sample of size
$10$ from Basic data, another sample of size $10$ from the Extended
data, and compute the difference in means.
The resulting bootstrap distribution is shown in
Figure~\ref{figure:bootTV2}.
The mean of the distribution is very close to the observed
difference in means, $2.34$; the bootstrap standard error is $0.76$,
and the 95\% bootstrap percentile confidence interval is
$(0.87, 3.84)$. The interval does not include zero, which suggests
that the difference between the samples is larger than can be
explained by random variation; this is consistent with
the permutation test above.

Recall that for the permutation test we resampled
in a way that was consistent with the null hypothesis of no difference
between populations, and the permutation distribution for the difference
in means was centered at zero.
Here we make no such assumption, and the bootstrap distribution
is centered at the observed statistic; this is used for confidence intervals
and standard errors.

\subsection{Pedagogical Value}

Like permutation tests, the bootstrap makes the abstract concrete.
Concepts like sampling distributions, standard errors, bias,
central limit theorem,
and confidence intervals are abstract, and hard for many students,
and this is usually compounded by a scary cookbook of formulas.

The bootstrap process, involving sampling,
reinforces the central
role that sampling from a population plays in statistics.
Sampling variability is visible, and it is natural to measure the
variability of the bootstrap distribution
using the interquartile range
or the standard deviation; the latter is the bootstrap standard error.
Students can see if the sampling distribution has a bell-shaped
curve.
It is natural to use the middle 95\% of the distribution as a 95\%
confidence interval.
Students can obtain the confidence interval by working directly
with the statistic of interest, rather than using a $t$~statistic.

The bootstrap works the same way with a wide variety of statistics.
This makes it easy for students to work with a
variety of statistics without needing to memorize
more formulas.

The bootstrap can also reinforce the understanding of formula methods,
and provide a way for students to check their work.
Students may know the formula $s/\sqrt{n}$ without understanding
what it really is; but they can compare it to the bootstrap standard error,
and see that it measures how the sample mean varies due to random sampling.

The bootstrap lets us do better statistics. In Stat 101 we talk
early on about means and medians for summarizing data,
but ignore the median later,
like a crazy uncle hidden away in a closet,
because there are no easy formulas for confidence intervals.
Students can bootstrap the median or trimmed mean as easily as the mean.
We can use robust statistics when appropriate, rather than only
using the mean.

You do not need to talk about $t$~statistics and $t$~intervals at all,
though you will undoubtedly want to do so later.
At that point you may introduce
another quick-and-dirty confidence interval, the
{\em $t$~interval with bootstrap standard error},
$\thetahat \pm t_{\alpha/2} s_B$
where $s_B$ is the bootstrap standard error.
(This is not to be confused with the {\em bootstrap~$t$ interval},
see Section~\ref{section:bootstrapT}.)

\boxx{Pedagogical Value of the Bootstrap}
{
\begin{squeezeitemize}
\item Make abstract concepts concrete---sampling distribution, standard error,
  bias, central limit theorem.
\item The process mimics the role of random sampling in real life.
\item Use familiar tools, like histograms and normal quantile plots.
\item Easy, intuitive confidence interval---bootstrap percentile interval.
\item Work with the statistic of interest, e.g.\ difference of means.
\item Generalizes to other statistics, don't need new formulas.
\item Can check answers obtained using formulas.
\end{squeezeitemize}%
}

\subsection{Teaching Tips}

For both bootstrapping and permutation tests, start small,
and let students do some samples by hand.
For permutation tests, starting with small groups like 2 and 3 allows
students to do all
${{n \choose {n_1}}}$ partitions exhaustively.

There is a nice visualization of the process of
permutation testing as part of the iNZight package
\url{<https://www.stat.auckland.ac.nz/~wild/iNZight>}. It demonstrates
the whole process: pooling the data, then repeatedly randomly splitting the
data and calculating a test statistic, to build up the permutation distribution.

\subsection{Practical Value}
Resampling is also important in practice, often providing the only
practical way to do inference.
I'll give some examples from Google, from my work or others.

In Google Search we estimate the average number of words per
query, in every country \citeP{cham14a}.
The data is immense, and is ``sharded''---stored
on tens of thousands of machines.
We can count the number of queries
in each country, and the total number of words in queries in each country,
by counting on each machine and adding across machines, using
MapReduce \citeP{dean2008mapreduce}.
But we also want to estimate the variance, for users in each country,
in words per query per user. The queries for each user are sharded,
and it is not feasible to calculate queries and words
for every user. But there is a bootstrap procedure we can use.
In the ordinary bootstrap, the number of times each observation is included
in a bootstrap sample is $\mbox{Binomial}(n, 1/n)$, which we approximate
as $\mbox{Poisson}(1)$. For each user, we generate $r$ Poisson
values, one for each resample. We don't actually save these, instead
they are generated on the fly, each time the user comes up on any machine,
using a random number seed generated from the user's cookie, so the user
gets the same numbers each time. We then compute weighted counts
using the Poisson weights, to estimate the variance across users of
words per query.

Also in Search,
we continually run hundreds of experiments, trying to improve search
results, speed, usability, and other factors;
each year there are thousands of experiments resulting in hundreds
of improvements%
\footnote{\url{http://www.businessweek.com/the_thread/techbeat/archives/2009/10/google_search_g.html}}
(when you search on Google you are probably in a dozen or more experiments).
The data are sharded, and we cannot combine results for each user.
We split the users into 20 groups, and analyze
the variability across these groups using the jackknife
(another resampling technique).
% Wording revised after checking with Mike and Hal.

In {\em Brand Lift}%
\footnote{\url{https://www.thinkwithgoogle.com/products/brand-lift.html}.
\cite{chan10} describe an earlier version.
}
we use designed experiments to
estimate the effectiveness of display advertisements.
We ask people brand awareness questions such as which brands they are
familiar with, to see whether the exposed (treatment)
and control groups differ.
There are four nested populations:
\begin{squeezedescription}
\item[  (1)] people who saw an ad (and control subjects who would have seen one),
\item[  (2)] those eligible for solicitation (they visit a website
where the survey can be presented),
\item[  (3)] those randomly selected for solicitation,
\item[  (4)] respondents.
\end{squeezedescription}
We use two logistic regression models:
\begin{squeezedescription}
\item[  (A)] data = (4), Y = actual response,
\item[  (B)] data = (4,3), Y = actual response or predictions from (A),
\end{squeezedescription}
with $X$'s such as age and gender to correct for random
differences in these covariates between exposed and controls.
We use predictions from model (A) to extrapolate to (2--3),
and predictions from model (B) to extrapolate to (1).
%We also handle missing data, and produce estimates for different slices.
The estimated average ad effect is the difference, across exposed people,
of $\hat p_1 - \hat p_0$, where
$\hat p_1 = \hat P(Y=1|x)$
is the usual prediction, and
$\hat p_0 = \hat P(Y=1|x \mbox{ except set to control})$
is the prediction if the person were a control.
Formula standard errors for this process
are theoretically possible but difficult to derive,
and would need updating when we change the model;
we bootstrap instead.

For the People Analytics gDNA%
\footnote{\url{http://blogs.hbr.org/2014/03/googles-scientific-approach-to-work-life-balance-and-much-more/}}
longitudinal survey,
we use 5-fold cross-validation (another resampling technique)
to evaluate a messy variable selection routine: multiple-imputation followed by backward-elimination in a linear mixed effects model. The models produced within each fold give an indication of the stability of the final result, and we calculate precision, recall, and accuracy on the holdout sets.
% Ben Ogorek
% Ben says OK.

\subsection{Idea behind Bootstrapping}
\label{section:ideaBehindBootstrapping}

At this point you may have a million questions, but foremost
among them is probably: why does this work?
We'll address that next by talking about the key
idea behind the bootstrap, saving other questions for later.

\begin{figure}[\figureplace]    % scripts/BootPix.R
\centerline{\includegraphics[width=\figurewidth]{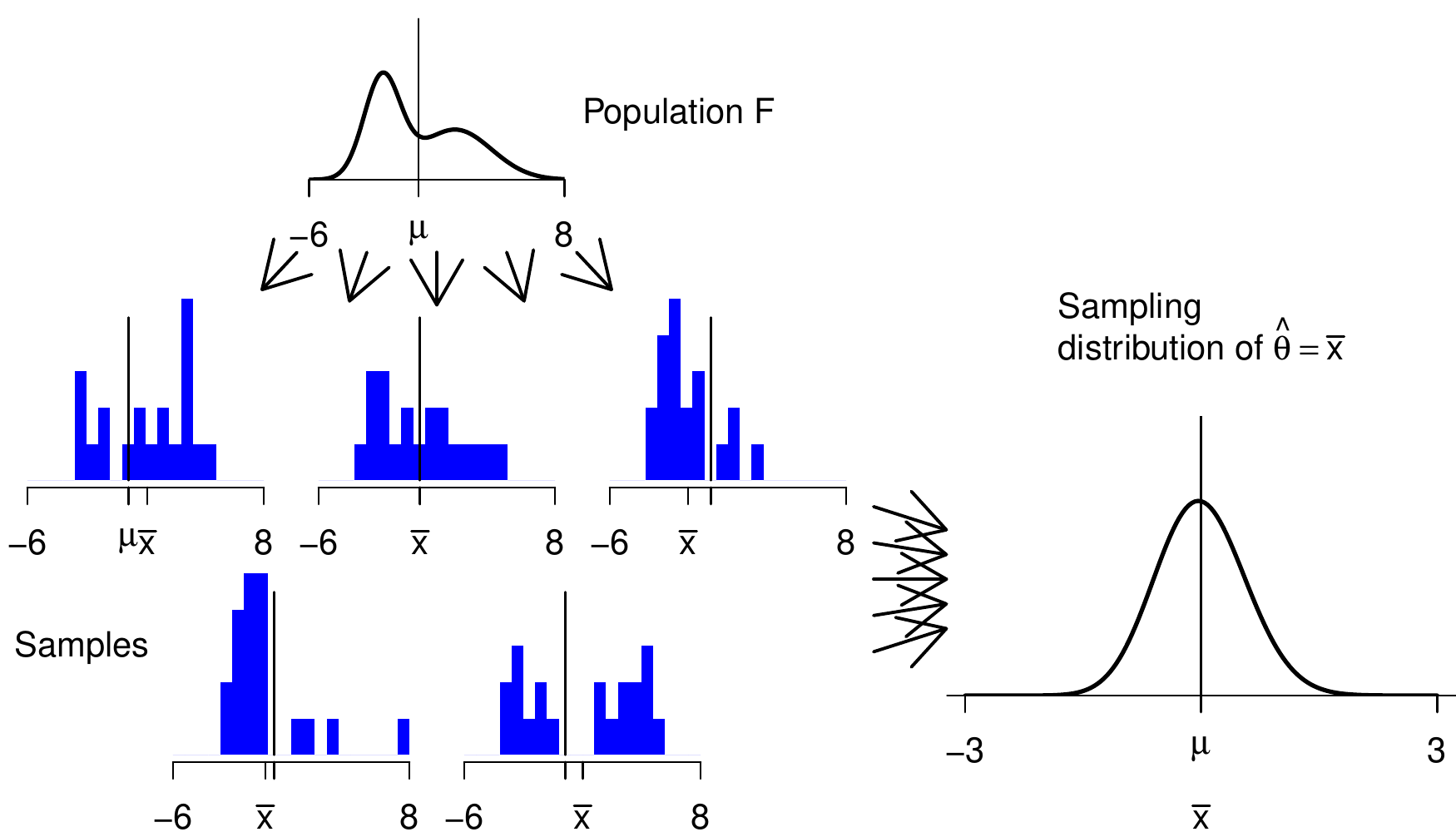}}
\renewcommand{\mycaption}{Ideal world.}
\caption[\mycaption]{\label{figure:idealWorld}
{\em \mycaption}
Sampling distributions are obtained by drawing repeated samples
from the population, computing the statistic of interest for each,
and collecting (an infinite number of) those statistics as the
sampling distribution.
}    \end{figure}

Much of inferential statistics requires estimating something about
the sampling distribution, e.g.\
standard error is an estimate of the standard deviation of that
distribution.
In principle, the sampling distribution is obtained by
\begin{squeezeitemize}
\item Draw samples from the {\em population}.
\item Compute the statistic of interest for each sample
(such as the mean, median, etc.)
\item The distribution of the statistics is the {\em sampling distribution}.
\end{squeezeitemize}
This is shown in Figure~\ref{figure:idealWorld}.

\begin{figure}[\figureplace]    % scripts/BootPix.R
\centerline{\includegraphics[width=\figurewidth]{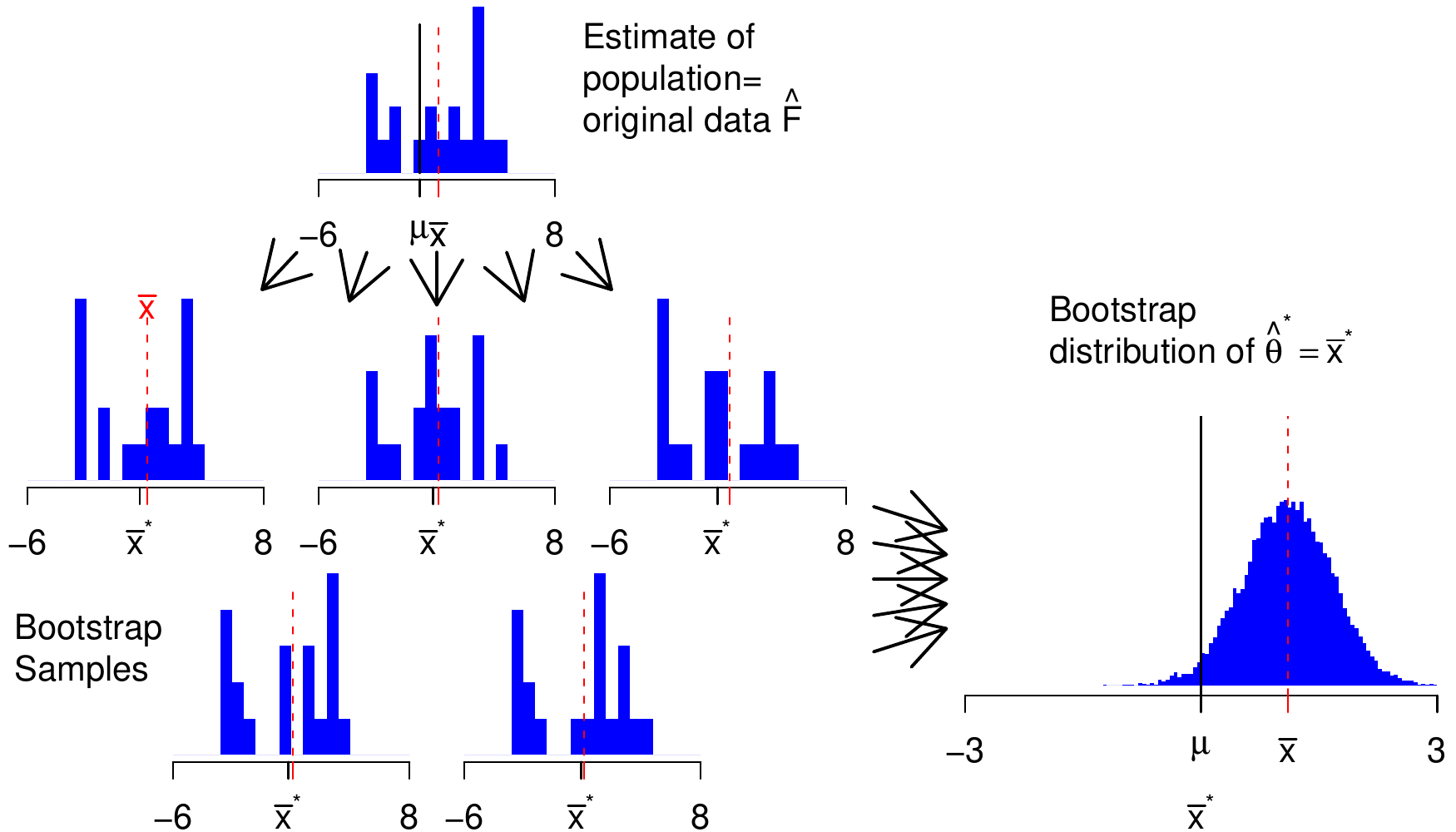}}
\renewcommand{\mycaption}{Bootstrap world.}
\caption[\mycaption]{\label{figure:bootstrapWorld}
{\em \mycaption}
The bootstrap distribution is obtained by drawing repeated samples
from an estimate of the population, computing the statistic of
interest for each, and collecting those statistics.
The distribution is centered at the observed statistic ($\xbar$),
not the parameter ($\mu$).
}    \end{figure}

The problem with this is that we cannot draw arbitrarily many
samples from the population---it is too expensive,
or infeasible because we don't know the population. Instead, we have
only one sample. The bootstrap idea is to draw samples from
an estimate of the population, in lieu of the population:
\begin{squeezeitemize}
\item Draw samples from {\em an estimate of} the population.
\item Compute the statistic of interest for each sample.
\item The distribution of the statistics is the {\em bootstrap distribution}.
\end{squeezeitemize}
This is shown in Figure~\ref{figure:bootstrapWorld}.

\paragraph{Plug-in Principle}
The bootstrap is based on the {\em plug-in principle}---if something
is unknown, then substitute an estimate for it.
This principle is very familiar to statisticians.
For example, the variance for the sample mean is $\sigma/\sqrt{n}$;
when $\sigma$ is unknown we substitute an estimate $s$, the sample
standard deviation. With the bootstrap we take this one step
farther---instead of plugging in an estimate for a single parameter,
we plug in an estimate for the whole distribution.

\paragraph{What to Substitute}
This raises the question of what to substitute for $F$.
Possibilities include:
\begin{squeezedescription}
\item[  Nonparametric bootstrap:] The common bootstrap procedure, the
{\em nonparametric bootstrap}, consists of drawing samples from the empirical
distribution $\Fhat_n$ (with probability $1/n$ on each observation),
i.e.\ drawing
samples with replacement from the data.
This is the primary focus of this article.
\item[  Smoothed Bootstrap:] When we believe the population is continuous,
we may draw samples from a smooth population, e.g.\ from a kernel
density estimate of the population.
\item[  Parametric Bootstrap:] In parametric situations we may estimate
parameters of the distribution(s) from the data, then draw samples from
the parametric distribution(s) with those parameters.
% \item[Other:] We discuss some other methods later, including
% finite-populations and regression.
\end{squeezedescription}
We discuss these and other methods below in
Section~\ref{section:samplingMethods}.

\paragraph{Fundamental Bootstrap Principle}
The fundamental bootstrap principle is that this substitution works.
In most cases, the bootstrap distribution tells us something useful
about the sampling distribution.

There are some things to watch out for, ways that the bootstrap
distribution cannot be used for the sampling distribution.
We discuss some of these below, but one is important enough
to mention immediately:
% in Section~\ref{section:problems} below.

\paragraph{Inference, Not Better Estimates}
{\em The bootstrap distribution is centered at the observed
statistic, not the population parameter},
e.g.~at $\xbar$, not $\mu$.

This has two profound implications. First, it means that we do not use the
bootstrap to get better estimates\footnote{
There are exceptions, where the bootstrap
is used to obtain better estimates, for example in random
forests. These are typically where a bootstrap-like procedure is used to
work around a flaw in the basic procedure.
For example, consider estimating $E(Y|X=x)$
where the true relationship is smooth, but you are limited to
using a step function with relatively few steps.
By taking bootstrap samples and applying
the step function estimation procedure to each, the step boundaries
vary between samples; by averaging across samples the few large
steps are replaced by many smaller ones, giving a smoother estimate.
This is {\em bagging} (bootstrap aggregating).
}.
For example, we cannot use the bootstrap to improve on $\xbar$;
no matter how many bootstrap samples
we take, they are always centered at $\xbar$, not $\mu$.
We'd just be adding random noise to $\xbar$.
Instead we use the bootstrap
to tell how accurate the original estimate is.

Some people are suspicious of the bootstrap, because they think
the bootstrap creates data out of nothing.
(The name ``bootstrap'' doesn't help, since it implies creating
something out of nothing.)
% The name originates in
% {\em The Surprising Adventures of Baron von Munchhausen},
% where at one point the baron is drowning in the bottom of a lake,
% until he thinks to save himself by pulling himself up by the
% bootstraps. OK, Wikipedia says the term doesn't originate there,
% but I like this story.
The bootstrap doesn't create all those bootstrap samples and use
them as if they were more real data; instead it uses them to tell how
accurate the original estimate is.

In this regard it is no different than formula methods that
use the data twice---once to compute an estimate,
and again to compute a standard error for the estimate.
The bootstrap just uses a different approach to estimating the
standard error.

The second implication is that
we do not use quantiles of the bootstrap distribution of $\thetahat^*$
to estimate quantiles of the sampling distribution of $\thetahat$.
% That would only make sense if the sampling distribution does not
% depend on $\theta$---in which case the estimator would be a poor estimator,
% indeed.
Instead, we use the bootstrap distribution to estimate the standard deviation
of the sampling distribution, or the expected value of $\thetahat - \theta$.
Later, in Sections~\ref{section:reversePercentile}
and~\ref{section:bootstrapT}, we will use the bootstrap to estimate
quantiles of $\thetahat - \theta$ and $(\thetahat - \theta) / \SE$.

% TODO: talk about the
% m out of n bootstrap - based on the examples I've seen, it reflects a
% basic misunderstanding, but need to confirm that in more of the literature.

\paragraph{Second Bootstrap Principle}
The second bootstrap principle is to sample with replacement from
the data.

Actually, this isn't a principle at all, but an implementation detail.
We may sample from a parametric distribution, for example.
And even for the nonparametric bootstrap, we sometimes avoid
random sampling.
There are $n^n$ possible samples, or ${2n-1 \choose n}$ if order doesn't
matter; if $n$ is small we could evaluate all of these.
In some cases, like binary data, the number of unique samples is smaller.
We'll call this a {\em theoretical bootstrap} or {\em exhaustive bootstrap}.
But more often this is infeasible, so we draw say 10000
random samples instead; we
call this the {\em Monte Carlo implementation} or {\em sampling implementation}.

We talk about how many samples to draw in
Section~\ref{section:howMany}.

\paragraph{How to Sample}

Normally we should draw bootstrap samples the same way
the sample was drawn in real life,
e.g.~simple random sampling, stratified sampling, or finite-population sampling.

There are exceptions to that rule,
see Section~\ref{section:samplingMethods}.
One is important enough to mention here---to {\em condition on the
observed information}. For example, when comparing samples of size $n_1$
and $n_2$, we fix those numbers, even if the original sampling process
could have produced different counts.

We can also modify the sampling to answer {\em what-if} questions.
Suppose the original sample size was 100, but we draw samples of size
200. That estimates what would happen with samples of that
size---how large standard errors and bias would be, and how wide
confidence intervals would be.
(We would not actually use the confidence intervals from this
process as real confidence intervals; they would imply more precision
than our sample of 100 provides.)
Similarly, we can bootstrap with and without stratification
and compare the resulting standard errors,
to investigate the value of stratification.

Hypothesis testing is another what-if question---if the
population satisfies $H_0$, what would the sampling distribution
(the null distribution) look like?
We may bootstrap in a way that matches $H_0$, by
modifying the population or the sampling method;
see Section~\ref{section:bootstrapHypothesisTesting}.

\boxx{Idea Behind the Bootstrap}
{
The idea behind the bootstrap is to estimate the population,
then draw samples from that estimate, normally sampling the same way as
in real life. The resulting {\em bootstrap distribution}
is an estimate of the sampling distribution.

We use this for inferences, not to obtain better estimates.
It is centered at the statistic (e.g.\ $\xbar$) not the parameter ($\mu$).%
}

\section{Variation in Bootstrap Distributions}
\label{section:variationBootstrapDistributions}

I claimed above that the bootstrap distribution usually tells us
something useful about the sampling distribution, with exceptions.
I elaborate on that now with a series of visual examples,
starting with one where things generally work well, and
three with problems.

% In this section we look at how bootstrap distributions vary,
% and how accurately they represent sampling distributions,
% focusing on how accurately the bootstrap
% represents the spread and shape of the sampling distribution,
% largely using pictures.
%
% This is part of a larger question---how accurate are resampling methods?
% Later we consider how accurate sampling methods are for use
% in hypothesis tests and confidence intervals.

% We consider two questions:
The examples illustrate two questions:
\begin{squeezeitemize}
\item How accurate is the theoretical (exhaustive) bootstrap?
\item How accurately does the Monte Carlo implementation approximate
the theoretical bootstrap?
\end{squeezeitemize}
Both reflect random variation:
\begin{squeezeitemize}
\item The original sample is chosen randomly from the population.
\item Bootstrap resamples are chosen randomly from the original sample.
\end{squeezeitemize}

% We'll look at a series of pictures that show both sources of randomness.

% There is a large literature that looks at the first question
% rigorously and asymptotically; we reference some of that work in other
% sections, particularly Section~\ref{section:confidenceIntervals}
% about confidence intervals, and also refer the reader to
% \citeP{hall92a,shao95} and some sections of \citeP{davi97}, and
% the references therein.

% We conclude this section with a discussion of
% cases where the theoretical
% bootstrap is not accurate, and remedies.
% We return to the question of Monte Carlo accuracy
% in Section~\ref{section:howMany}.

\begin{figure}[\figureplace]    % copied from Chihara/Hesterberg
\centerline{\includegraphics[width=\figurewidth]{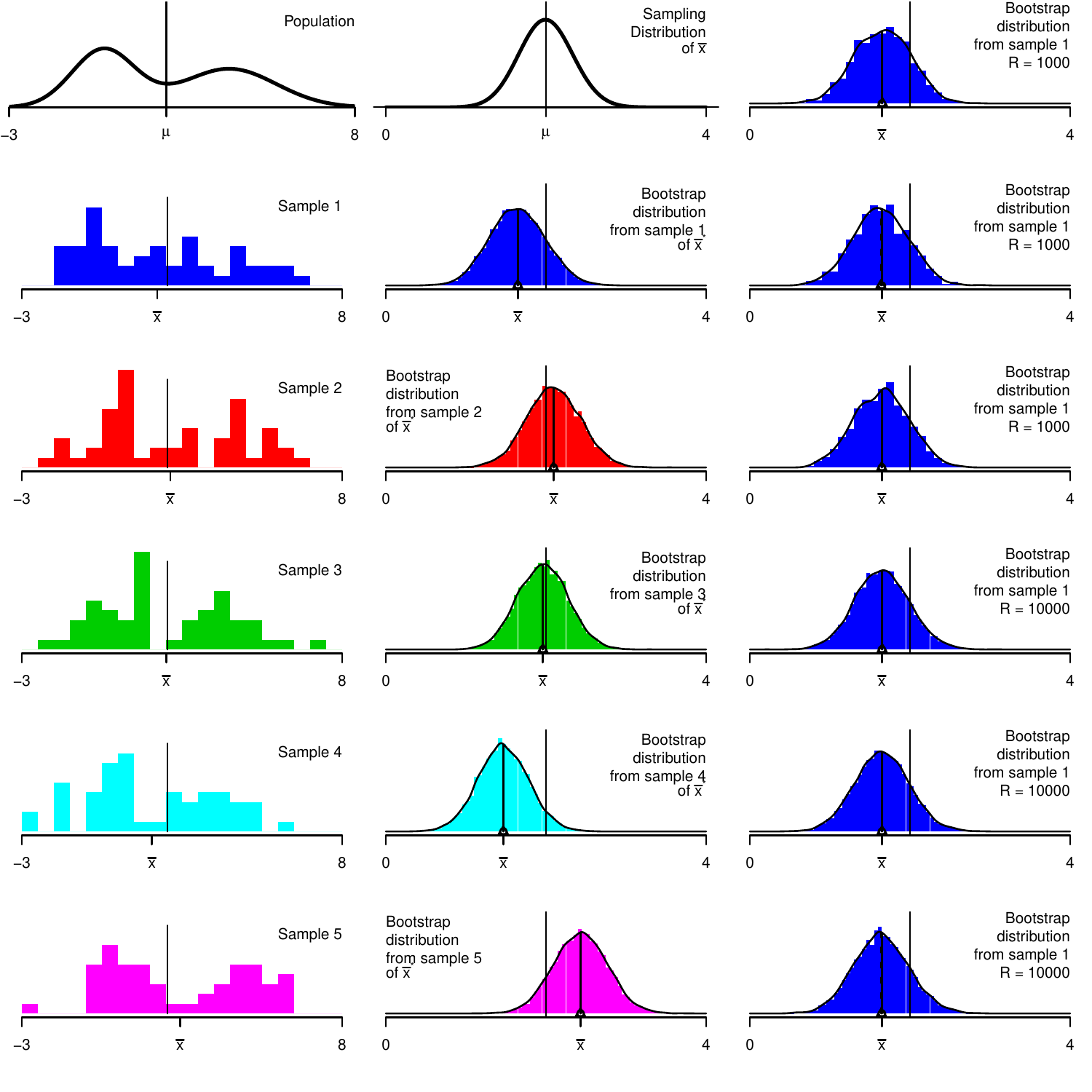}}
\renewcommand{\mycaption}{Bootstrap distribution for the mean, $n=50$.}
\caption[\mycaption]{\label{figure:distributionGraph1}
{\em \mycaption}
The left column shows the population and five samples.
The middle column shows the sampling distribution for $\Xbar$, and bootstrap
distributions of $\Xbar^*$ from each sample, with $r=10^4$.
The right column shows more bootstrap distributions from the
first sample, three with $r=1000$ and three with $r=10^4$.
}    \end{figure}

\subsection{Sample Mean, Large Sample Size:}
Figure~\ref{figure:distributionGraph1} shows
a population and five samples of size 50 from the population in the left
column.
The middle column shows the sampling distribution for the mean
and bootstrap distributions from each sample, based on $r=10^4$
bootstrap samples.
Each bootstrap distribution is centered at the statistic ($\xbar$)
from the corresponding sample rather than being centered at the population
mean $\mu$.  The spreads and shapes of the bootstrap distributions
vary a bit but not a lot.

This informs what the bootstrap distributions may be used for.
The bootstrap does not provide a better estimate of the population
parameter, because no matter how many bootstrap samples
are used, they are centered at $\xbar$,
not $\mu$.  Instead, the bootstrap distributions are useful
for estimating the spread and shape of the sampling distribution.

The right column shows additional bootstrap distributions from the
first sample, with $r=1000$ or $r=10^4$ resamples.
% ; the first four using $r=1000$ resamples,
% and the final using $r=10^4$.
Using more resamples reduces random Monte Carlo variation,
but does not fundamentally change the bootstrap
distribution---it still has the same approximate center, spread,
and shape.

The Monte Carlo variation is much smaller than the variation
due to different original samples.  For many uses, such as
quick-and-dirty estimation of standard errors
or approximate confidence intervals, $r=1000$ resamples is adequate.
However, there is noticeable variability, particularly in the tails
of the bootstrap distributions, so when accuracy matters, $r=10^4$
or more samples should be used.

% Note the difference between using $r=1000$ and $r=10^4$ bootstrap
% samples.  These correspond to drawing samples of size
% $1000$ or $10^4$ observations, with replacement,
% from the theoretical bootstrap distribution.
% Using more samples reduces random Monte Carlo variation,
% but does not fundamentally change the bootstrap
% distribution---it still has the same approximate center, spread,
% and shape.

\begin{figure}[\figureplace]    % copied from Chihara/Hesterberg
\centerline{\includegraphics[width=\figurewidth]{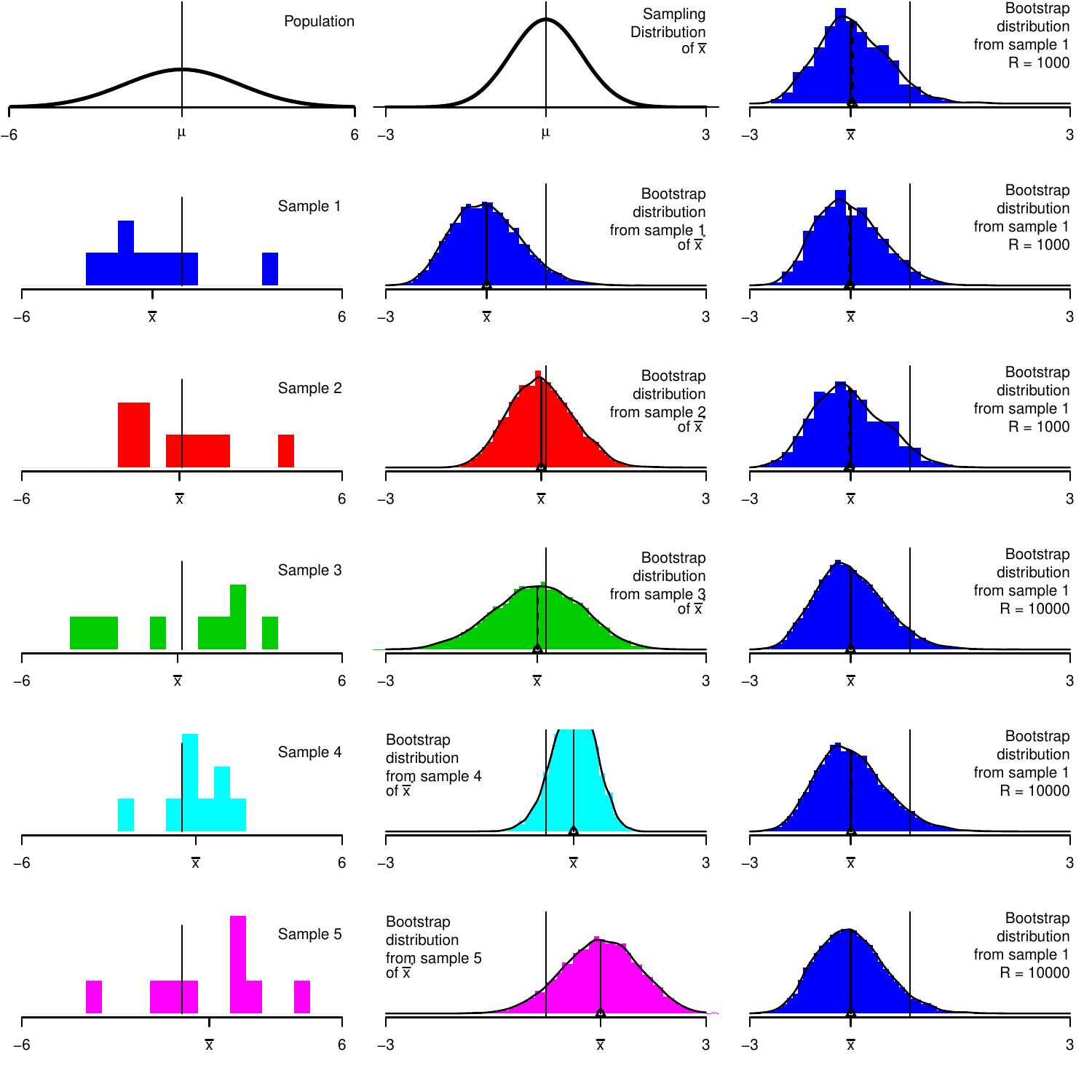}}
\renewcommand{\mycaption}{Bootstrap distributions for the mean, $n=9$.}
\caption[\mycaption]{\label{figure:distributionGraph2}
{\em \mycaption}
The left column shows the population and five samples.
The middle column shows the sampling distribution for $\Xbar$, and bootstrap
distributions of $\Xbar^*$ from each sample, with $r=10^4$.
The right column shows more bootstrap distributions from the
first sample, three with $r=1000$ and three with $r=10^4$.
}    \end{figure}

\subsection{Sample Mean: Small Sample Size}
\label{section:smallSamplePicture}
Figure~\ref{figure:distributionGraph2} is similar
to Figure~\ref{figure:distributionGraph1}, but
for a smaller sample size, $n=9$ (and a different population).
As before, the bootstrap distributions are centered at the
corresponding sample means, but now the spreads and shapes
of the bootstrap distributions vary substantially, because
the spreads and shapes of the samples vary substantially.
As a result, bootstrap confidence interval widths vary substantially
(this is also true of non-bootstrap confidence intervals).
As before, the Monte Carlo variation is small and may be reduced
with more samples.
% using $r=10^4$ or

While not apparent in the pictures, bootstrap distributions
tend to be too narrow, by a factor of $\sqrt{(n-1)/n}$
%(exact for linear statistics, approximate otherwise).
for the mean;
the theoretical bootstrap standard error is
$s_B = \sigmahat/\sqrt{n} = \sqrt{(n-1)/n}(s/\sqrt{n})$.
The reason for this goes back to the plug-in principle;
the empirical distribution has variance
$\Var_\Fhat(X) = \sigmahat^2 = (1/n)\sum (x_i-\xbar)^2$,
not $s^2$.
For example, the
bootstrap standard error for the TV Basic mean is $0.416$,
while $s/\sqrt{10} = 0.441$.

In two-sample or stratified sampling situations,
this {\em narrowness bias}
depends on the individual sample or strata sizes, not the
combined size. This can result in severe narrowness bias.
For example, the first bootstrap short course I ever taught
was for the U.K.~Department of Work and Pensions, who wanted to
bootstrap a survey they had performed to estimate welfare cheating.
They used a stratified sampling procedure that resulted in two
subjects in each stratum---then the bootstrap standard error
would be too small by a factor of $\sqrt{1/2}$.
There are remedies, see
Section~\ref{section:samplingMethods}.
For Stat 101 I recommend warning students about the issue;
for higher courses you may discuss the remedies.

The narrowness bias and the variability in spread affect
confidence interval coverage badly in small samples, see
Section~\ref{section:stat101}.

\begin{figure}[\figureplace]    % copied from Chihara/Hesterberg
\centerline{\includegraphics[width=\figurewidth]{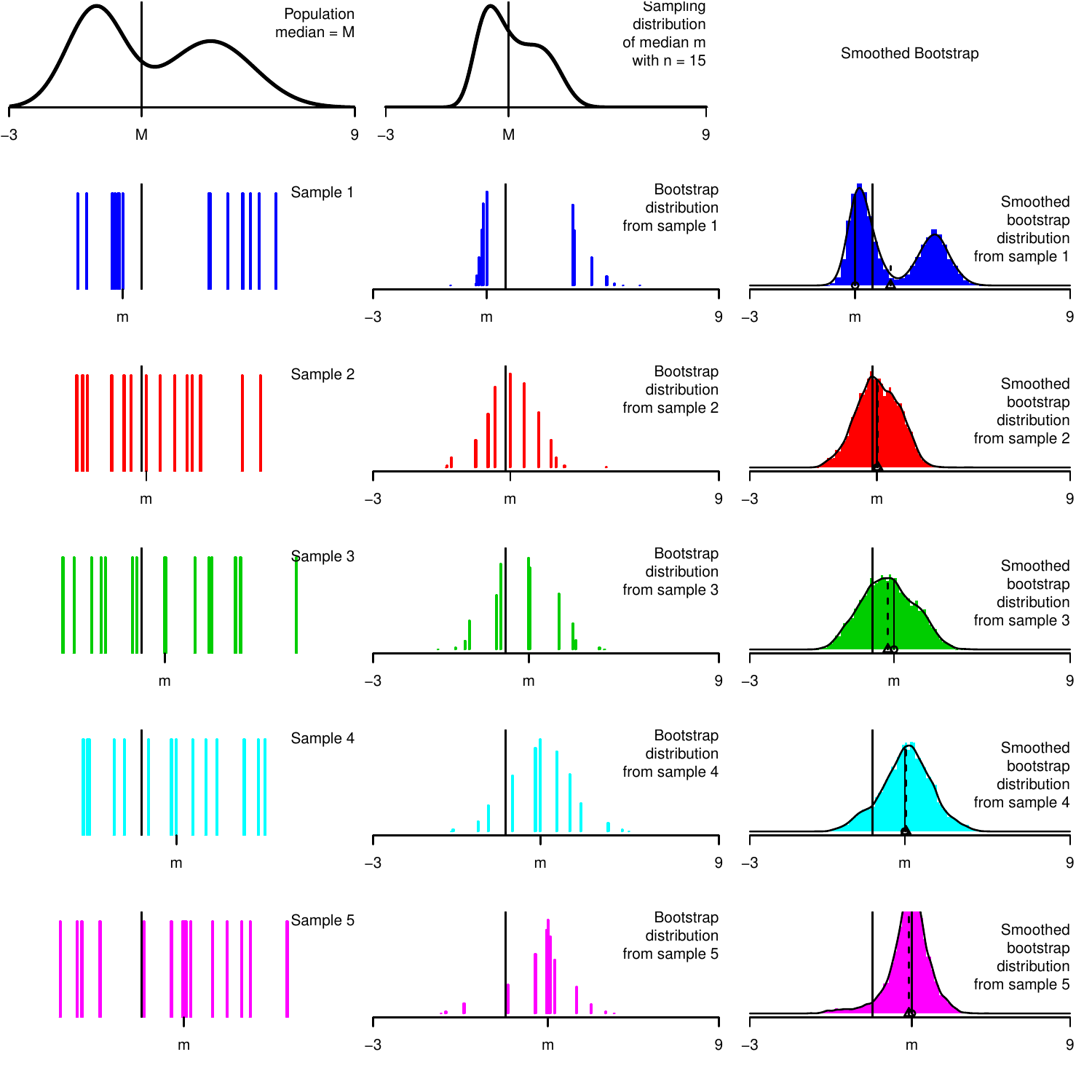}}
\renewcommand{\mycaption}{Bootstrap distributions for the median, $n=15$.}
\caption[\mycaption]{\label{figure:distributionGraph3}
{\em \mycaption}
The left column shows the population and five samples.
The middle column shows the sampling distribution, and bootstrap
distributions from each sample, with $r=10^4$.
The right column shows smoothed bootstrap distributions,
with kernel sd $s/\sqrt{n}$ and $r=10^4$.
}    \end{figure}

%\vspace*{-6pt}
\subsection{Sample Median}
Now turn to Figure~\ref{figure:distributionGraph3} where the
statistic is the sample median.  Here the bootstrap distributions
are poor approximations of the sampling distribution.
The sampling distribution is continuous, but
the bootstrap distributions are discrete---since $n$ is odd,
the bootstrap sample median is always one of the original data points.
The bootstrap distributions are very sensitive to the sizes of
gaps among the observations near the center of the sample.

The ordinary bootstrap tends not to work well for statistics such as the
median or other quantiles in small samples,
that depend
heavily on a small number of observations out of a larger sample;
the bootstrap distribution in turn depends heavily on a small number
of observations (though different ones in different bootstrap samples,
so bootstrapping the median of large samples works OK).
The shape and scale of the bootstrap distribution may be very different
than the sampling distribution.

Curiously, in spite of the ugly bootstrap distribution,
the bootstrap percentile interval for the median is not bad
\citeP{efro82}.
For odd $n$, percentile interval endpoints fall on one of the
observed values. Exact interval endpoints also fall on one of the
observed values (order statistics), and for a 95\% interval
those are typically the same or adjacent order statistics as the
percentile interval.

The right column shows the use of a
{\it smoothed bootstrap} \citeP{silv87,hall89b},
drawing samples from a density estimate based on the data, rather than
drawing from the data itself. See Section~\ref{section:smoothedBootstrap}.
It improves things somewhat, though it is still not great.

The bootstrap fails altogether for estimating the sampling
distribution for $\mbox{max}(x)$.

\begin{figure}[\figureplace]   % scripts/distributionGraph4.R
\centerline{\includegraphics[width=\figurewidth]{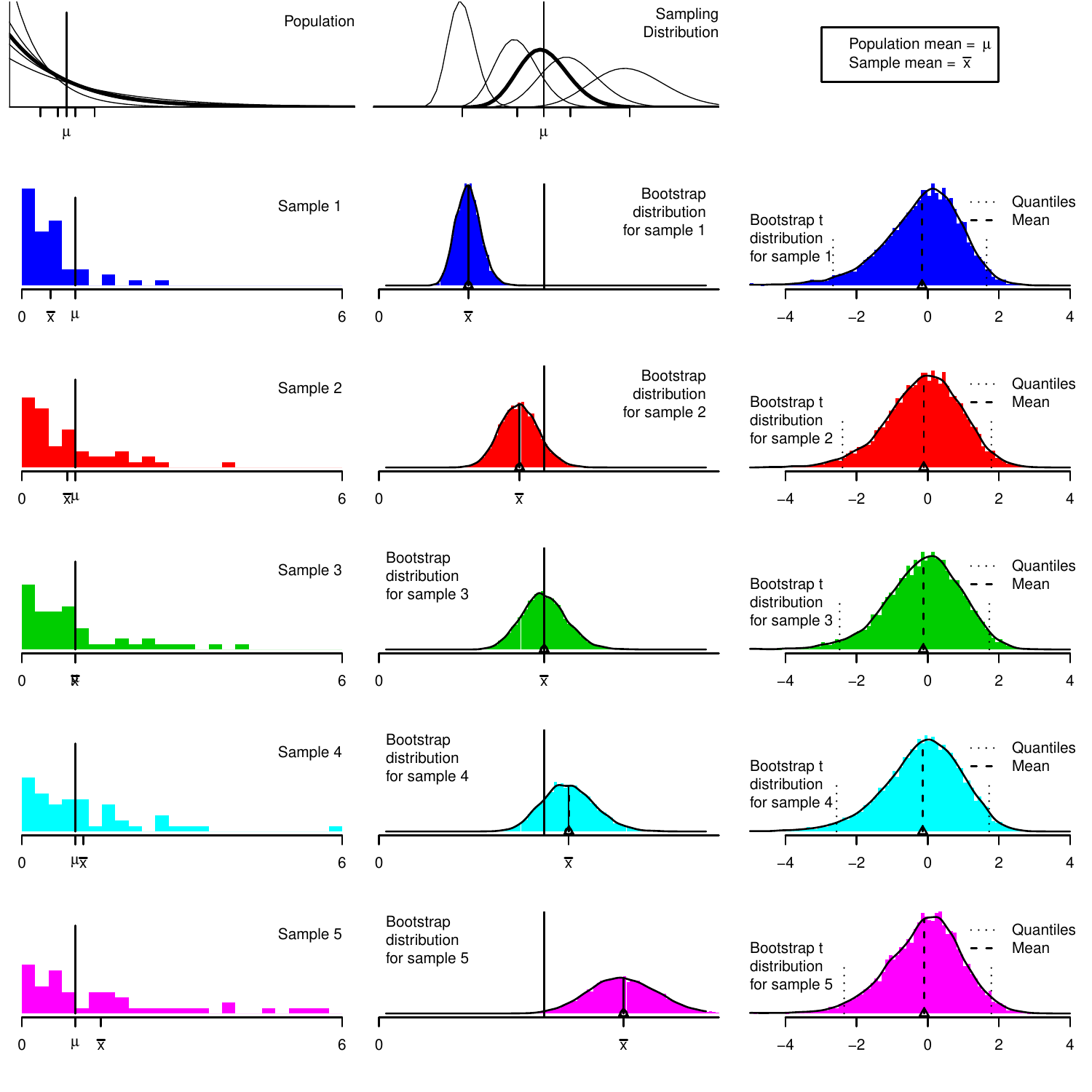}}
\renewcommand{\mycaption}{Bootstrap distributions for the mean, $n=50$, exponential population.}
\caption[\mycaption]{\label{figure:distributionGraph4}
{\em \mycaption}
The left column shows the population and five samples.
%The thin lines in the top figure are exponential distributions with means
%equal to the sample means.
(These samples are selected from a larger set of random samples, to
have means spread across the range of sample means, and
average standard deviations conditional on the means.)
The middle column shows the sampling distribution and bootstrap
distributions from each sample.
%The thin lines in the top figure are sampling distributions corresponding
%to the different exponential distributions.
The thinner curves in the top two figures are populations and sampling
distributions with means equal to the sample means.
Each bootstrap distribution is like the sampling distribution
with $\mu$ set to match $\xbar$.
The right column shows bootstrap~$t$ distributions.
}    \end{figure}

\subsection{Mean-Variance Relationship}
\label{section:acceleration}
In many applications, the spread or shape of the sampling distribution
depends on the parameter of interest.
For example, the binomial distribution spread and shape depend on $p$.
Similarly, for the mean from an exponential
distribution, the standard deviation of
the sampling distribution is proportional to the population mean.

This is reflected in bootstrap distributions.
Figure~\ref{figure:distributionGraph4} shows samples
and bootstrap distributions from an exponential population.
There is a strong dependence between $\xbar$ and the
corresponding bootstrap SE.

This has important implications for confidence intervals;
good confidence intervals need to reach many (short) SE's
to the right to avoid missing $\theta$ too often in that direction,
and should reach fewer (long) SE's to the left.
We discuss this more in
Section~\ref{section:confidenceIntervals}.

This mean-variance relationship in samples normally corresponds
to a similar mean-variance relationship between the
parameter and variance of the sampling distribution.
For example, see the five sampling distributions in the top
middle of Figure~\ref{figure:distributionGraph4}.
We call such a relationship {\em acceleration}.

The right column of Figure~\ref{figure:distributionGraph4} shows
bootstrap distributions of the $t$~statistic,
defined in Section~\ref{section:bootstrapT}.
These distributions are much less sensitive to the original sample.

% Note the difference between using $r=1000$ and $r=10^4$ bootstrap
% samples.  These correspond to drawing samples of size
% $1000$ or $10^4$ observations, with replacement,
% from the theoretical bootstrap distribution.
% Using more samples reduces random Monte Carlo variation,
% but does not fundamentally change the bootstrap
% distribution---it still has the same approximate center, spread,
% and shape.

There are other applications where sampling distributions
depend strongly on the parameter;
for example sampling distributions for chi-squared statistics
depend on the non-centrality parameter. Similarly for statistics
for estimating the number of modes of a population
Use caution when bootstrapping in these applications; the bootstrap
distribution may be very different than the sampling distribution.

\subsection{Summary of Visual Lessons}

The bootstrap distribution reflects the original sample.
If the sample is narrower than the population,
the bootstrap distribution is narrower than the sampling distribution.

Typically for large samples the data represent the population
well; for small samples they may not.
%
% For most statistics, almost all the variation in bootstrap
% distributions comes from the selection of the original sample from the
% population.  Reducing this variation requires collecting a larger
% original sample.
%
% For small samples, the data may not provide an accurate estimation
% of the population, for use in bootstrapping.
Bootstrapping does not overcome the weakness of small samples as a
basis for inference.

Indeed, for the very smallest samples, you may not want to bootstrap;
it may be
better to make additional assumptions such as smoothness or
a parametric family.  When there is a lot of data
(sampled randomly from a population)
we can trust the
data to represent the shape and spread of the population; when
there is little data we cannot.

\boxx{Visual Lessons about Bootstrap Distributions}
{
The bootstrap distribution reflects the data.
For large samples the data represent the population
well; for small samples they may not.

The bootstrap may work poorly when the statistic, and sampling
distribution, depend on a small number of observations.

Using more bootstrap samples reduces the variability of bootstrap
distributions, but does not fundamentally change the center,
spread, or shape of the bootstrap distribution.%
}

Looking ahead, two things matter for accurate inferences:
\begin{squeezeitemize}
\item
how close the bootstrap distribution is to the sampling distribution
(in this regard, the bootstrap~$t$ has an advantage,
judging from Figure~\ref{figure:distributionGraph4});
\item
some procedures better allow for the fact that there
is variation in samples. For example, the usual formula $t$ tests
and intervals allow for variation in $s$ by using $t_{\alpha/2}$
in place of $z_\alpha$; we discuss a bootstrap analog
in Section~\ref{section:expandedPercentile}.
\end{squeezeitemize}
%
% Some bootstrap procedures are more accurate than
% common non-bootstrap procedures.
% We return to these points later.

It appears that the bootstrap resampling process using 1000 or more
resamples introduces little additional variation, but for good
accuracy use 10000 or more.
Let's consider this issue more carefully.

\subsection{How many bootstrap samples?}
\label{section:howMany}

We suggested above that using 1000 bootstrap samples for
rough approximations, or $10^4$ or more for better accuracy.
This is about Monte Carlo accuracy---how well
the usual random sampling implementation of the bootstrap
approximates the theoretical bootstrap distribution.

A bootstrap
distribution based on $r$ random samples corresponds to drawing
$r$ observations with replacement from the theoretical bootstrap distribution.

Brad Efron, inventor of the bootstrap,
suggested in 1993 that $r=200$, or even as few as $r=25$,
suffices for estimating standard errors and
that $r=1000$ is enough for confidence intervals \citeP{efro93}.

We argue that more resamples are appropriate, on two grounds.
First, those criteria were developed when computers were much slower;
with faster computers it is easier to take more resamples.

Second, those criteria were developed using arguments that combine
the random variation due to the original random sample
with the extra variation due to the Monte Carlo implementation.
We prefer to treat the data
as given and look just at the variability due to the implementation.
Data is valuable, and computer time is cheap. Two people analyzing the
same data should not get substantially
different answers due to Monte Carlo variation.

\paragraph{Quantify accuracy by formulas or bootstrapping}
We can quantify the Monte Carlo error in two ways---using
formulas, or by bootstrapping.
For example, in permutation testing we need to estimate the fraction
of observations that exceed the observed value; the
Monte Carlo standard error
is approximately $\sqrt{\phat(1-\phat)/r}$, where $\phat$ is the
estimated proportion. (It is a bit more complicated because we add
1 to the numerator and denominator, but this is close.)

In bootstrapping, the bias estimate
depends on $\overline{\thetahat^*}$, a sample average of $r$ values;
the Monte Carlo standard error for this is $s_B/\sqrt{r}$ where $s_B$ is the
sample standard deviation of the bootstrap distribution.

We can also bootstrap the bootstrap!
We can treat the $r$ bootstrap replicates
like any old sample, and bootstrap from that sample.
For example, to estimate the Monte Carlo SE for the 97.5\% quantile of the
bootstrap distribution
(the endpoint of a bootstrap percentile interval),
we draw samples of size $r$ from the $r$ observed values
in the bootstrap distribution, compute the quantile for each,
and take the standard deviation of those quantiles as the Monte Carlo SE.

For example, the 95\% percentile interval for the mean of the CLEC
data is $(10.09, 25.40)$ (from $r=10^4$ resamples);
the Monte Carlo standard errors for those endpoints
are $0.066$ and $0.141$. The syntax for this using the {\em resample}
package \citeP{hest14} is
\begin{verbatim}
bootCLEC <- bootstrap(CLEC, mean, B = 10000)
bootMC <- bootstrap(bootCLEC$replicates,
              quantile(data, probs = c(.025, .975), type = 6))
\end{verbatim}
(The resample package uses type=6 when computing quantiles, for
more accurate confidence intervals.)

\paragraph{Need $r \ge 15000$ to be within 10\%}
Now, let's use those methods to determine how large $r$ should
be for accurate results.
We consider two-sided 95\% confidence intervals and
tests with size 5\%.

Consider tests first.
We'll determine the $r$ necessary to have a 95\% chance
that the Monte Carlo estimate of the $P$-value
is within 10\% when the exhaustive
one-sided $P$-value is 2.5\%,
i.e.\ 95\% chance that the estimated $P$-value is between
2.25\% and 2.75\%.

For a permutation test, let $q = G^{-1}(0.025)$ be the true
2.5\% quantile of the permutation distribution.
Suppose we observe $\thetahat = q$, so the true (exhaustive)
$P$-value is $0.025$.
The standard deviation for the estimated $P$-value is
$\sqrt{0.025 \cdot 0.975/r}$, so we solve
$1.96 \sqrt{0.025\cdot 0.975/r} \le 0.025/10$, or
$r \ge 14982$.
% Calculations in scripts/coverage.R

Similar results hold for a bootstrap percentile or bootstrap~$t$ confidence
interval. If $q$ is the true 2.5\% quantile of the theoretical
bootstrap distribution (for $\thetahat^*$ or $t^*$, respectively),
for $\Ghat(q)$ to fall between 2.25\% and 2.75\%
with 95\% probability requires $r \ge 14982$.

For a $t$ interval with bootstrap SE, $r$ should be large enough
that variation in $s_B$ has a similar small effect on coverage.
This depends on $n$ and the shape of the bootstrap distribution,
but for a rough approximation we assume that
(1) $n$ is large and
hence we want 95\% central probability that
$z_{0.0275}/\zSub < s_B/\sigma_B < z_{0.0225}/\zSub$
where $\sigma_B$ is the standard deviation of the theoretical bootstrap
distribution, and
(2) the bootstrap distribution is approximately normal,
so $(r-1) (s_B/\sigma_B)^2$ is approximately
chi-squared with $r-1$ degrees of freedom.
By the delta method, $s_B/\sigma_B$ has approximate variance $1/(2r)$.
For the upper bound, we set
$1.96 / \sqrt{2r} < |z_{0.0275}/\zSub - 1|$;
this requires $r \ge 4371$. The calculation for the lower bound is similar,
and a slightly smaller $r$ suffices.
% (1.96 / abs(qnorm(.0225)/qnorm(.025) - 1))^2 / 2 = 3694

% The expanded percentile interval would need more.
% The reverse percentile method would need the same as the percentile method.

Rounding up, we need $r \ge 15000$ for simulation variability to
have only a small effect on the percentile and bootstrap~$t$,
and $r \ge 5000$ for the $t$ with bootstrap SE.
While students may not need this level of accuracy, it is good to
get in the habit of doing accurate simulations.
Hence I recommend $10^4$ for routine use.
And, for statistical practice, if the results with $r=10^4$ are
borderline, then increase $r$ to reduce the Monte Carlo error.
We want decisions to depend on the data, not Monte Carlo variability
in the resampling implementation.

We talk below about coverage accuracy of confidence intervals.
Note that a large $r$ isn't necessary for an interval to have the
right coverage probability. With smaller $r$, sometimes an
interval is too short, sometimes too long, and it roughly balances out
to give the same coverage as with larger $r$.
But that is like flipping a coin---if heads then compute a
96\% interval and if tails a 94\% interval; while it may have the right
overall coverage, the endpoints are variable in a way that does not
reflect the data.

\boxx{Use 10000 or More Resamples}
{
When the true one-sided permutation test $P$-value is 2.5\%,
we need $r \ge 15000$ to have a 95\% chance that the estimated
$P$-value is between 2.25\% and 2.75\% (within 10\% of the true value).

Similarly, we need $r \ge 15000$ to reduce Monte Carlo variability
in the percentile interval endpoints to 10\%,
and $r \ge 5000$ for a $t$ interval with bootstrap SE.

We suggest $r=10000$ for routine use, and more when accuracy matters.

These recommendations are much larger than previous recommendations.
Statistical decisions should depend on the data,
not Monte Carlo variability.%
}

% The effect of small $r$ on coverage probability is smaller.
% Small r results in intervals sometimes being too long, sometimes
% too short, and this largely cancels out.
% For the percentile interval, if we use 999 resamples,
% and for endpoints use the \alpha/2*(r+1) and r+1-\alpha/2)*(r+1)
% order statistics
% then fraction \alpha/2 of the bootstrap distribution is outside those
% order statistics regardless how small r is.

\section{Transformation, Bias, and Skewness}
\label{section:biasSkewness}

Three important issues for estimation, confidence intervals,
and hypothesis tests are
{\em transformations},
{\em bias} (of the statistic) and
{\em skewness} (of the population, and the sampling distribution).
We'll look at these in this section, and how they affect the
accuracy of permutation tests and $t$~tests,
and take a first look at how they affect confidence intervals,
with a more complete look in the next section.
We also discuss {\em functional statistics}, and how non-functional
statistics can give odd results when bootstrapping.

\subsection{Transformations}

Table~\ref{table:relativeRisk} gives rates of cardiovascular disease
for subjects with high or low blood pressure.
The high-blood pressure group was 2.12 times as likely to develop
the disease.

\begin{table}[\tableplace]
 \centering
\begin{tabular}{cc}\hline
  Blood Pressure & Cardiovascular Disease \\
  High  & 55/3338 = 0.0165 \\
  Low   & 21/2676 = 0.0078 \\
  Relative risk & 2.12 \\\hline
\end{tabular}
\renewcommand{\mycaption}{Relative risk of cardiovascular disease.}
\caption[\mycaption]{\label{table:relativeRisk}
  {\em \mycaption}
}
\end{table}

\begin{figure}[\figureplace]    % scripts/relativeRisk.R
\centerline{\includegraphics[width=\figurewidth]{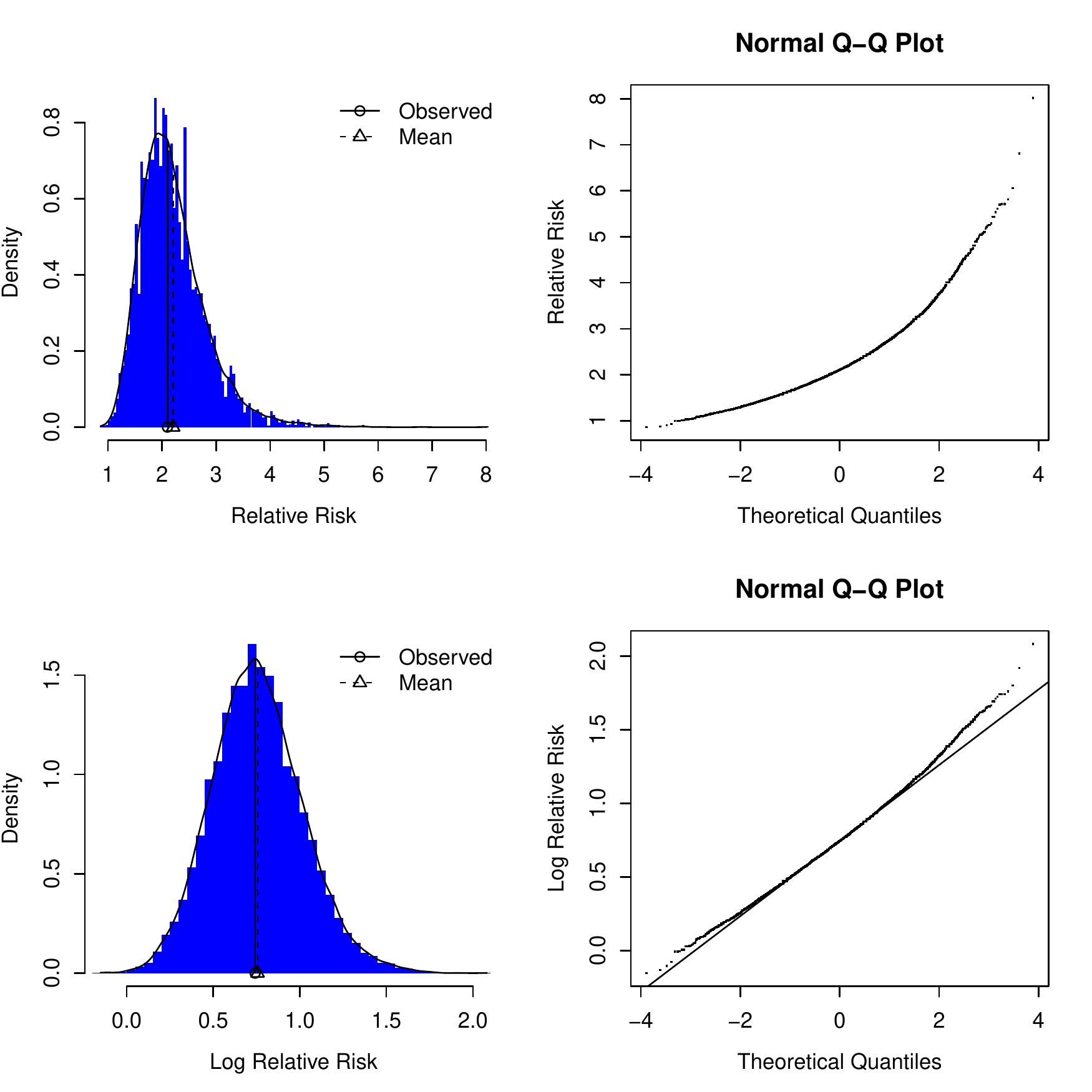}}
\renewcommand{\mycaption}{Bootstrap relative risk.}
\caption[\mycaption]{\label{figure:bootRelativeRisk}
{\em \mycaption}
The top panel shows the bootstrap distribution for relative risk.
The bottom panel shows the bootstrap distribution for log relative risk.
}    \end{figure}

Figure~\ref{figure:bootRelativeRisk} shows the bootstrap distribution
for relative risk.
The distribution is highly skewed, with a long right
tail. Also shown is the bootstrap distribution for log relative risk;
this is less skewed. Both distributions exhibit bias;
the summary statistics are:
\begin{verbatim}
                   Observed     SE   Mean   Bias
 Relative Risk       2.0996 0.6158 2.2066 0.1070
 Log Relative Risk   0.7417 0.2625 0.7561 0.0143
\end{verbatim}
% I lined those up manually, and removed some digits (without rounding).

One desirable property for confidence intervals is
{\em transformation invariance}---if
$h$ is a monotone transformation and $\psi = h(\theta)$, a procedure
is transformation invariant if the endpoints of the confidence interval
for $\psi$ are $h(L)$ and $h(U)$, where $(L, U)$ is the
interval for $\theta$.
Transformation invariance means that people taking different
approaches get equivalent results.

The bootstrap percentile interval is transformation invariant.
If one student does a confidence interval for
$\theta =$ relative risk, and another
for
$\psi =$ log relative risk, they get equivalent answers;
the percentile interval for relative risk is
$(1.31, 3.70)$, and for log-relative-risk is
$(0.273, 1.31) = (\log(1.31), \log(3.70))$.

In contrast, a $t$ interval is not transformation invariant.
The $t$ interval with bootstrap SE for relative risk is
$2.0996 \pm 1.96 \cdot 0.6185 = (0.893, 3.306)$;
taking logs gives $(-.113, 1.196)$.
Those differ from $t$ endpoints for log relative risk,
$0.7417 \pm 1.96 \cdot 0.2625 = (0.227, 1.256)$.

Using an interval that is not transformation invariant means that
you can choose the transformation to get the answer you want.
Do it one way and the interval includes zero;
do it the other way and the interval excludes zero.

\subsection{Bias}
\label{section:bias}

The bootstrap estimate of bias derives from the plug-in principle.
The bias $B$ of a statistic is
\begin{equation}
  B = E(\thetahat) - \theta = E_F(\thetahat) - \theta(F)
  \label{eqn.bias}
\end{equation}
where $E_F$ indicates sampling from $F$, and $\theta(F)$ is
the parameter for population $F$.
The bootstrap substitutes $\Fhat$ for $F$, to give
\begin{equation}
\hat B = E_\Fhat(\thetahat^*) - \theta(\Fhat)
   = \overline{\thetahat^*} - \thetahat.
  \label{eqn.biasBoot}
\end{equation}
The bias estimate is the mean of the bootstrap
distribution, minus the observed statistic.

The relative risk and
log relative risk statistics above are biased
(see the summary statistics above).

\begin{figure}[\figureplace]    % scripts/regression.R
\centerline{\includegraphics[width=\figurewidth]{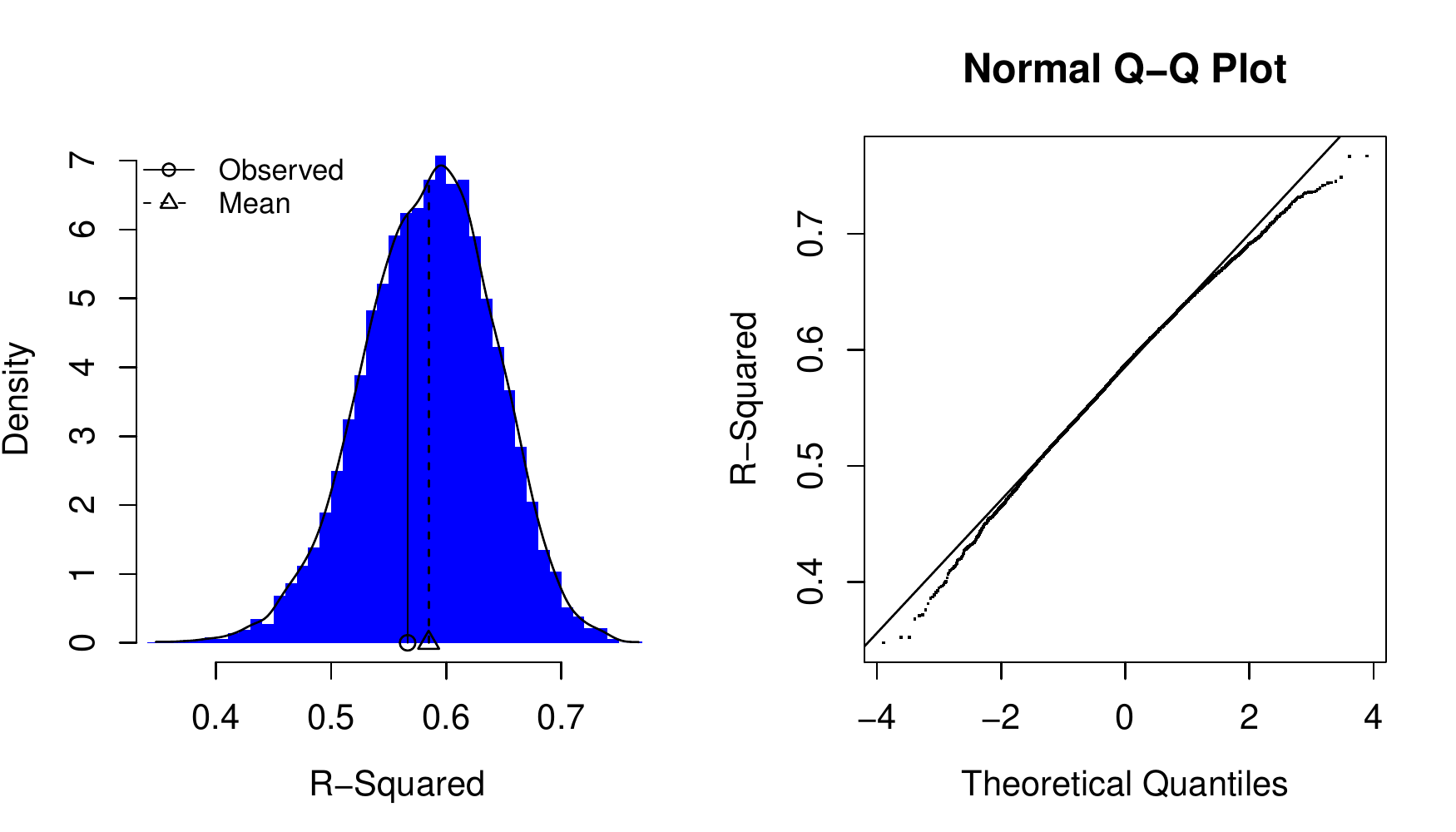}}
\renewcommand{\mycaption}{Bootstrap distribution for R-Squared in regression.}
\caption[\mycaption]{\label{figure:bootstrapRSquared}
{\em \mycaption}
Artificial data, $n=100$ and $p=5$.
}    \end{figure}

\paragraph{Regression R-Squared}
Another example of bias is unadjusted R-squared in regression.
Figure~\ref{figure:bootstrapRSquared} shows a bootstrap
for unadjusted $R^2$ for an artificial dataset
with $n=100$ and $p=5$.
The summary statistics are:
\begin{verbatim}
     Observed         SE      Mean       Bias
R^2 0.5663851 0.05678944 0.5846771 0.01829191
\end{verbatim}
% Here we resampled observations; this is defined
% in Section~\ref{section:bootstrapRegression}.

\subsubsection{Bias-Adjusted Estimates}

We may use the bias estimate to produce a bias-adjusted estimate,
$\thetahat - \hat B = 2 \thetahat - \overline{\thetahat^*}$.

We generally do not do this---bias estimates can have high
variability, see \citeP{efro93}.
%, who also give a bias estimate that is
%less sensitive to Monte Carlo variability.
Instead, just being aware that a statistic is biased may help us proceed
more appropriately.

Bias is another reason that we do not use the bootstrap
average $\overline{\thetahat^*}$ in place of $\thetahat$---it would
have double the bias of $\thetahat$.

The bootstrap BCa confidence interval \citeP{efro87} makes use
of another kind of bias estimate, the fraction of the bootstrap
distribution that is $\le \thetahat$. This is not sensitive to
transformations.
It is related to {\em median bias}---a statistic is median unbiased
if the median of the sampling distribution is $\theta$.

\subsubsection{Causes of Bias}
\label{section:causesOfBias}
There are three common causes of bias.
In the first of these bias correction would be harmful, in the second
it can be helpful, and in the third the bias would not be apparent
to the bootstrap.
The differences are also important for confidence intervals.

One cause of bias relates to nonlinear transformations, as in the
relative risk example above;
$E(\hat p_1/\hat p_2) = E(\hat p_1) E(1/\hat p_2) \neq
E(\hat p_1) /E(\hat p_2)$.
In this case the median bias is near zero, but the mean bias
estimate $\overline{\thetahat^*} - \thetahat$
can be large and have high variability,
and is strongly dependent on how close the denominator is to zero.
Similarly,
$E(\log(\hat p_1/\hat p_2)) =
 E(\log(\hat p_1) - \log(\hat p_2)) \neq
 \log(E(\hat p_1)) - \log(E(\hat p_2))$.

Similarly, $s^2$ is unbiased but $s$ is not;
$E(s) \neq \sqrt{E(s^2)} = \sigma$.

Another cause is bias due by optimization---when one or more parameters
are chosen to optimize some measure, then the estimate of that measure
is biased. The $R^2$ example falls into this category, where the
regression parameters are chosen to maximize $R^2$;
the estimated $R^2$ is higher than if we used the true unknown parameters,
and the unadjusted $R^2$ is biased upward.
Another example is sample variance.
The population variance is $E((X-\mu)^2)$.
An unbiased estimate of that is
$(1/n) \sum (X_i - \mu)^2$.
Replacing $\mu$ with the value that
minimizes that quantity, $\xbar$, gives
a biased estimate $\sigmahat^2 = (1/n) \sum (X_i - \xbar)^2$.

The optimization bias can be large in stepwise regression, where both the
variable selection and parameter estimates optimize.

% Remove the following, at least for now.
% The optimization bias can be enormous
% in portfolio optimization.
% If the loadings on many stocks or other portfolio components are
% chosen to optimize some criterion, like minimizing volatility subject
% to a given expected gain, or maximizing expected gain subject
% to a given volatility, then the resulting measures are biased,
% often severely so. The whole estimate of the
% {\em efficient frontier} \citeP{mark52}
% is biased.%
% \footnote{
% This makes a nice picture that your students interested in finance
% will enjoy; plot the estimated efficient frontier,
% then add curves from bootstrap samples.
% I'm not including this because of a patent.
% The idea of applying the bootstrap to portfolio optimization
% is obvious; it is just another application of the bootstrap.
% I did it long ago, then heard about the patent.
% In my opinion this should not be patentable.
% }

\begin{figure}[\figureplace]    % scripts/regression.R
\centerline{\includegraphics[width=\figurewidth]{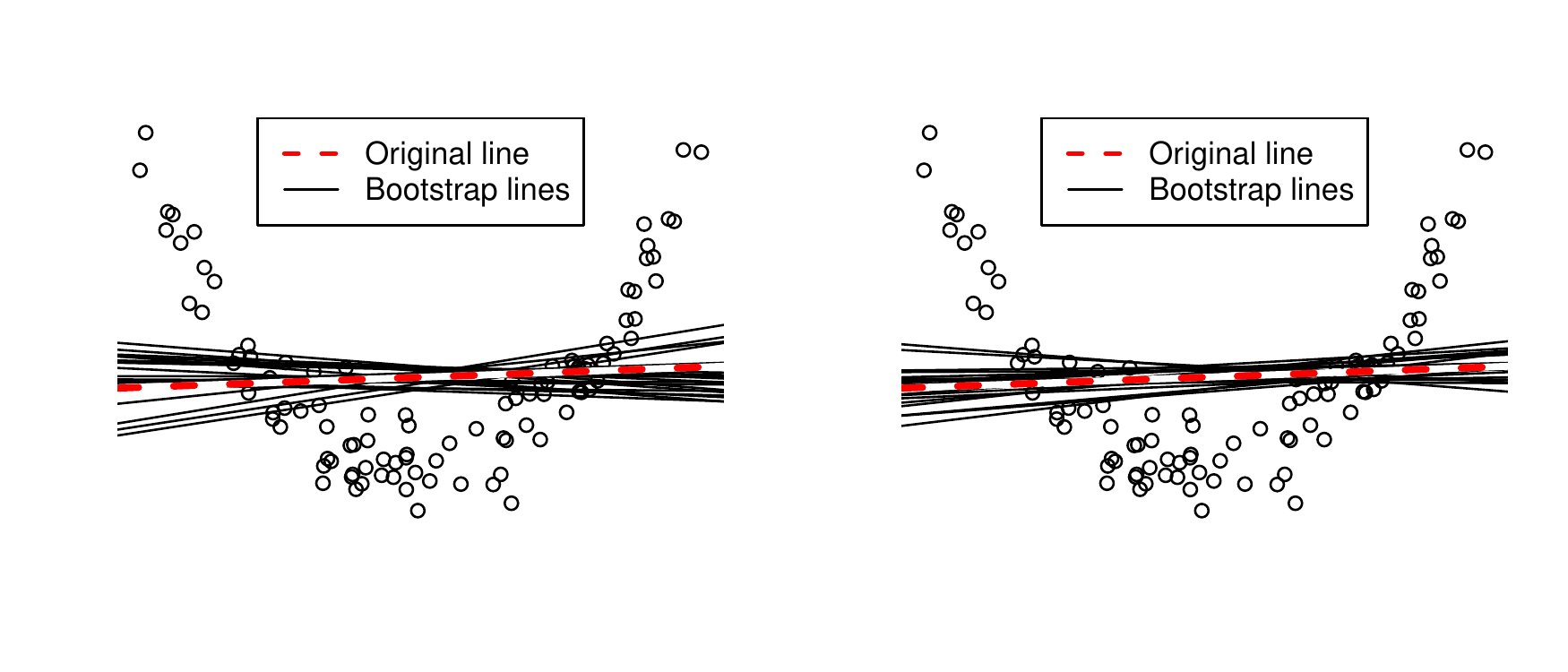}}
\renewcommand{\mycaption}{Bootstrapping a model that does not fit.}
\caption[\mycaption]{\label{figure:bootstrapBadFit}
{\em \mycaption}
The left panel shows resampling observations; the right panel
shows resampling residuals. See Section~\ref{section:bootstrapRegression}
}    \end{figure}

The third cause of bias is lack of model fit.
Here the bootstrap may not even show that there is bias.
It can only quantify the performance of the procedure you actually
used, not what you should have used.
For example, Figure~\ref{figure:bootstrapBadFit}
shows the result of fitting a line to data with obvious curvature.
The bootstrap finds no bias---for any $x$, the bootstrap lines are
centered vertically around the original fit.
We discuss this further when we consider regression,
in Section~\ref{section:bootstrapRegression}.

\boxx{The Bootstrap Does Not Find Bias Due to Lack of Fit}
{The bootstrap does not show bias due to a poor model fit.

Bootstrap bias estimates for non-functional statistics may be wrong.%
}

% Typical causes of Bias:
% Optimization: when (many) parameters are chosen to optimize some criterion
% Multiple regression (unadjusted R2 )
% Stepwise regression
% Nonlinear transformation or relationship
%
% Possible examples:
% kyphosis (library(rpart))
% relative risk

\subsection{Functional Statistics}

There is a subtle point in the bootstrap bias estimate
(equation \ref{eqn.biasBoot})---it assumes that $\theta = \theta(F)$
and $\thetahat = \theta(\Fhat)$---in other words, that the
statistic is {\em functional},
that it depends solely on the empirical distribution,
not on other factors such as sample size.
A functional statistic gives the same answer if each observation
is repeated twice (because that does not change the empirical distribution).
We can get odd results if not careful when bootstrapping non-functional
statistics.

For example, the sample variance $s^2 = (n-1)^{-1}\sum (x_i-\xbar)^2$
is not functional, while
$\sigmahat^2 = n^{-1}\sum (x_i-\xbar)^2$ is.
If $\theta(F)$ is the variance of the population $F$, then
$\sigmahat^2 = \theta(\Fhat_n)$,
the variance of the empirical distribution $\Fhat_n$,
with probability $1/n$ on each of the $x_i$.
$s^2$ is $n/(n-1)$ times the functional statistic.

Say the sample size is 10; then to the bootstrap, $s^2$ looks
like $\psi(\Fhat_n) = (10/9) \theta(\Fhat_n)$, which
it treats as an estimate for $\psi(F) = (10/9) \theta(F) = (10/9) \sigma^2$.
The bootstrap doesn't question why we want to
analyze such an odd statistic; it just does it.
Here is the result of bootstrapping
$s^2$ for the Basic TV data:
\begin{verbatim}
      Observed        SE     Mean       Bias
stat1 1.947667 0.5058148 1.762771 -0.1848956
\end{verbatim}
The observed value is $\hat\psi = s^2 = 1.95$,
while the average of the bootstrap values of
$\hat\psi^*$ is $1.76$; the bias is negative.
It concludes that $\hat\psi$ is negatively biased for $\psi$,
i.e.\ that $s^2$ is downward biased for $(10/9)\sigma^2$.
If we're not aware of what happened,
we might think that the bootstrap says that $s^2$ is biased for $\sigma^2$.

% Essentially, the bootstrap bias estimate treats finite-population
% factors as if the same factors will apply as the sample size increases
% to infinity.

Other non-functional statistics include adjusted R-squared in
regression, scatterplot smoothing procedures,
stepwise regression, and regularized regression.

Bootstrap SE estimates are not affected the same way as bias estimates,
because they are calculated solely from the bootstrap statistics,
whereas the bias estimate compares the bootstrap statistics to the observed
statistic.
Confidence intervals are affected---bootstrap procedures typically
provide confidence bounds for functional statistics.

\subsection{Skewness}

Another important issue for the bootstrap, and inference in general,
is skewness---skewness of the data for the mean, or more generally
skewness of the empirical influence of the observations \citeP{efro93}.

\paragraph{Verizon Example}
I consulted on a case before the New York Public Utilities
Commission (PUC). Verizon was an {\it Incumbent Local Exchange Carrier} (ILEC),
responsible for maintaining land-line phone service in certain areas.
Verizon also sold long-distance service, as did a number of competitors,
termed {\it Competitive Local Exchange Carrier} (CLEC). When something
would go wrong, Verizon was responsible for repairs, and was supposed
to make repairs as quickly for CLEC long-distance customers as for their
own.
The PUC monitored this by comparing repair times for Verizon and the
various CLECs, for many different classes of repairs, and many different
time periods. In each case a hypothesis test was performed at the 1\%
significance level, to determine whether repairs for a CLEC's customers
were significantly slower than for Verizon's customers. There were hundreds
of such tests. If substantially more than 1\% of the tests were significant,
then Verizon would pay a large penalty.
These tests were performed using $t$~tests; Verizon
proposed using permutation tests instead.

The data for one combination of period, class of service, and CLEC
is shown in Table~\ref{table:VerizonData},
and Figure~\ref{figure:VerizonData}. Both datasets are positively skewed.
There are odd bends in the normal quantile plot, due to 24-hour
periods (few repairs are made outside of normal working hours).

\begin{table}[\tableplace]
 \centering
\begin{tabular}{lrrr}\hline
       & n    & mean & sd\\
  ILEC & 1664 &  8.41 & 16.5 \\
  CLEC &   23 & 16.69 & 19.5 \\ \hline
\end{tabular}
\renewcommand{\mycaption}{Verizon repair times.}
\caption[\mycaption]{\label{table:VerizonData}
{\em \mycaption}
}    \end{table}

\begin{figure}[\figureplace]    % scripts/Verizon.R
\centerline{\includegraphics[width=3in]{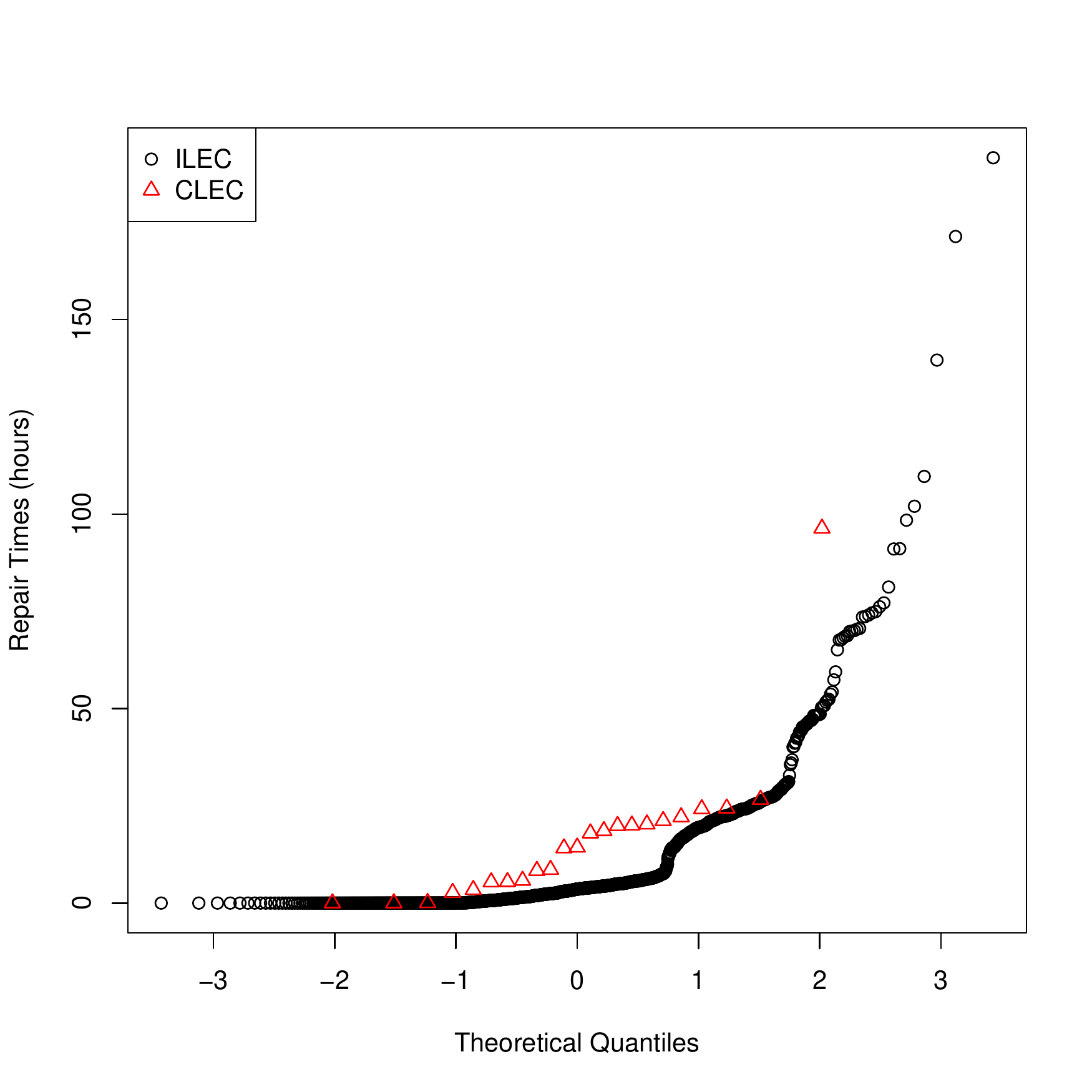}}
\renewcommand{\mycaption}{Normal quantile plot of ILEC and CLEC repair times.}
\caption[\mycaption]{\label{figure:VerizonData}
{\em \mycaption}
}    \end{figure}

The mean CLEC repair time is nearly double that for ILEC---surely this
must be evidence of discrimination? Well maybe not---the CLEC distribution
contains one clear outlier, the difference would be less striking
without the outlier.
But even aside from the outlier, the CLEC repair times tend to be larger
than comparable quantiles of the ILEC distribution.

\paragraph{Permutation Test}
The permutation distribution for the difference in means is shown in
Figure~\ref{figure:permVerizon}. The one-sided $P$-value is $0.0171$,
well above the 1\% cutoff for these tests, see
Table~\ref{table:VerizonTests}.
In comparison, the pooled $t$~test $P$-value is $0.0045$, about four times
smaller. The unpooled $P$-value is $0.0300$.
Under the null hypothesis that the two distributions are the same,
pooling is appropriate.
In fact, the PUC mandated the use of a $t$~statistic
$(\xbar_2 - \xbar_1) / (s_1 \sqrt(1/n_1 + 1/n_2))$ with standard error
calculated solely from the ILEC sample,
to prevent large CLEC repair times from contaminating the denominator;
this $P$-value is even smaller.

\begin{table}[\tableplace]
 \centering
\begin{tabular}{lrrr}\hline
                 &$t$   & $P$-value \\
Permutation test &      & 0.0171\\
Pooled $t$ test  &2.61  & 0.0045\\
Welch $t$ test   &1.98  & 0.0300\\
PUC $t$ test     &2.63  & 0.0044\\ \hline
\end{tabular}
\renewcommand{\mycaption}{Permutation and $t$~tests for the difference between CLEC and ILEC mean repair times.}
\caption[\mycaption]{\label{table:VerizonTests}
{\em \mycaption}
}    \end{table}

So, given the discrepancy between the permutation test result and the
various $t$~tests, which one is right?
Absolutely, definitely, the permutation test.
Sir Ronald Fisher originally argued for $t$~tests by
describing them as a computationally-feasible approximation
to permutation tests (known to be the right answer),
given the computers of the time.% ---young women.
We should not be bound by that limitation.
% TODO(find reference)

\boxx{Permutation and $t$~tests}
{Permutation tests are accurate.
$t$~tests are a computationally feasible approximation to permutation
tests, given the computers of the 1920's---young women.%
}
%We are no longer bound by that limitation.}

$t$~tests assume normal populations, and are quite sensitive to
skewness unless the two sample sizes are nearly equal.
Permutation test make no distributional assumptions,
and don't care about biased statistics.
Permutation test are ``exact''---when populations are the same,
the $P$-value is very close to uniformly distributed; if there are no ties
(different samples that give the same value of the statistic),
then the exhaustive permutation distribution has
mass $1/{n \choose {n_1}}$ on each of the ${n \choose {n_1}}$
possible values of the statistic, given the combined data.
With random sampling the test is not quite exact, but with
large $r$ is close.

\begin{figure}[\figureplace]
% Scripts/Verizon.R
\centerline{\includegraphics[width=\figurewidth]{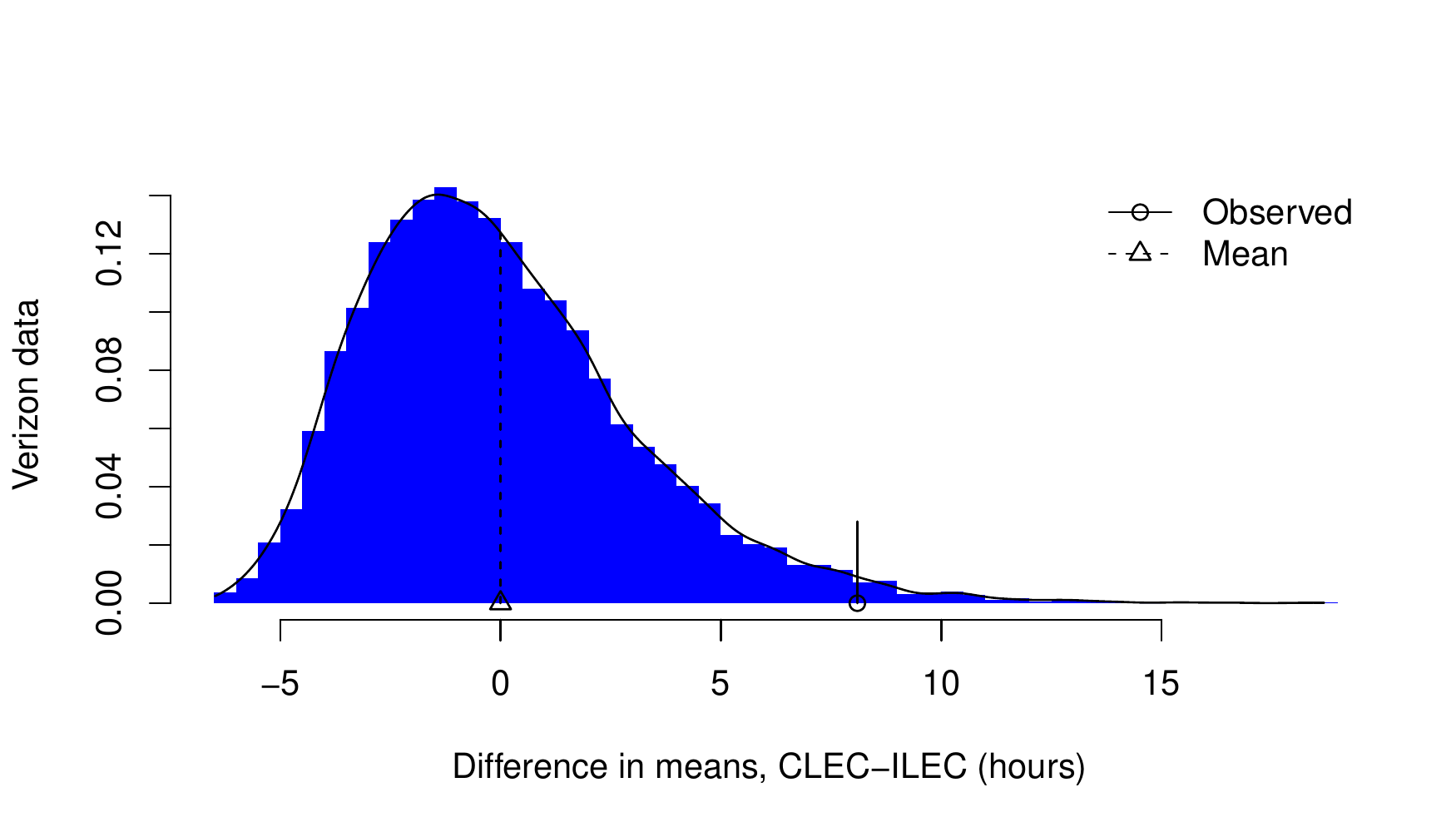}}
\renewcommand{\mycaption}{Permutation test of the difference in CLEC and ILEC repair times.}
\caption[\mycaption]{\label{figure:permVerizon}
{\em \mycaption}
The observed difference in means is $8.1$, and the $P$-value is $0.0171$.
}    \end{figure}

\paragraph{Bootstrap}

\begin{figure}[\figureplace]    % scripts/Verizon.R
\centerline{\includegraphics[width=\figurewidth]{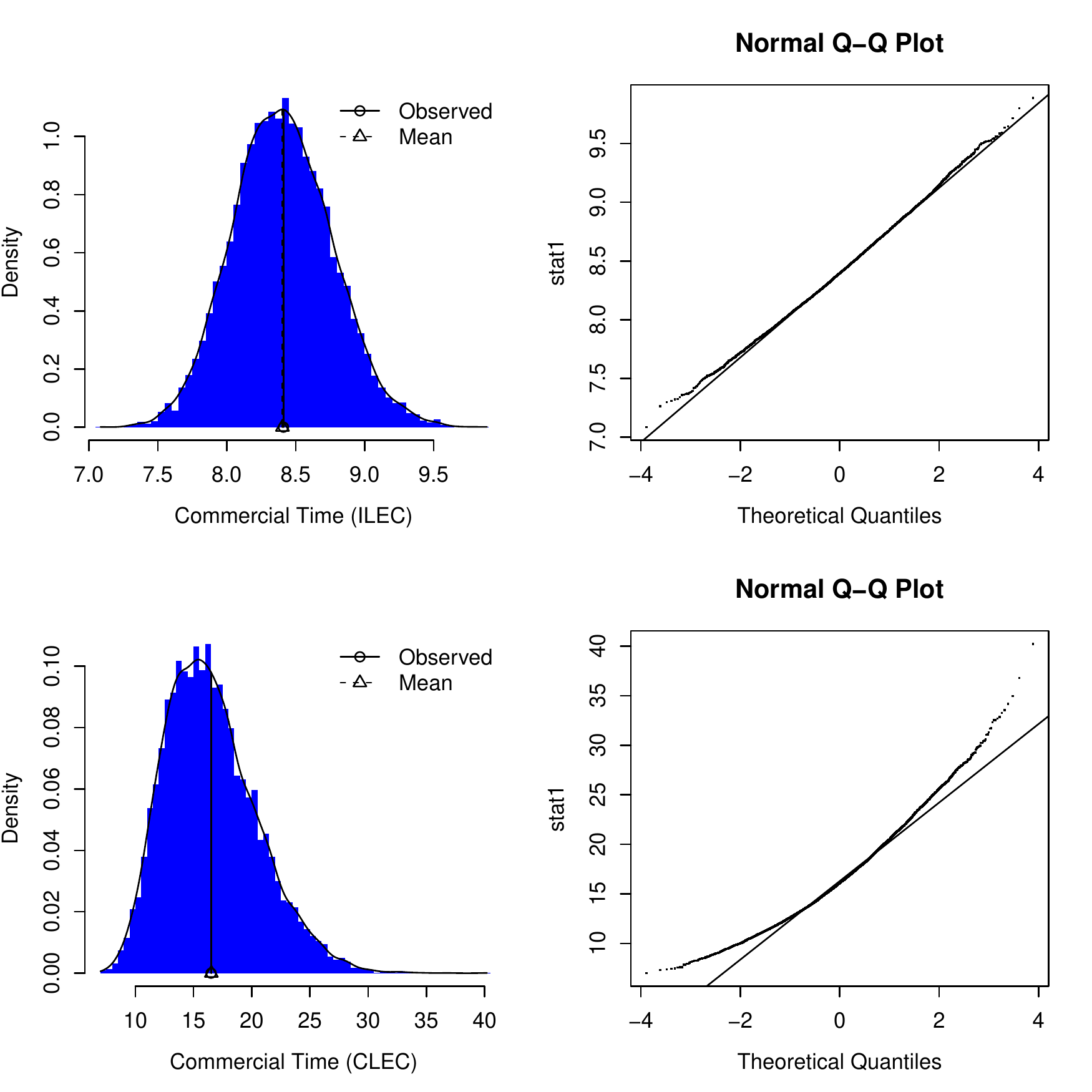}}
\renewcommand{\mycaption}{Bootstrap distributions for ILEC and CLEC data.}
\caption[\mycaption]{\label{figure:bootVerizon1}
{\em \mycaption}
}    \end{figure}

Figure~\ref{figure:bootVerizon1} shows the bootstrap distributions
for the ILEC and CLEC data.
Each is centered at the corresponding observed mean.
The CLEC distribution is much wider,
reflecting primarily the much smaller sample size (a valuable
lesson for students), and the larger sample standard
deviation.

\begin{figure}[\figureplace]    % scripts/Verizon.R
\centerline{\includegraphics[width=\figurewidth]{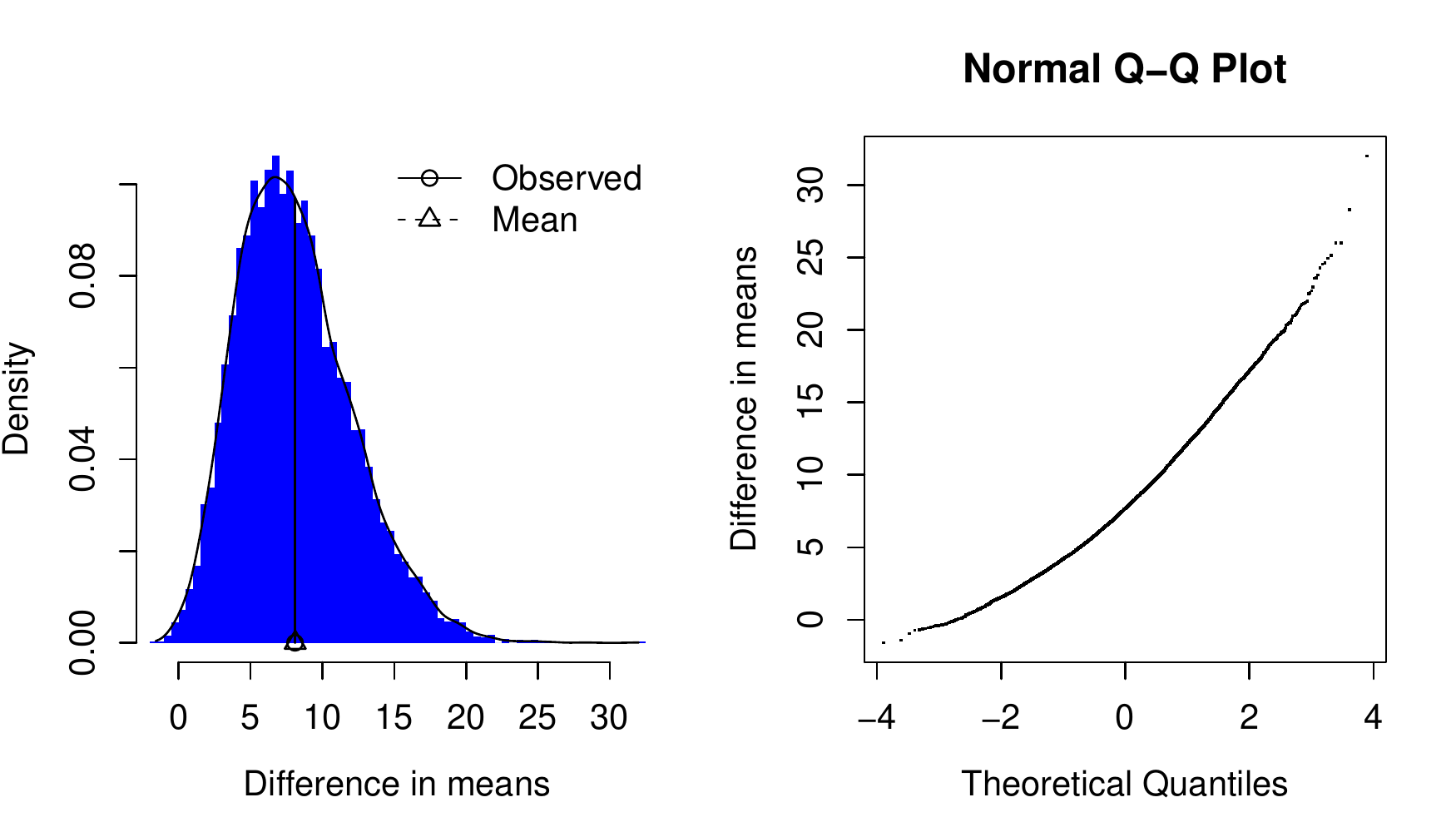}}
\renewcommand{\mycaption}{Bootstrap distribution for difference of means, CLEC - ILEC.}
\caption[\mycaption]{\label{figure:bootVerizon2}
{\em \mycaption}
}    \end{figure}

Figure~\ref{figure:bootVerizon2} shows the bootstrap distributions
for difference of means, CLEC - ILEC.
This is centered at the observed difference in means.
The SE reflects the contributions to the SE from both samples.

There is skewness apparent in the bootstrap distribution for the difference
in means. Does that amount of skewness matter?

Before answering, I'll share a story.
%  I had an NSF grant to develop
% resampling software, and got a supplemental grant to write a book chapter
% for resampling for introductory statistics,
% \citeP{hest03}.\footnote{For different versions downloadable
% versions of that chapter see
% \url{http://www.timhesterberg.net}.}
% Two college students, and high school teacher, David Moore and I set
% about to write this.
I co-authored \citeP{hest03}%
\footnote{A resampling chapter for an introductory statistics text;
this and similar chapters can be freely downloaded, see
\url{http://www.timhesterberg.net/bootstrap}.
};
one homework question included a bootstrap distribution similar to
Figure~\ref{figure:bootVerizon2}, and asked if the skewness mattered.
The publisher had someone else write the first draft of the solutions,
and his answer was that it did not.

That is dead wrong. His answer was based on his experience, using
normal quantile plots to look at data. But this is a sampling
distribution, not raw data. The Central Limit Theorem has already had
its one chance to make things more normal. At this point, any
deviations from normality will have bad effects on any procedure that
assumes normal sampling distributions.

He's not alone. I often ask that question during talks and
courses, and typically over half of the audience answers that it
is no problem.

That points out  a common flaw in statistical practice---that
we don't often use effective ways to judge whether the CLT is really working,
and how far off it is.
To some extent the bootstrap distributions above provide this;
the bootstrap~$t$ distributions below are even more effective.

Even the skewness in the ILEC distribution, with 1664 observations,
has a measurable effect on the accuracy of a $t$~interval for that data.
A 95\% $t$~interval misses by being too low about 39\% too often
(3.5\% instead of 2.5\%).
Similarly, a percentile interval is too low about 28\% too often.
% Verizon.R, section "What does the bootstrap t say about the actual coverage of the other two?"
To reduce the 39\% to a more
reasonable 10\% would require about 16 times as many observations.
The Central Limit Theorem operates on glacial time scales.
We return to this issue below.

\paragraph{Permutation test, not Pooled Bootstrap}

We could perform a permutation test by pooling the data,
then drawing bootstrap samples of size $n_1$ and $n_2$
with replacement from the pooled data. This sampling would be
consistent with the null hypothesis.

It is not as accurate as the permutation test.
Suppose, for example, that the data contain three outliers.
The permutation test tells how common the observed statistic is,
given that there are a total of three outliers.
With a pooled bootstrap the number of outliers would vary,
and the $P$-value would not as accurately reflect the data we have.

\subsection{Accuracy of the CLT and t Statistics}
\label{section:accuracyCLT}

In the Verizon example the two-sample pooled-variance
$t$~test was off by a factor of four, and the one-sample $t$~interval
with $n=1664$ missed 39\% too often on one side.
These are not isolated examples.

When there is skewness, the standard $t$~test and $t$~interval
converge to the correct size and coverage very slowly,
at the rate $O(1/\sqrt{n})$, with a large constant.
(The corresponding constant for the percentile interval is about $2/3$
as large.)
We can demonstrate this using simulation or asymptotic
methods, see Figures~\ref{figure:coverage2}--\ref{figure:coverage4}
and
Section~\ref{section:asymptotics}.

\boxx{The CLT requires $n \ge 5000$ for a moderately skewed population}
{
  For $t$~tests and confidence intervals to be reasonably accurate
  (off by no more than 10\% on each side)
  requires $n \ge 5000$ for a 95\% interval or two-sided $\alpha=0.05$
  test, for an exponential population.

The central limit theorem acts over glacial time scales, when skewness
is present.%
}

The inaccuracy of $t$ procedures when there is skewness has
been known since at least 1928 \citeP{sutt93},
and a number of more accurate alternatives have been proposed,
see e.g.~\citeP{john78,klei86,sutt93,meed99},
a number of bootstrap procedures discussed below, and undoubtedly others.
Unfortunately, these have
had little impact on statistical practice.

I think the reason is a combination of historical practicality
and momentum.
The simulations needed to accurately estimate error probabilities
used to be too costly.
\cite{klei86} noted
\begin{quotation}
Unfortunately, Monte Carlo experimentation requires much computer time.
Obviously the number of replications $R$ needed to estimate the
actual $\alpha$-error within 10\% with 90\% probability, is
$R = 100 (1.6449)^2 (1-\alpha)/\alpha$.
Hence if $\alpha$ is 0.1, 0.05, 0.01 then $R$ is 2435, 5140,
26786 respectively. Such high $R$ values are prohibitive,
given our computer budget
\end{quotation}
I once estimated that some simulations I did
related to confidence interval coverage
would have taken about 20000 hours of computer time in 1981.
%Figure~\ref{figure:JohnsonInterval}

Then, there is momentum---the statistics profession got in the habit
of using $t$~tests, with each person following the example of their
instructors, and only perform what is provided in software.
I plead guilty to that same inertia---it was not until I developed
examples for \citeP{hest03} that I started to pay attention
to this issue.
% And software vendors are conservative---I used to work for the
% vendor of S-PLUS, the predecessor of R, and tried to add
% adjusted $t$ procedures to the $t.test$ function, without success.

The usual rule in statistics, of using classical $t$ methods
if $n \ge 30$ and the data are not too skewed,
is imprecise and inadequate.
For a start, we should look at normal (or $t$) quantile plots for
bootstrap distributions; next, look at bootstrap~$t$ distributions
rather than $\xbar$, because $t$ is twice as skewed in the opposite
direction and is biased.
Finally, our eye can't accurately judge effects on coverage probabilities
from quantile plots, so we need to calculate rather than
eyeball the effect on coverage or $P$-values.

\section{Confidence Intervals}
\label{section:confidenceIntervals}

We begin with some introductory material, then turn
in Section~\ref{section:ConfidenceIntervalPictures}
to a pair of pictures that
help explain how confidence intervals should behave
in the easy case (normal with no bias), and
in harder cases (bias, and skewness/acceleration).

We then discuss different intervals.
In Section~\ref{section:stat101}
we recommend two easy intervals for Stat 101,
the bootstrap percentile interval
and the $t$~interval with bootstrap standard error.

Section~\ref{section:expandedPercentile} has an adjusted version
of the percentile interval, to correct for too-low coverage.

In Sections~\ref{section:reversePercentile}~and~\ref{section:bootstrapT}
we turn to intervals for other courses and for practice,
starting with an interval with a natural derivation
for statisticians---that turns out to be terrible but with
pedagogical value---then a better interval.
We summarize in Section~\ref{section:CIAccuracy}, including
simulation and asymptotic results showing how well the intervals
actually perform.

\paragraph{Accurate Confidence Intervals}
A accurate confidence interval procedure includes the true value 95\% of the
time, and misses 2.5\% of the time on each side.
Different intervals, both formula and bootstrap,
have trouble achieving this or even coming close,
in different applications and with different sample sizes.

It is not correct for a 95\% interval to miss 4\% of the time
on one side and 1\% of the time on the other---in practice almost
all statistical findings are ultimately
one-sided, so
making an error on one side does not compensate for an error on
the other.
It would be rare indeed to find a report that says,
``my new way of teaching was significantly different in
effectiveness than the control'' without also reporting the
direction!

Say that the right endpoint of an interval is too low, so the interval
misses 4\% of the time on that side.
I'd rather have an interval that is correct on the other side
than one that is too low---because the combination of being too
low on both sides gives an even more biased picture about
the location of $\theta$.
A {\em biased confidence interval} has endpoints that are too low, or too high,
on both sides.

I am not arguing against two-sided tests, or two-sided confidence
intervals. In most cases we should be receptive to what the data tell us,
in either direction.
My point is that those two-sided procedures should have the
correct probabilities on both sides,
so that we correctly understand what the data says.

As for so-called ``shortest intervals'', that intentionally
trade under-coverage on one side for over-coverage on the other,
to reduce the length---that is statistical malpractice, and anyone
who uses such intervals should be disbarred from Statistics and
sentenced to 5 years of listening to Justin Bieber crooning.
% In addition to giving incorrect coverage on each side, the overall
% coverage is too low, due to the optimization bias
% discussed in Section~\ref{section:causesOfBias}.

\boxx{Accurate Confidence Intervals}
{An accurate 95\% confidence interval misses 2.5\% of the time
on each side.

An interval that under-covers on one side and over-covers on the other
is biased.%
}

\paragraph{First and Second Order Accurate}
A hypothesis test or confidence interval is
{\em first-order accurate} if
the one-sided actual rejection probabilities or
one-sided non-coverage probabilities differ from the nominal values
by $O(n^{-1/2})$.
They are {\em second-order accurate} if the differences are $O(n^{-1})$.

The usual $t$~intervals and tests, percentile, and $t$~interval
with bootstrap standard errors are first-order accurate.
The bootstrap~$t$ and skewness-adjusted $t$~interval
(see Section~\ref{section:Johnson}) are second-order accurate.

These statements assume certain regularity conditions,
and apply to many common statistics,
e.g.\ smooth functions of sample moments
(for example, the correlation coefficient can be written
as a function of
$(\xbar, \ybar, \bar{x^2}, \bar{xy}, \bar{y^2})$),
smooth functions of solutions to smooth estimating equations
(including most maximum likelihood estimators), and generalized linear
models.
For details see \citeP{efro93,davi97,dici88,hall88,hall92a}.

There are many other bootstrap confidence intervals; in the
early days of the bootstrap there was quite a cottage industry,
developing second-order accurate or even higher order intervals.
Some are described in \citeP{efro93,davi97};
for a review see \cite{dici96}.

To be second-order accurate, a procedure needs to handle
bias, skewness, and transformations.

But just being second-order accurate isn't enough in practice; an interval
should also have good small-sample performance.
A first-order accurate interval can be better in small samples
than a second-order accurate interval, if it handles the ``little things''
better---things that have an $O(1/n)$ effect on coverage,
with little effect for large $n$, but that matter for small $n$.
We'll see below that the bootstrap percentile interval is poor
in this regard, and has poor accuracy in small samples.

\boxx{First and Second-Order Accurate Inferences}
{
  A hypothesis test or confidence interval is
{\em first-order accurate} if
the one-sided actual rejection probabilities or
one-sided non-coverage probabilities differ from the nominal values
by $O(n^{-1/2})$.
They are {\em second-order accurate} if the differences are $O(n^{-1})$.

To be second-order accurate, a procedure needs to handle
bias, skewness, and transformations.%
}

\subsection{Confidence Interval Pictures}
\label{section:ConfidenceIntervalPictures}

Here are some pictures that show how confidence intervals
should behave in different circumstances.

\begin{figure}[\figureplace]     % scripts/CIs.R
\centerline{\includegraphics[width=\figurewidth]{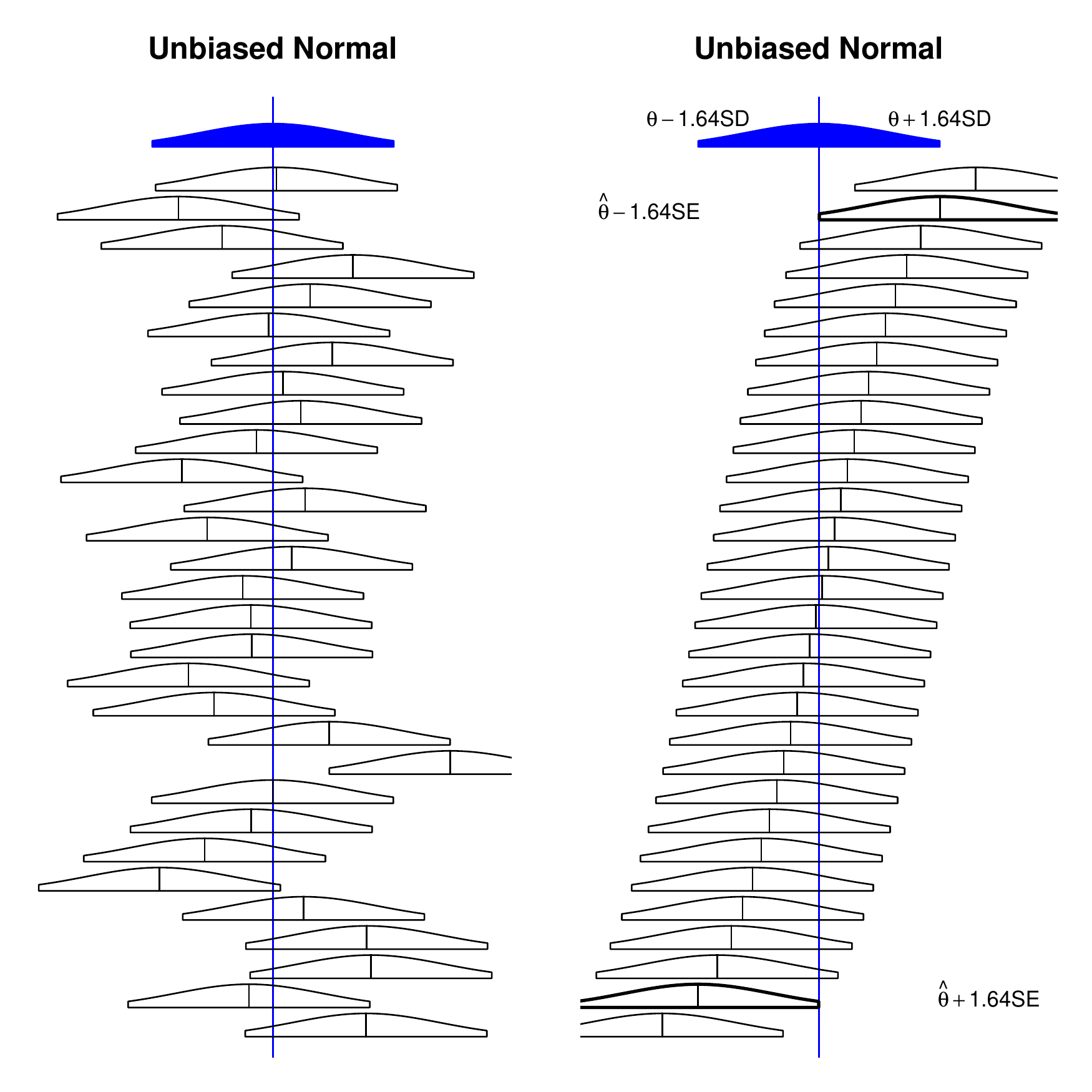}}
\renewcommand{\mycaption}{Confidence intervals for normal with no bias.}
\caption[\mycaption]{\label{figure:CIcurves1}
{\em \mycaption}
The vertical lines correspond to true values of the parameter.
The solid figures are the normally-distributed sampling distributions
with no bias,
truncated at the middle 90\%.
To have correct 90\% coverage,
a sample with $\thetahat$ in that middle range should result in
a confidence interval that includes $\theta$, and
others should miss $\theta$.
For simplicity, we assume that
$\SD^2=\Var(\thetahat)=\SE^2=\Var(\thetahat^*)$.
On the left are truncated bootstrap distributions, each for one random sample,
centered at the corresponding $\thetahat$.
In this case, the bootstrap percentile interval and a $z$ interval
coincide, and both have the correct coverage; both CIs include $\theta$
when their $\thetahat$ is in the middle 90\% of the
sampling distribution.
On the right are bootstrap distributions, ordered by the $\thetahat$,
scaled so the bold distributions should just touch $\theta$.
A correct interval is
$(\thetahat - 1.64\SE, \thetahat + 1.64\SE)$.
}    \end{figure}

\begin{figure}[\figureplace]     % scripts/CIs.R
\centerline{\includegraphics[width=\figurewidth]{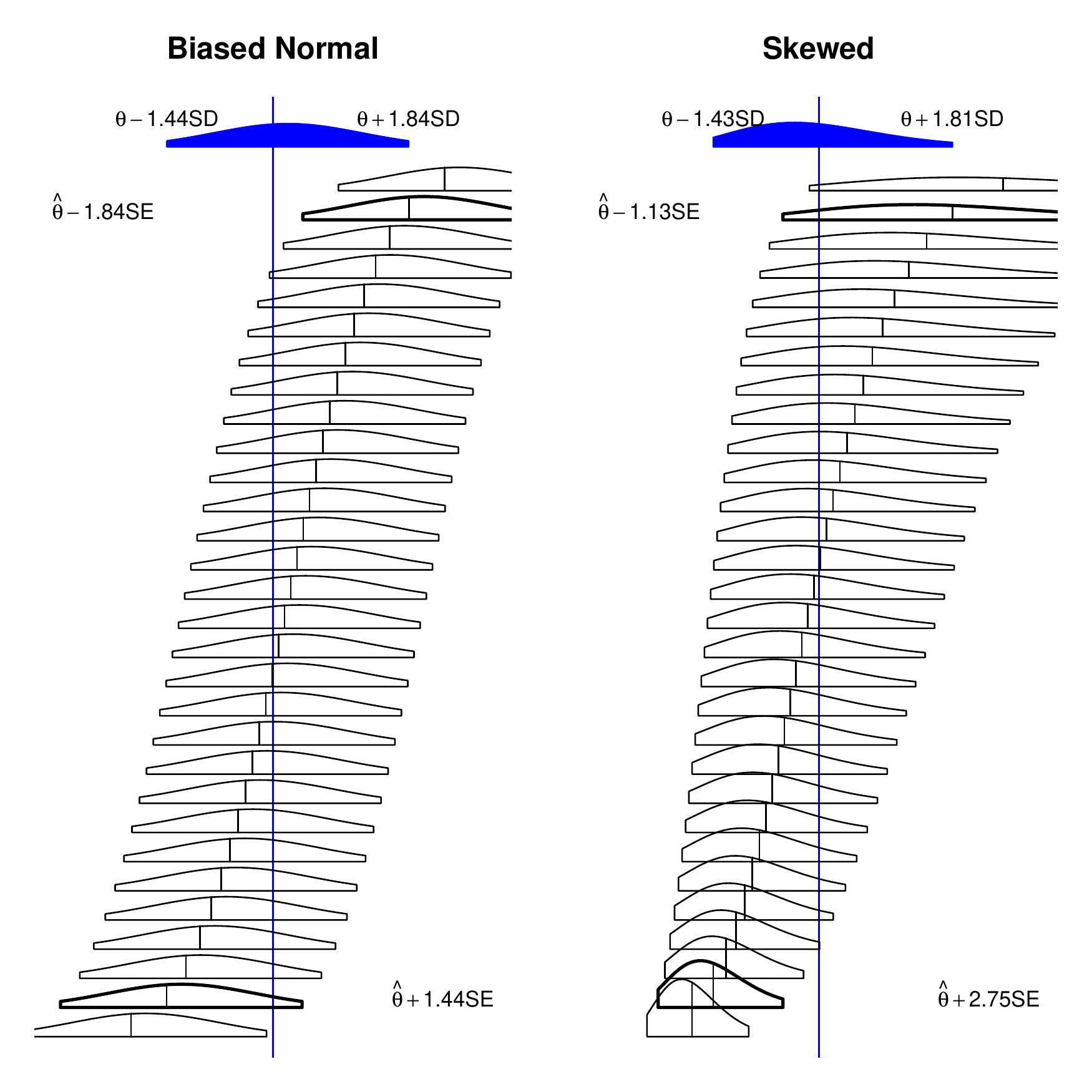}}
\renewcommand{\mycaption}{Confidence intervals for bias, and acceleration.}
\caption[\mycaption]{\label{figure:CIcurves2}
{\em \mycaption}
The vertical lines correspond to true values of the parameter.
The solid curves are sampling distributions, truncated at the middle 90\%.
On the left the sampling distribution, and bootstrap distributions,
are normal with bias $0.2\SD = 0.2\SE$.
For correct coverage, an interval should be
$(\thetahat - 1.84\SE, \thetahat + 1.44\SE)$
(see the text beside the bold distributions).
The bootstrap percentile interval is asymmetrical in the wrong
direction: $(\thetahat - 1.44\SE, \thetahat + 1.84\SE)$.
On the right the sampling distribution, and bootstrap distributions,
are unbiased
with skewness $2/3$ (the distributions are gamma with shape = 9).
For correct coverage, an interval should be
$(\thetahat - 1.13\SE, \thetahat + 2.75\SE)$.
The bootstrap percentile interval
$(\thetahat - 1.43\SE, \thetahat + 1.81\SE)$
is not asymmetrical enough.
A $t$~interval
$(\thetahat - 1.64\SE, \thetahat + 1.64\SE)$
is even worse.
}    \end{figure}

In all cases the parameter is shown with a vertical line,
the sampling distribution is on the top, and below that are
bootstrap distributions.
In the right side of Figure~\ref{figure:CIcurves1}
and both sides of Figure~\ref{figure:CIcurves2},
the confidence intervals for the samples that are
second from top and from bottom should just touch the parameter, those in between
should include the parameter, and the top and bottom ones should miss
the parameter.

Figure~\ref{figure:CIcurves1} shows what happens in the nice
case, of normally-distributed sampling distributions with no bias
(and to make things simple, with known variance).
Each bootstrap distribution is centered about the statistic
for its sample. The bootstrap percentile interval and $t$-interval
coincide, and each misses exactly the right fraction of the time
on each side.

\paragraph{Simple Bias}
The left side of Figure~\ref{figure:CIcurves2} shows what
happens when there is simple bias,
similar to that of unadjusted R-squared.
The statistic is positively biased; the bootstrap distributions
are similarly biased. The bias is $b=0.2\SD$ where
$\SD = \sqrt{\Var(\thetahat)}$.
A correct interval would be $\thetahat - b \pm \zSub \SE$,
or $(\thetahat - 1.84\SE, \thetahat + 1.44\SE)$.

$z$~intervals are symmetric about the
corresponding statistic, so end up with one copy of the bias
(from the bias in the original statistic).
The intervals miss too often by being above $\theta$, and not often
enough below $\theta$.

Bootstrap percentile intervals are even worse,
because they get a second copy of the bias (the original bias, and
bootstrap bias).
A bias-corrected percentile interval would subtract twice the bias
from the percentile interval endpoints.

\paragraph{Skewness}
The right side of Figure~\ref{figure:CIcurves2} shows what happens
for unbiased statistics when the distribution is skewed;
in this case, the mean of a gamma distribution with shape 9.
The sampling distribution has roughly the same asymmetry
$(\theta - 1.43\SD, \theta + 1.81\SD)$ as in
the bias example.

The bootstrap distributions show the same asymmetry; the middle
90\% (the 90\% bootstrap percentile interval) is
$(\thetahat - 1.43\SE, \thetahat + 1.81\SE)$.

A correct interval is
$(\thetahat - 1.13\SE, \thetahat + 2.75\SE)$
(see the text beside the bold curves).
A correct interval needs to reach many (short) standard errors to the right
to avoid missing too often when $\thetahat < \theta$ and
standard errors are small.

This time the bootstrap percentile misses too often by being
below $\theta$, and not often enough by being above.
Even though the interval is asymmetrical, it is not asymmetrical
enough.

A $t$ interval is even worse.

See Section~\ref{section:Johnson} below
for a skewness-corrected
$t$~interval obtained using asymptotic methods;
the percentile interval has only about one-third the asymmetry
of this interval (asymptotically, for 95\% intervals).

\boxx{Confidence Intervals for Skewed Data}
{When the data are skewed, a correct interval is even
more asymmetrical than the bootstrap percentile interval---reaching
farther toward the long tail.%
}

\paragraph{Failure of intuition}
This runs counter to our intuition. If we observe data with
large observations on the right, our intuition may be to downweight
those observations, and have the confidence interval reach
farther left, because the sample mean may be much larger than the true mean.
In fact, when the data show that the population has a long right tail,
a good confidence interval must protect against the possibility
that we observed {\em fewer} than average observations from
that tail, and especially from the far right tail.
If we're missing those observations, then $\xbar$ is too small,
and $s$ is also too small, so the interval must reach many standard errors
to the right.

Conversely, we may have gotten more observations from the right tail
than average, and the observed mean is too large---but in that case
the standard error is inflated, so we don't need to reach so many
standard errors to reach the parameter.

\paragraph{Transformations}

The 90\% endpoints of the bootstrap distributions had roughly
the same asymmetry in the bias and skewness examples:
$(-1.44, 1.84)$ vs $(-1.43, 1.81)$.
We could get the same asymmetry by applying a nonlinear transformation
$\psi = \exp(0.15 \mu)$ to the case of normal with no bias,
with $\hat\psi = \exp(0.15 \mu)$.
This gives sampling distributions and bootstrap distributions with
the same asymmetry as the bias example, $0.28 / 0.22 \approx 1.84 / 1.44$.
In this case a bootstrap percentile interval would be correct,
and a $t$~interval would not.

\paragraph{Need more information}

We can't tell just from the asymmetry of the
endpoints whether a correct interval should be asymmetrical
to the right or left.
The correct behavior depends on whether the asymmetry is caused by
bias, skewness, transformations, or a combination.
We need more information---and second-order
accurate bootstrap confidence interval procedures collect and use that
information, explicitly or implicitly.

But lacking that information, the percentile interval is a good
compromise, with transformation
invariance and a partial skewness correction.

% But if we're choosing between a $t$~interval or a percentile interval,
% I'd pick the percentile interval. By making a partial skewness correction,
% it occupies the middle ground between long-left to handle
% bias and far-long-right to handle skewness.
% And in my experience, skewness occurs more often than simple bias.

\boxx{Need More Information for Accurate Intervals}
{Asymmetric bootstrap distributions could be caused by
bias, skewness, transformations, or a combination.
The asymmetry of a correct confidence interval differs, depending
on the cause.
Second-order accurate bootstrap confidence intervals are based on additional
information.
In the absence of more information, a percentile interval is
a reasonable compromise.%
}

\subsection{Statistics 101---Percentile, and T with Bootstrap SE}
\label{section:stat101}

For Stat 101 I would stick with the two
quick-and-dirty intervals mentioned earlier: the
bootstrap percentile interval, and the
$t$~interval with bootstrap standard error
$\thetahat \pm t_{\alpha/2} \SE_b$.
If using software that provides it, you may also use the expanded
bootstrap percentile interval, see Section~\ref{section:expandedPercentile}.

The percentile interval will be more intuitive for students.
The $t$~with bootstrap standard error helps them learn formula methods.

Students can compute both and compare. If they are similar,
then both are probably OK. Otherwise, if their software computes
a more accurate interval they could use that.
If the data are skewed, the percentile interval has an advantage.
If $n$ is small, the $t$~interval has an advantage.

Both intervals are poor
in small samples---they tend to be too narrow.
The bootstrap standard error is too small, by a factor $\sqrt{(n-1)/n}$
so the $t$~interval with bootstrap SE is too narrow by
that factor;
this is the narrowness bias discussed in
Section~\ref{section:smallSamplePicture}.

The percentile interval suffers the same narrowness and more---for
symmetric data
it is like using $\zSub \sigmahat/\sqrt{n}$ in place of $\tSub s/\sqrt{n}$.
It is also subject to random variability
in how skewed the data is. This adds random variability to the interval
endpoints, similar to the effect of randomness in the
sample variance $s$, and reduces coverage.

These effects are $O(n^{-1})$ (effect on coverage probability) or smaller,
so they become negligible fairly quickly
as $n$ increases.
For larger $n$, these effects are overwhelmed by the effect of skewness,
bias, and transformations.
But they matter for small $n$, see
Table~\ref{table:narrowness}, and the confidence interval coverage
in Figures~\ref{figure:coverage2} and~\ref{figure:coverage3}.

% file coverage.R
%latex.default(cbind(round(cbind(nn, sqrt((nn - 1)/nn), qnorm(0.975)/qt(0.975,     nn - 1)), 3), round(pnorm(sqrt(nn/(nn - 1)) * qt(0.025, nn -     1)), 4)), file = "")%
\begin{table}[\tableplace]
\begin{center}
\begin{tabular}{rrrrr}
\hline
$n$ & $\sqrt{(n-1)/n}$ & $z_{0.025} / t_{0.025,n-1}$ & size & $\alpha'/2$\\
\hline
$ 5$&$0.894$&$0.706$&$0.077$&$0.0010$\\
$10$&$0.949$&$0.866$&$0.048$&$0.0086$\\
$20$&$0.975$&$0.936$&$0.036$&$0.0159$\\
$40$&$0.987$&$0.969$&$0.030$&$0.0203$\\
$80$&$0.994$&$0.985$&$0.028$&$0.0226$\\
\hline
\end{tabular}\end{center}
\renewcommand{\mycaption}{Narrowness bias, and z/t, and adjusted quantiles.}
\caption[\mycaption]{\label{table:narrowness}
{\em \mycaption}
Column~2 shows the narrowness bias.
Column~3 shows narrowness due to using $\zSub$ instead of
$\tSub$.
Column~4 shows the combined effect of columns 2--3
on coverage, corresponding to an interval
$\xbar \pm \zSub \sigmahat/\sqrt{n}$.
Column~5 shows the nominal $\alpha'/2$ to use
to correct for the two effects, see
Section~\ref{section:expandedPercentile}.
}
\end{table}

In practice, the $t$ with bootstrap standard error offers no
advantage over a standard $t$ procedure, for the sample mean.
Its advantages are pedagogical, and that it can be used for
statistics where there are no easy standard error formulas.

In Stat 101 it may be best to avoid the small-sample problems by
using examples with larger $n$.
% You've perhaps been meaning to use larger, more meaningful datasets in
% that class anyway; now is the time.

Alternately, you could use software that corrects for the
small-sample problems. See the next section.

% For the sample mean from skewed data, the
% bootstrap percentile interval stretches farther to the right than to
% the left. In this regard it falls between
% \begin{squeezeitemize}
% \item symmetric intervals like $t$~intervals (using either a formula or
% bootstrap SE)
% \item more accurate (but more complicated)
% intervals like BCa \citeP{efro87} and bootstrap~$t$ \citeP{efro82}.
% \end{squeezeitemize}
%
% Hence, the percentile interval is better than symmetric intervals
% (except for really small samples, where the percentile interval has
% other problems), though not as good as the really accurate intervals.
%
% Our intuition suggests that if the data is positively skewed, that
% maybe the percentile interval reaches too far to the right because it
% is "affected by outliers". That intuition is wrong. When the data has
% a long right tail, a good interval will actually reach farther to the
% right.
%
% The reason is that a confidence interval is supposed to have a high
% probability of including the real mean. When the data are positively
% skewed, the population is probably positively skewed. The data could
% have
% * more observations than average from the long right tail,
% * fewer observations than average from the long right tail.
% To protect against the second case the interval needs to reach longer
% to the right. When the data doesn't have enough observations from that
% tail, we're kind of blind - we don't know that our mean is too small,
% and our standard deviation is too small too. So we need to be
% conservative in that direction.

\boxx{Simple Intervals for Stat 101; Poor Coverage for Small $n$}
{
I recommend two intervals for Stat 101---the bootstrap percentile interval
provides an intuitive introduction to confidence intervals, and
the $t$~interval with bootstrap standard error as a bridge to formula
$t$~intervals.

However, these intervals are too short in small samples, especially
the percentile interval. It is like using
$\xbar \pm \zSub \sqrt{(n-1)/n} s/\sqrt{n}$ as a confidence interval for $\mu$.

People think of the bootstrap (and bootstrap percentile interval)
for small samples, and classical methods for large samples.
That is backward, because the percentile interval is too narrow
for small samples.
The $t$ interval is more accurate than the percentile interval
for $n \le 34$, for exponential populations.%
}

\subsection{Expanded Percentile Interval}
\label{section:expandedPercentile}

The bootstrap percentile interval performs poorly in small samples,
because of the narrowness bias, and because it lacks a fudge factor
to allow for variation in the standard error.
The standard $t$~interval handles both, using $s$ in place of $\sigmahat$
to avoid narrowness bias,
and $\tSub$ in place of $\zSub$ as a fudge factor to allow for variation in $s$.
We can interpret the $t$~interval as multiplying the length of
a reasonable interval, $\xbar \pm \zSub \sigmahat$, by
$a_{\alpha/2,n} = (\tSub/\zSub) (s/\sigmahat)$,
to provide better coverage.
This multiplier is the inverse of
the product of columns 2--3 of Table~\ref{table:narrowness}.

The fact that the $t$~interval is exact for normal populations is a
bit of a red herring---real populations are never exactly normal,
and the multiplier isn't correct for other populations.
% with positive kurtosis, for example, it would be better to use a
% $t$~quantile with degrees of freedom lower than $(n-1)$,
% to allow for the greater uncertainty in $s$
Yet we continue to use it, because it helps in practice.
Even for long-tailed distributions, where the fudge factor
should be larger, using at least a partial fudge factor helps.
(For binomial data we do use $\zSub$ instead of $\tSub$, because
given $p$ there is zero uncertainty in the variance.)

Similarly, we may take a sensible interval, the percentile interval,
and adjust it to provide better coverage for normal populations,
and this will also help for other populations.

A simple adjustment is to multiply both sides of a percentile interval
by $a_{\alpha/2,n}$.
% giving
% $\thetahat + a_{\alpha/2,n}(\Ghat^{-1}(\alpha/2) - \thetahat),
%  \thetahat + a_{\alpha/2,n}(\Ghat^{-1}(1-\alpha/2) - \thetahat)$
% where $\Ghat^{-1}(p)$ is the $p$ quantile of the bootstrap distribution
% and $(\Ghat^{-1}(\alpha/2), \Ghat^{-1}(1-\alpha/2))$ is the usual
% bootstrap percentile interval.
But that would not be transformation invariant.
% For a statistic like the correlation coefficient,
% the confidence interval could go outside $(-1,1)$.

We can achieve the same effect, while not losing transformation
invariance, by adjusting the percentiles.
If the bootstrap distribution is approximately normal then
$$
  \Ghat^{-1}(\alpha/2) \approx \thetahat - \zSub \sigmahat/\sqrt{n}.
$$
We want to find an adjusted value $\alpha'$ with
\begin{eqnarray*}
  \Ghat^{-1}(\alpha'/2)
  &\approx& \thetahat - z_{\alpha'/2} \sigmahat/\sqrt{n} \\
  &=&  \thetahat - \tSub s/\sqrt{n}
\end{eqnarray*}
This gives $z_{\alpha'/2} = \sqrt{n/(n-1)} \tSub$,
or $\alpha'/2 = \Phi(-\sqrt{n/(n-1)} \tSub)$.
The values of $\alpha'/2$ are given in Table~\ref{table:narrowness}.
For a nominal one-sided level of $0.025$, the adjusted values range from
$0.0010$ at $n=5$ to $0.0226$ for $n=80$.

Coverage using the adjusted levels is dramatically
better, see \citeP{hest99b} and Figure~\ref{figure:coverage2},
though is still poor with $n=5$.

This adjustment has no terms for bias or skewness; it only counteracts
the narrowness bias and provides a fudge factor for uncertain width.
Still, we see in Figure~\ref{figure:coverage3} that it also helps
for skewness.

This technique of using modified quantiles of the bootstrap distribution
is motivated by the bootstrap BCa confidence interval \citeP{efro87},
that uses modified quantiles to handle skewness and median bias.
However it has no adjustment for narrowness or variation in SE,
though these could be added.

I plan to make expansion the default for both percentile intervals
and BCa intervals in a future version of the
{\em resample} package \citeP{hest14}.

\boxx{Expanded Percentile Interval}
{
The expanded percentile interval corrects for the poor coverage
of the common percentile interval using adjusted quantiles of the
bootstrap distribution. This gives much better coverage in small samples.
For exponential populations, this is better than the $t$ interval for
$n \ge 7$.%
}

\subsection{Reverse Bootstrap Percentile Interval}
\label{section:reversePercentile}

The bootstrap percentile interval has no particular derivation---it just
works.
This is uncomfortable for a mathematically-trained statistician,
and unsatisfying for a mathematical statistics course.

The natural next step is the {\em reverse bootstrap percentile interval},
called ``basic bootstrap confidence limits'' in \citeP{davi97}.
We assume that the bootstrap distribution of
$\deltahat^* = \thetahat^* - \thetahat$
can be used to approximate the distribution of
$\deltahat = \thetahat - \theta$.
For comparison, in the bootstrap estimate of bias we used
$E(\thetahat^* - \thetahat)$ to estimate
$E(\thetahat - \thetahat)$.

We estimate the CDF for $\deltahat$ using the bootstrap distribution
of $\deltahat^*$.
Let $q_\alpha$ be the $\alpha$ quantile of the bootstrap
distribution, i.e.~$\alpha = P(\deltahat^* \le q_\alpha)$.
Then
\begin{eqnarray*}
  1-\alpha
  &=&       P(q_{\alpha/2} < \thetahat^* - \thetahat < q_{1-\alpha/2}) \\
  &\approx& P(q_{\alpha/2} < \thetahat - \theta < q_{1-\alpha/2}) \\
  &=&       P(-q_{\alpha/2} > \theta - \thetahat > -q_{1-\alpha/2}) \\
  &=&       P(\thetahat -q_{\alpha/2} > \theta > \thetahat -q_{1-\alpha/2})
\end{eqnarray*}
Hence the confidence interval is of the form
\begin{eqnarray*}
  (\thetahat - q_{1-\alpha/2}, \thetahat - q_{\alpha/2})
 &=& (2 \thetahat - \Ghat^{-1}(1-\alpha/2), 2 \thetahat - \Ghat^{-1}(\alpha/2)).
\end{eqnarray*}

This is the mirror image of the bootstrap percentile interval;
it reaches as far above $\thetahat$ as the bootstrap
percentile interval reaches below.
For example, for the CLEC mean, the
sample mean is $16.5$, the percentile interval is
$(10.1, 25.4) = 16.5 + (-6.4, 8.9)$,
and the reverse percentile interval is
$16.5 + (-8.9, 6.4) = 2\cdot 16.5 - (25.4, 10.1) = (7.6, 22.9)$.

For applications with simple bias, like the left side
of Figure~\ref{figure:CIcurves2}, this interval behaves well.
But when there is skewness, like for the CLEC data or
the right side of Figure~\ref{figure:CIcurves2}, it does
exactly the wrong thing.

% Our intuition suggests that if the data is positively skewed, that
% maybe the percentile interval reaches too far to the right because it
% is "affected by outliers".
% If the bootstrap distribution
% reaches farther to the right, then the sampling distribution
% probably did too; the observed statistic could be a lot larger than
% the parameter, so a confidence interval for the parameter
% should reach farther left.
%
% Except it doesn't work that way.
% That intuition is wrong. When the data has
% a long right tail, a good interval will actually reach farther to the
% right.

The reason is worth discussing in a Mathematical Statistics
class---that the sampling distribution is not one constant thing,
but depends very strongly on the parameter, and the bootstrap distribution
on the observed statistic.
% For a skewed population, there is {\em acceleration}
% (Section~\ref{section:acceleration}).
When sampling from a skewed population, the distribution of
$\deltahat = \hat\theta - \theta$
depends strongly on $\theta$;
similarly the bootstrap
distribution of $\deltahat^*$
is strongly dependent on $\hat\theta$.
Hence the bootstrap distribution of $\deltahat^*$
is a good approximation for the distribution of $\deltahat$
only when $\thetahat = \theta$. That isn't very useful for
a confidence interval.

The interval also does exactly the wrong thing for nonlinear transformations.

The reverse percentile interval is almost the worst of everything:
\begin{squeezeitemize}
\item the same small-sample problems as the percentile interval,
\item asymmetrical in the wrong direction for skewed data,
\item asymmetrical in the wrong direction for nonlinear transformations.
\end{squeezeitemize}
Its coverage accuracy in
Figures~\ref{figure:coverage2} and~\ref{figure:coverage3}
below is terrible.

\boxx{Reverse Percentile Interval}
{
The reverse percentile interval has pedagogical value, but don't
use it in practice.

See Figure~\ref{figure:coverage3} to see how badly it performs.%
}

\cite{hall92a} calls the bootstrap percentile interval
``the wrong pivot, backwards''; the reverse percentile interval
is that same wrong pivot, forward.
The moral of the story is that you are going to use the wrong pivot,
to do it backwards.

But better yet is to use the right pivot.
This leads us to the next interval.
$\deltahat$ is the wrong pivot because it isn't even close to
{\em pivotal}---a pivotal statistic is one whose distribution
is independent of the parameter.
We can use a statistic that is closer to pivotal, namely
a $t$~statistic.

\subsection{Bootstrap T}
\label{section:bootstrapT}

Standard normal theory says that when the population is normal,
that $\Xbar$ and $s$ are independent, and the $t$~statistic
$t = (\Xbar - \mu)/(s/\sqrt{n})$ has a $t$~distribution.

Reality says otherwise. When the population is positively skewed,
then $\Xbar$ and $s$ are positively correlated, the correlation
doesn't get smaller with large $n$, and the $t$~statistic does not
have a $t$~distribution.
In fact, while $\Xbar$ is positively skewed,
$t$ is twice as skewed in the opposite direction and is has a negative mean,
because the denominator $s$ is
more affected by large
observations than the numerator $\Xbar$ is.

\begin{figure}[\figureplace]     % scripts/Verizon.R
\centerline{\includegraphics[width=\figurewidth]{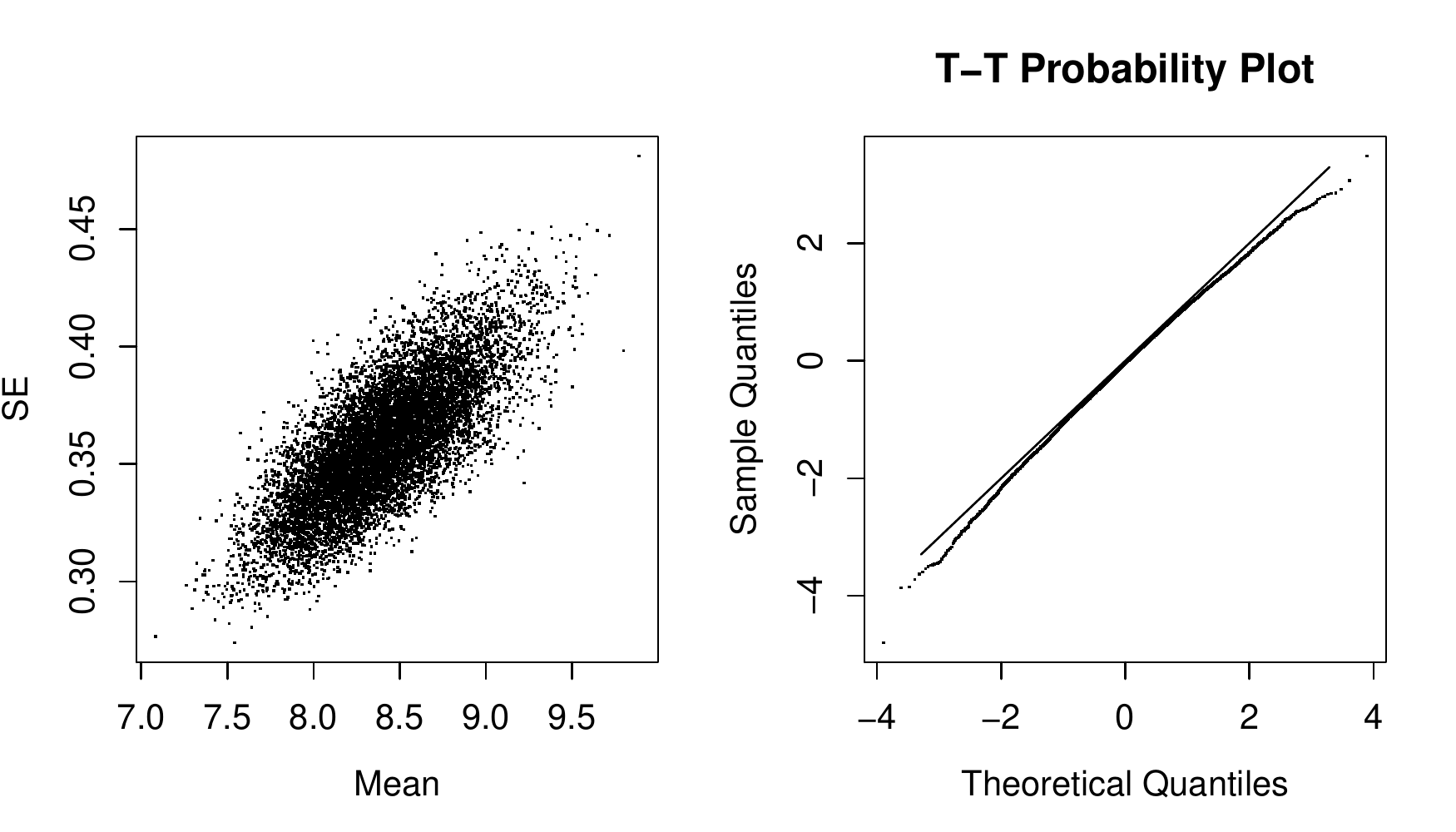}}
\renewcommand{\mycaption}{CLT with n=1664.}
\caption[\mycaption]{\label{figure:bootVerizonT}
{\em \mycaption}
Left: scatterplot of bootstrap means and standard errors, ILEC data.
Right: bootstrap~$t$ distribution.
}    \end{figure}

Figure~\ref{figure:bootVerizonT} shows the correlation of $\Xbar$ and $s$
and the skewness of the $t$~statistic, with $n=1664$.
Compare the right panel,
showing negative skewness in the $t^*$ statistic,
to the top right panel of Figure~\ref{figure:bootVerizon1},
showing smaller positive skewness in $\xbar^*$.

So, the $t$ statistic does not have a $t$~distribution.
We can bootstrap to estimate the actual distribution, then use quantiles
of that distribution in the confidence interval.
\cite{efro93} call this ``Confidence intervals based on bootstrap
tables''---the bootstrap is used to generate the right table for an
individual dataset, rather than using a table from a book.

In general, the $t$~statistic is
\begin{equation}
  t = \frac{\thetahat - \theta}{\Shat}
  \label{eqn.tstat}
\end{equation}
where $\Shat$ is a standard error calculated from the original sample.
The bootstrap~$t$ substitutes
\begin{equation}
  t^* = \frac{\thetahat^* - \thetahat}{\Shat^*}
  \label{eqn.bootTstat}
\end{equation}
where the $^*$ quantities are from each bootstrap sample.
Then, assuming that the distribution of $t^*$ is approximately
the same as the distribution of $t$, we perform a similar calculation
as for the reverse bootstrap percentile interval.
Let $q_\alpha$ be the $\alpha$ quantile of the bootstrap~$t$
distribution, then
\begin{eqnarray*}
  1-\alpha
  &=&       P(q_{\alpha/2} < t^* < q_{1-\alpha/2}) \\
  &\approx& P(q_{\alpha/2} < t < q_{1-\alpha/2}) \\
  &=&       P(q_{\alpha/2} \Shat < \thetahat - \theta < q_{1-\alpha/2} \Shat) \\
  &=&       P(-q_{\alpha/2}\Shat > \theta - \thetahat > -q_{1-\alpha/2} \Shat) \\
  &=&       P(\thetahat -q_{\alpha/2}\Shat > \theta > \thetahat -q_{1-\alpha/2}\Shat)
\end{eqnarray*}
Hence the confidence interval is of the form
$$(\thetahat - q_{1-\alpha/2}\Shat, \thetahat - q_{\alpha/2}\Shat).$$
The upper quantile of the bootstrap~$t$ distribution is used for the
lower endpoint, and vice versa.

The right panel of Figure~\ref{figure:bootVerizonT}
shows the bootstrap distribution of the $t$~statistic for the
ILEC data. Even with a large sample, $n=1664$, the distribution
is far enough from a $t$~distribution to make the standard $t$~interval
inaccurate.
This table shows how far the endpoints for the $t$, percentile,
and bootstrap~$t$ intervals are above and below the sample mean:
%\begin{table}[\tableplace]
\begin{center}
\begin{tabular}{rrrrr}
\hline
      &t & percentile& bootstrapT& tSkew\\
\hline
 2.5\% &$-0.701$ & $-0.683$ &$-0.646$ &$-0.648$\\
97.5\% &$ 0.701$ & $ 0.718$ &$ 0.762$ &$ 0.765$\\
\hline
\end{tabular}\end{center}
%\end{table}
The bootstrap~$t$ is more than three times as asymmetrical
as the percentile interval; in other words,
the percentile intervals makes one-third of a skewness correction.
``tSkew'' is an asymptotic skewness-adjusted $t$ interval,
(equation \ref{equation:adjustedTInterval}) in Section~\ref{section:asymptotics}; it closely matches the bootstrap~$t$.

In Figures~\ref{figure:coverage2} and~\ref{figure:coverage3}
below, the bootstrap~$t$ does the best
of all intervals in overall coverage accuracy.

\boxx{The bootstrap~$t$ doesn't pretend}{
$t$~statistics do not have $t$~distributions when populations are skewed.

Bootstrap~$t$ confidence intervals and tests use a $t$~statistic,
but estimate its actual distribution by bootstrapping
instead of pretending that it has a $t$~distribution.

They have pedagogical value, and
are second-order accurate.%
}

% And this table shows the estimate probability for the interval to miss
% on each side (caveat, with probabilities estimated by the bootstrap~$t$):
% %latex.default(round(temp4[, c(2, 3, 1)], 3), file = "")%
% %\begin{table}[\tableplace]
% \begin{center}
% \begin{tabular}{rrr}
% \hline
%     t & percentile& bootstrapT\\
% \hline
% $0.017$&$0.020$&$0.025$\\
% $0.035$&$0.032$&$0.025$\\
% \hline
% \end{tabular}\end{center}
% %\end{table}

%\paragraph{Need a Standard Error}

To use the bootstrap~$t$ interval you need standard errors---for
the original sample, and each bootstrap sample.
When formula standard errors are not available, we can use the bootstrap
to obtain these standard errors \citeP{efro93}.
This involves an {\em iterated bootstrap}, in which
a set of second-level bootstrap samples is drawn from each top-level
bootstrap sample, to estimate the standard error for that bootstrap sample.
If $r_1$ bootstrap samples are drawn from the original data, and $r_2$
second-level samples from each top-level sample, there are a total
of $r_1 + r_1r_2$ samples.

%\paragraph{Other Applications}

\cite{efro93} note that the bootstrap~$t$ is particularly suited to
location statistics like the sample mean, median, trimmed mean,
or percentiles, but performs poorly for a correlation coefficient;
they obtain a modified version by using a bootstrap~$t$ for
a transformed version of the statistic $\psi = h(\theta)$,
where $h$ is a {\em variance-stabilizing transformation}
(so that $\Var(\hat\psi)$ does not depend on $\psi$) estimated
using a creative use of the bootstrap.
The same method improves the reverse percentile interval \citeP{davi97}.

\subsection{Confidence Intervals Accuracy}
\label{section:CIAccuracy}

\begin{figure}[\figureplace]     % scripts/coverage.R
\centerline{\includegraphics[width=\figurewidth]{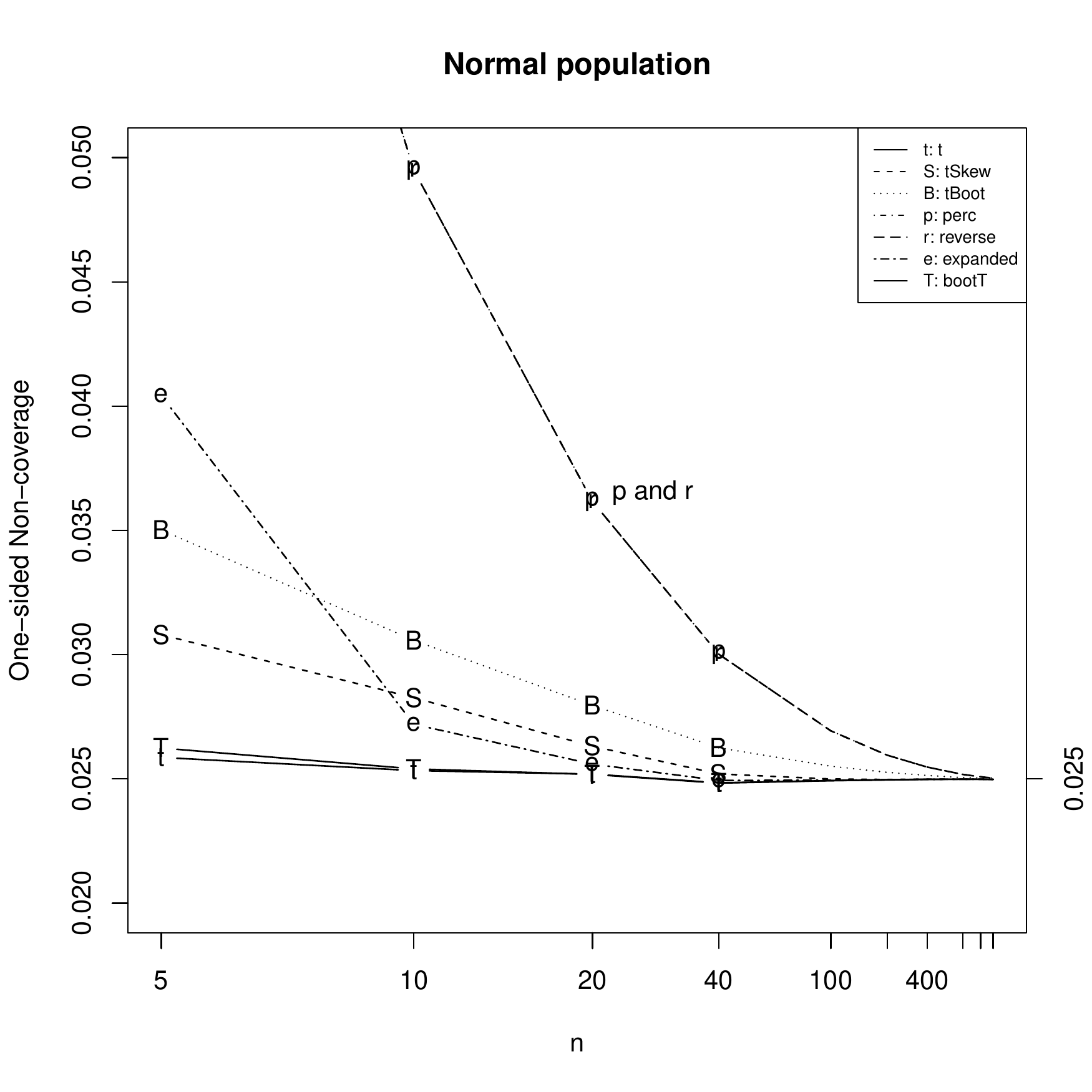}}
\renewcommand{\mycaption}{Confidence interval one-sided miss probabilities for normal populations.}
\caption[\mycaption]{\label{figure:coverage2}
{\em \mycaption}
The intervals are described
at the beginning of Section~\ref{section:CIAccuracy}.
Only one side is shown, because non-coverage probabilities are
the same on both sides.
}    \end{figure}

\begin{figure}[\figureplace]     % scripts/coverage.R
\centerline{\includegraphics[width=\figurewidth]{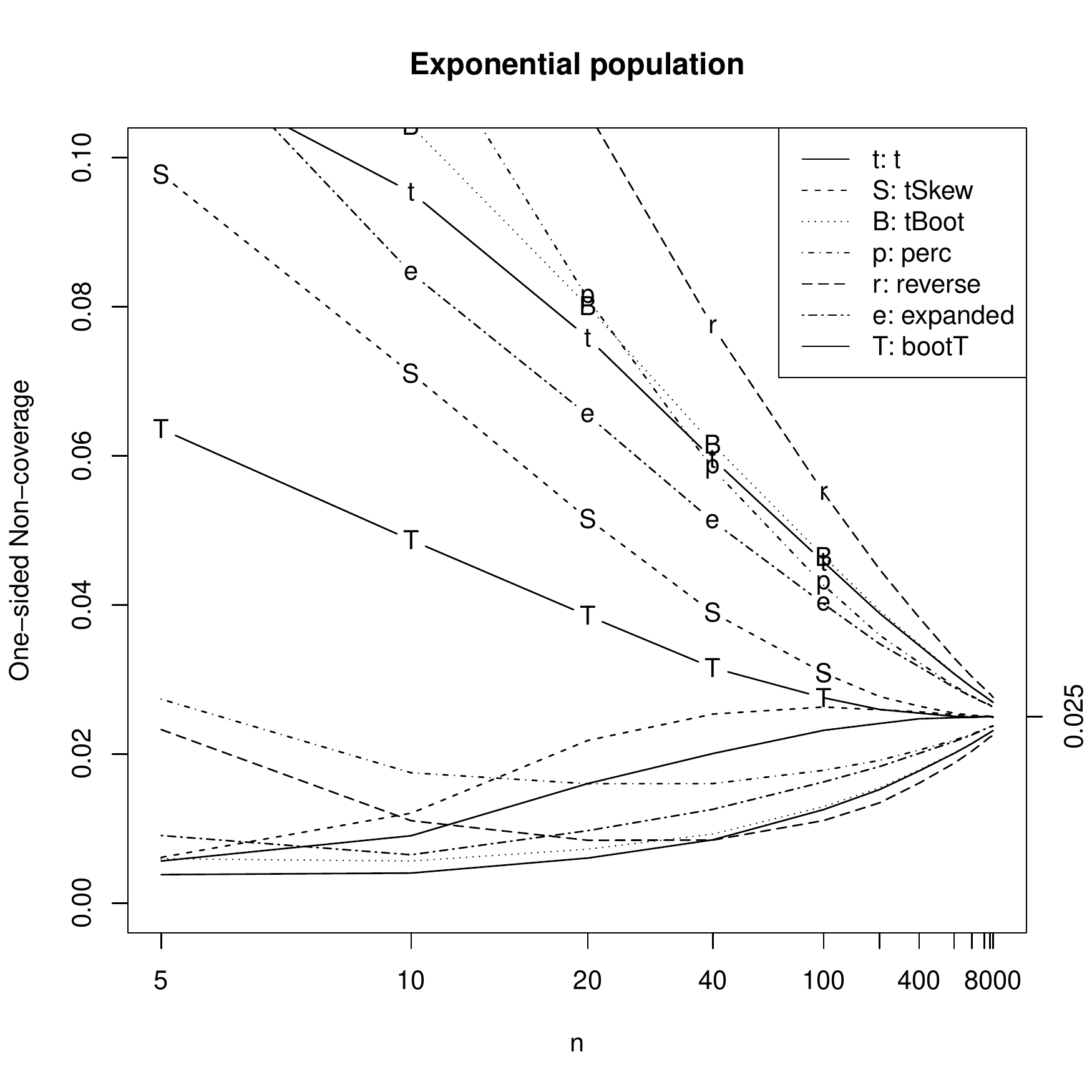}}
\renewcommand{\mycaption}{Confidence interval one-sided miss probabilities for gamma populations.}
\caption[\mycaption]{\label{figure:coverage3}
{\em \mycaption}
The intervals are described
at the beginning of Section~\ref{section:CIAccuracy}.
The lines with codes are non-coverage probabilities on the right,
where the interval is below $\theta$.
The lines without codes correspond to the left side.
}    \end{figure}

Figures~\ref{figure:coverage2} and~\ref{figure:coverage3}
show estimated non-coverage probabilities
for normal and exponential populations, respectively.
The intervals are:
\begin{squeezedescription}
\item[t = t:] ordinary $t$~interval;
\item[S = tSkew:] $t$~interval with skewness correction,
(equation \ref{equation:adjustedTInterval}) in Section~\ref{section:asymptotics};
\item[B = tBoot:] $t$~interval with bootstrap standard error;
\item[p = perc:] bootstrap percentile interval;
\item[r = reverse:] reverse percentile interval;
\item[e = expanded:] expanded percentile interval;
\item[T = bootT:] Bootstrap~$t$.
\end{squeezedescription}

\paragraph{Normal population}
The percentile and reverse percentile (``p'' and ``r'' on the plot) do poorly.
For normal data, that interval corresponds to
\begin{squeezeitemize}
\item using $z$ instead of $t$
\item using a divisor of $n$ instead of $n-1$ when calculating SE,
\item doing a partial correction for skewness
\item add some extra variability because it pays attention to skewness.
\end{squeezeitemize}
For normal data the skewness correction doesn't help.
For small samples, the other three things kill them.

The expanded percentile interval (plot label ``e'')
(Section~\ref{section:expandedPercentile})
does much better.
It is still poor for $n=5$, due to extra variability from estimating
skewness.

The $t$~interval (``t'') and bootstrap~$t$ (``T'') interval do very well.
That is not surprising for the $t$~interval, which is optimized for this
population, but the bootstrap~$t$ does extremely well, even for
very small samples.

The $t$~interval with skewness correction
(``S'', equation \ref{equation:adjustedTInterval}),
does a bit worse than an ordinary~$t$ interval,
and the $t$~interval with bootstrap SE (``B'')
a bit worse yet.

\paragraph{Exponential population}

This is a much harder problem.  All of the intervals badly under-cover
on the right---the intervals are not long
enough on the right side.
And most over-cover (by smaller amounts) on the left.

The bootstrap~$t$ interval (``T'') does best, by a substantial margin.
Next best is the $t$~interval with skewness correction (``S'').
Those are the two second-order accurate intervals.

The other intervals are all quite poor.
The expanded percentile (``e'') is the best of the bunch, and
the reverse percentile interval (``r'')
is the worst.
The percentile interval (``p'') is poor for small samples,
but better than the ordinary $t$ (``t'') for $n \ge 35$.

\begin{figure}[\figureplace]     % scripts/coverage.R
\centerline{\includegraphics[width=\figurewidth]{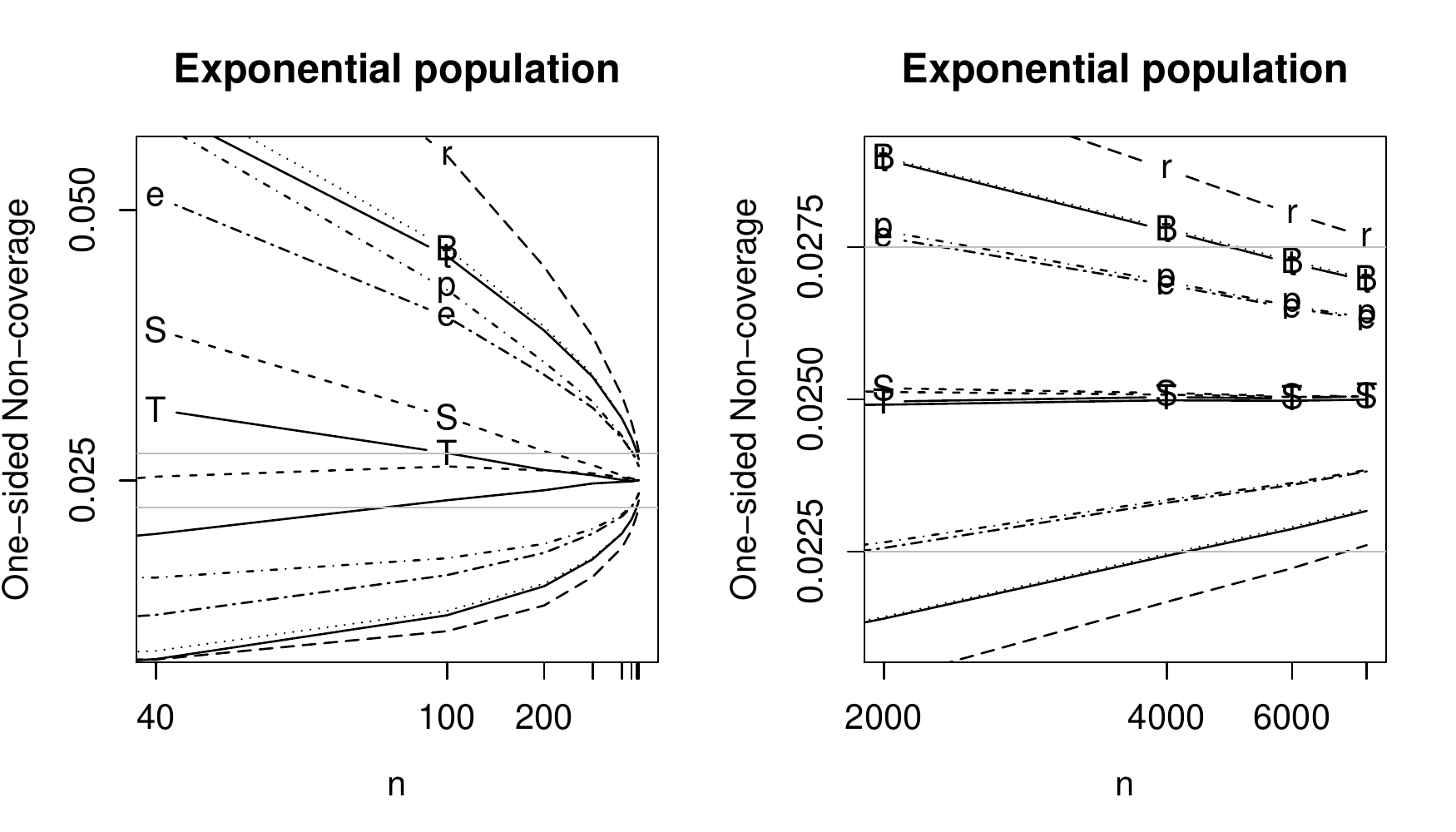}}
\renewcommand{\mycaption}{Zoom in: confidence interval one-sided miss probabilities for gamma populations.}
\caption[\mycaption]{\label{figure:coverage4}
{\em \mycaption}
Labels and line types are the same as the previous figures.
The intervals are described
at the beginning of Section~\ref{section:CIAccuracy}.
Here we zoom in. In the left panel, the $x$ axis scaling
is $x=-n^{-1}$, so that second-order accurate intervals appear to converge
linearly. In the right panel, the scaling is $x=-n^{-1/2}$.
The estimated sample sizes necessary for one-sided coverage errors to be
within 10\% of the true value (i.e.\ between $0.0225$ and $0.0275$
are $n \ge 101$ for bootT, 220 for tSkew, 2235 for expanded, 2383 for perc,
4815 for t, 5063 for tBoot, and over 8000 for reverse.
}    \end{figure}

\boxx{Confidence Interval Accuracy}
{
Accurate coverage for skewed populations is hard.
The bootstrap~$t$ interval is the best of the intervals considered
here, with the skewness-adjusted~$t$ next best
(see Section~\ref{section:Johnson}).
These are second-order accurate, and give coverage within 10\%
for $n \ge 101$ and $n \ge 220$, respectively, for exponential populations.
The other intervals are only first-order accurate, and
require $n \ge 2235$ or more, including roughly $n \ge 5000$
for standard $t$~intervals.%
}

Table~\ref{table:effects} summarizes some of the effects that
can make confidence intervals inaccurate, the order of the
effects, and which intervals are affected.

\begin{table}[\tableplace]
\begin{center}
\begin{tabular}{lllllll}
\hline
Effect & Size & t   & tBoot   & perc & reverse&bootT\\
\hline
bias                    & $O(n^{-1/2})$& yes & yes& $\times 2$ &no &no \\
skewness                & $O(n^{-1/2})$& yes & yes& partial& $\times 2$&no \\
transformations         & $O(n^{-1/2})$& yes & yes& no & $\times 2$& partial\\
narrowness bias         & $O(n^{-1})$  & no  & yes& yes & yes& no \\
$z$ vs $t$ (random $s$) & $O(n^{-1})$  & partial & partial & yes &yes &no \\
random skewness         & $O(n^{-3/2})$& no  & no & yes & $\times 2$ &no\\
\hline
\end{tabular}\end{center}
\renewcommand{\mycaption}{Confidence Interval Issues.}
\caption[\mycaption]{\label{table:effects}
{\em \mycaption}
How various issues affect different confidence intervals.
``Yes'' indicates the interval is affected, ``no'' that it is not,
``$\times 2$ that it is affected twice as much as other intervals,
and ``partial'' that it is partially affected.
The $t$ methods make the right correction for random $s$ for normal
populations, but not for other distributions.
The bootstrap~$t$ interval is not exactly transformation invariant,
but is close enough to have no $O(n^{-1/2})$ effect.
}
\end{table}

\paragraph{Simulation details}
Figures~\ref{figure:coverage2}, \ref{figure:coverage3},
and~\ref{figure:coverage3}
were produced using $10^4$ samples
(except $5\cdot 10^3$ for $n \ge 6000$),
with $r=10^4$ resamples for bootstrap intervals, using a variance reduction
technique based on conditioning. For normal data, $\Xbar$ and
$V = (X_1-\Xbar, \ldots, X_n-\Xbar)$ are independent,
and each interval is translation-invariant (the intervals for
$V$ and $V+a$ differ by $a$. Let $U$ be the upper endpoint of an interval,
and
$P(U < \mu) = E_V(E(U < \mu|V))$.
The inner expected value is a normal probability:
$E(U < \mu|V) = P(\Xbar + U(V) < \mu|V) = P(\Xbar < \mu - U(V)|V)$.
This increased the accuracy by a factor ranging from $9.6$
(for $n=5$) to over 500 (for $n=160$).
Similarly, for the exponential distribution, $\Xbar$ and
$V = (X_1/\Xbar, \ldots, X_n/\Xbar)$ are independent,
and we use the same conditioning procedure.
This reduces the Monte Carlo variance by a factor ranging from $8.9$
(for $n=5$) to over 5000 (for $n=8000$).
The resulting accuracy is as good as using 89000 or more samples
without conditioning.

\subsubsection{Asymptotics}
\label{section:asymptotics}
Here are asymptotic approximations for the mean,
% one-sample and two-sample means,
including estimates of the actual rejection/noncoverage probabilities
for $t$ procedures,
and skewness-adjusted $t$ inferences.
% for the one-sample and two-sample means.

Let $X_1, \ldots, X_n$ be $\iidB$ with mean $\mu$, variance $\sigma^2$,
and third central moment $\iota = E((X-\mu)^3)$.
Let $\gamma = \iota/\sigma^3$ be the skewness of $X$,
then the skewness of $\Xbar$ is $\gamma/\sqrt{n}$.

The first-order Edgeworth approximation for the distribution of $\Xbar$
is
\begin{eqnarray*}
  P(\Xbar \le x)
  &=& \Phi(z) - \frac{\gamma}{6\sqrt{n}}(z^2-1)\phi(z) + O(n^{-1}) \\
  &=& \Phi(z) - \kappa\  (z^2-1)\phi(z) + O(n^{-1})
\end{eqnarray*}
where $\Phi$ and $\phi = \Phi'$ are the standard normal cdf and pdf,
$z = (x-\mu)/(\sigma/\sqrt{n})$,
and $\kappa = \gamma / (6\sqrt{n})$.

The first three moments of the $t$~statistic $(\Xbar-\mu)/(s/\sqrt{n})$
are:
\begin{eqnarray*}
 E(t) &=& \frac{-\gamma}{2\sqrt{n}} + O(n^{-3/2}) \\ % -3A
 E(t^2) &=& 1 + O(n^{-1}) \\
 E(t^3) &=& \frac{-7\gamma}{2\sqrt{n}} + O(n^{-3/2})\\ % -21A
 E((t-E(t))^3) &=& \frac{-2\gamma}{\sqrt{n}} + O(n^{-3/2}) % -12A
\end{eqnarray*}
(for continuous distributions with enough finite moments)
% are the terms for E(t) and E(t^3) smaller?
so the skewness of $t$ is
twice as large as the skewness of $\Xbar$, and in the opposite direction.

The first-order Edgeworth approximation for the distribution of $t$
is
\begin{eqnarray*}
  P(t \le x)
%  &=& \Phi(x+3\kappa) + 2\kappa\  ((x+3\kappa)^2-1) \phi(x+3\kappa) + O(n^{-1}) \\
  &=& \Phi(x) + \kappa\  (2x^2 + 1) \phi(x) + O(n^{-1}).
\end{eqnarray*}
We can use this to estimate the rejection probabilities for a hypothesis
test.
Plugging in one-sided critical values
gives
$$ P(t \ge t_{\alpha,n-1})
  = \alpha - \kappa\  (2 t_{\alpha,n-1}^2 + 1) \phi(t_{\alpha,n-1}) + O(n^{-1})
$$
The error is the difference between the probability and $\alpha$.
For large $n$ the error is approximately
$\kappa\  (2 \zSub^2 + 1) \phi(\zSub)$.
To reduce this to 10\% of the desired value
(so the actual rejection probabilities are between 0.0225 and 0.275)
requires
\begin{equation}
n \ge \left( \frac{\gamma}{6} \frac{10}{\alpha}
  (2 \zSub^2 + 1) \phi(\zSub) \right)^2
\end{equation}

For an exponential distribution with skewness 2, that
requires $n > 4578$. Simulation results suggest that the
actual requirement is closer to 5000.
The usual ``$n \ge 30$'' rule isn't even close.

% \begin{figure}[\figureplace]
% \centerline{
%   \includegraphics[width=2.5in]{n30-1}
%   \includegraphics[width=2.5in]{n30-2}}
% \renewcommand{\mycaption}{Errors in coverage for $t$ and skewness-adjusted $t$~intervals.}
% \caption[\mycaption]{\label{figure:JohnsonInterval}
% {\em \mycaption}
% Mean for an exponential population.
% Unadjusted intervals require $n \ge 5000$ to achieve reasonable
% accuracy (within 10\% of the desired value).
% }    \end{figure}

\boxx{The CLT requires $n \ge 5000$; $n \ge 30$ isn't even close.}
{
  For $t$~tests and confidence intervals to be reasonably accurate
  (off by no more than 10\% on each side)
  requires $n \ge 5000$ for a 95\% interval or two-sided $\alpha=0.05$
  test, for an exponential population.

The central limit theorem acts over glacial time scales, when skewness
is present.

A corrected statistic for tests is
\begin{equation}
  t_1 = t + \kappa\  (2t^2+1),
  \label{equation:t1}
\end{equation}
where $\kappa = \mbox{skewness}/(6\sqrt{n})$.

A corrected confidence interval is
\begin{equation}
  \xbar + {s \over \sqrt{n}}(\kappa\  (1 + 2 t_{\alpha/2}^2) \pm t_{\alpha/2}).
  \label{equation:adjustedTInterval}
\end{equation}

The corrected procedures, and the bootstrap~$t$, are
second-order accurate, with errors $O(n^{-1})$.
$t$ procedures, and the bootstrap percentile interval,
are first-order accurate, with errors $O(n^{-1/2})$.%
}

\subsubsection{Skewness-Adjusted t Tests and Intervals}
\label{section:Johnson}

\cite{john78} gave a skewness-corrected $t$~statistic
\begin{equation}
  t_1 = t + \kappa\  (2 t^2 + 1)
  \label{equation:JohnsonTest}
\end{equation}
for use in hypothesis tests; with rejection if $|t_1| \ge \tSub$.
The confidence interval given there drops terms that are
needed for a second-order accurate interval;
\cite{klei86} obtains an interval by solving $t_1$ for $\mu$
(a quadratic equation),
but a simpler interval is
\begin{equation}
  \xbar + {s \over \sqrt{n}}(\kappa\  (1 + 2 t_{\alpha/2}^2) \pm t_{\alpha/2}).
  \label{equation:JohnsonInterval}
\end{equation}
We term this a {\em skewness-adjusted $t$~interval}.
% This wasn't in any of the articles I looked at. It would be good
% to do a more complete search.

A simple estimate for $\gamma$ (needed in $\kappa = \gamma/(6\sqrt{n})$) is
$(1/n)\sum((x_i-\xbar)^3) / s^3$.
% A less-biased estimate is
% $(n/((n-1)(n-2)) \sum((x_i-\xbar)^3) / s^3$.
This is biased toward zero, which makes the results a bit closer to $t$
results in small samples.

Equations (\ref{equation:JohnsonTest}) and (\ref{equation:JohnsonInterval})
are non-monotone (in $t$ and $t_{\alpha/2}$, respectively),
and for small samples with large skewness should be tweaked by
flattening the curve beyond the max or min.

For comparison, the endpoints of a bootstrap percentile interval
are
\begin{equation}
  \xbar + {s \over \sqrt{n}}(\kappa\  (\zSub^2 - 1) \pm \zSub)
   + O_P(n^{-3/2}).
  \label{equation:percentileCornishFisher}
\end{equation}
For large $n$, with $\tSub \approx \zSub \approx 2$,
this has about a third of the asymmetry of the skewness-adjusted
$t$~interval.

% Possibly give my gamma transformation as an alternative.

% The first-order Edgeworth approximation is
% $$ P(t < x) =  \Phi(x) - \phi(x) c_1/2
%               - \frac{-c_1 - 9 c_2)}{12}(x^2 - 1) \phi(x) + O(n^{-1})$$
% where
% $c_1 = (\frac{\mu_{31}}{n_1^2} - \frac{\mu_{32}}{n_2^2}) / \sigma^3_\delta$
% and $c_2 = (\frac{\mu_{31}}{n_1^2} + \frac{\mu_{32}}{n_2^2})
%                      (\frac{\sigma_1^2}{n_1} - \frac{\sigma_2^2}{n_2}) /
%             \sigma^5_\delta$,
%a corrected $t$~statistic for hypothesis testing is
%$$ t_2 =  t +- \frac{c_2}{2} + \frac{c_1 + 9 c_2}{12}(t^2-1)

\section{Bootstrap Sampling Methods}
\label{section:samplingMethods}

In this section we discuss a number of bootstrap sampling procedures
for different applications.

The general rule is to
sample in the same way the data were drawn, except
to condition on the observed information, and any constraints.

For example, when comparing samples of size $n_1$
and $n_2$, we fix those numbers and do a two-sample bootstrap
with sizes $n_1$ and $n_2$, even if the original sampling procedure
could have produced different counts.

In permutation testing to compare two samples,
we sample in a way that is consistent with the
null hypothesis that the distributions are the same;
we condition on combined data,
letting only the assignment of labels be random.

Conditioning on the observed information
comes up in more subtle ways in other contexts,
most notably regression.

\boxx{General Rule for Sampling}
{ In general, we
sample in the same way the data were drawn, except
to condition on the observed information, and satisfy any constraints.%
}

\subsection{Bootstrap Regression}
\label{section:bootstrapRegression}

Suppose we have $n$ observations, each with $Y$ and some number of $X$'s,
with each observation stored as a row in a data set.

The two basic procedures when bootstrapping regression are:
\begin{squeezeitemize}
\item bootstrap observations, and
\item bootstrap residuals.
\end{squeezeitemize}
The latter is a special case of a more general rule:
\begin{squeezeitemize}
\item sample $Y$ from its estimated conditional distribution given $X$.
\end{squeezeitemize}

In bootstrapping observations, we sample with replacement from the
rows of the data; each $Y$ comes with the corresponding $X$'s.
In any bootstrap sample some observations may be repeated multiple times,
and others not included.
We use this in bootstrapping R-squared, Figure~\ref{figure:bootstrapRSquared},
and in the left panel of Figure~\ref{figure:bootstrapBadFit}.

In bootstrapping residuals, we fit the regression model,
compute predicted values $\Yhat_i$ and residuals $e_i = Y_i - \Yhat_i$,
then create a bootstrap sampling using the same $X$ values as in
the original data, but with new $Y$ values obtained using the
prediction plus a random residual, $Y_i^* = \Yhat_i + e_i^*$,
where the residuals $e_i^*$ are sampled randomly with replacement
from the original residuals.
We use this
in the right panel of Figure~\ref{figure:bootstrapBadFit}.

Bootstrapping residuals corresponds to a designed experiment,
where the $x$ values are fixed and only $Y$ is random,
and bootstrapping observations  to randomly sampled data
where both $X$ and $Y$ are sampled. By the principle of sampling the
way the data were drawn, we would bootstrap observations if the $X$'s
were random. But we don't have to.

Consider the usual formula for the standard error in simple linear
regression, $\SE(\betahat_1) = s_r/\sqrt{\sum((x_i-\xbar)^2)}$,
where $s_r = \sqrt{(n-p)^{-1}\sum e_i^2}$ is the residual standard deviation.
The derivation of this SE assumes that the $X$'s were fixed, and in practice
we use it even if the $x$'s were random. In doing so, we condition
on the observed information---here given by $\sum((x_i-\xbar)^2)$;
the larger this is, the more accurate $\betahat$ is.

Similarly, in bootstrapping, we may resample the residuals, conditioning
on the observed information.

This can make a huge difference in multivariate regression, where
bootstrapping observations can be just plain dangerous.
For example, suppose one of the $X$'s is a factor variable with a rare
level, say only 5 observations. When resampling observations,
a bootstrap sample could omit those five observations entirely; the
regression software
would be unable to estimate a coefficient for that level.
Worse, there could be just one or two observations from that level
in a bootstrap sample; then the software would silently produce garbage,
estimates with high variance.
The same problem occurs if there are multiple factors in a model with
interactions and there are rare combinations of interactions.
And it occurs with continuous variables, when some bootstrap samples
may have linear combinations of the $X$'s with low variability.
We avoid these problems by bootstrapping residuals.

Bootstrapping residuals is a special case of a more general rule, to
sample $Y$ from its estimated conditional distribution given $X$.
For example, when bootstrapping logistic regression,
we fit the model, and calculate predicted values
$\Yhat_i = \hat E(Y | X=x_i) = \hat P(Y = 1|X = x_i)$.
To generate a bootstrap sample, we keep the same $X$'s, and let
$Y_i = 1$ with probability $\Yhat_i$, otherwise $Y_0 = 0$.

The more general rule is also helpful in some cases that
bootstrapping residuals behaves poorly---lack of fit, and
heteroskedasticity.
Refer back to Figure~\ref{figure:bootstrapBadFit},
where there was lack of fit.
The residuals are inflated---they have systematic bias in addition to
random variation---so the vanilla bootstrap residuals procedure will
overstate the variance of the regression coefficients.

Bootstrapping observations is also affected by poor fit;
see Figure~\ref{figure:bootstrapBadFit}, where
both bootstrapping observations
residuals show similar variability. When there are relative many
observations with large $x$ in a bootstrap sample, the resulting slope
is large positive; when there are relatively many with small $x$,
the resulting slope is negative; and the height of the curve at $\xbar$
depends on how many observations come from values of $x$ in the middle.

An alternative is to draw the random residual $e_i^*$ for observation $i$
from nearby observations (with similar $x$ values).

Similarly, if there is heteroskedasticity, with greater residual
variance for larger predictions, we may draw $e_i^*$ from observations
with similar predicted values.

The narrowness bias factor for bootstrapping residuals in
multiple linear regression
is $\sqrt{(n-p)/n}$ where $p$ is the number
of coefficients in the model.

\boxx{Resampling for Regression}
{
  The two common ways of resampling for regression are to
sample observations, and
sample $Y$ from its conditional distribution given $X$
(including the special case of resampling residuals).
The latter conditions on the observed information,
and avoids nasty small-sample problems.%
}

\paragraph{Pedagogical Value}

There are two ways that bootstrapping in regression
is particularly useful pedagogically.
The first is to help students understand
the variability of regression predictions
by a graphical bootstrap.
For example, in Figure~\ref{figure:bootstrapBadFit} we bootstrapped
regression lines; those lines help students understand the variability
of slope and intercept coefficients, and of predictions at each value of
$x$. The more we extrapolate in either direction, the more variable
those predictions become.

The second is to help students understand the difference between
a confidence interval and a prediction interval. For large datasets,
the regression lines won't vary much and the confidence intervals
are narrow, but the variability of individual observations above and
below those lines remains constant regardless of how much data there is.

% For higher level courses,
% it helps them see why estimates for the slope and intercept are negatively
% correlated when $\xbar$ is positive---most lines will go near
% $(\xbar, \ybar)$, and those lines with larger slope hit the $y$ axis lower.

\boxx{Pedagogical Value of the Bootstrap in Regression}
{
The bootstrap shows the variability of regression predictions
including the effect of extrapolation,
and helps students understand the difference between confidence intervals
and prediction intervals.%
}

\subsection{Parametric Regression}

An alternative to nonparametric regression is parametric regression,
where we assume a model (e.g. a gamma distribution with unknown shape
and scale), estimate parameters for that model, then draw
bootstrap samples from the parametric model with the estimated parameters.

The procedure mentioned above for bootstrapping in logistic regression
could be called a parametric regression.

Assuming a parametric structure can reduce the variance of estimates,
at the cost of introducing bias if the model does not fit.
In bootstrapping, for small $n$
we may prefer a parametric bootstrap, and for large $n$ use
a nonparametric bootstrap and rely on the data to reflect the population.

\subsection{Smoothed Bootstrap}
\label{section:smoothedBootstrap}

The smoothed bootstrap is a compromise between parametric and nonparametric
approaches; if we believe the population is continuous, we may sample
from a continuous $\Fhat$ rather than the empirical
distribution $\Fhat_n$.

A convenient way to do this is to sample from a {\em kernel density estimate},
e.g.~from density $\hat f(v) = n^{-1}\sum_{i=1}^n \phi(x_i-v;h)$
where $\phi(\cdot;h)$ is the density for a normal distribution with mean 0
and standard deviation $h$.
To generate a bootstrap sample from this distribution, we draw an
ordinary bootstrap sample with replacement from the data,
then add a random normal variate (with $\sigma=h$) independently to
each observation.

The choice $h = s/\sqrt{n}$ corrects for the narrowness bias
in the case of the sample mean \citeP{hest04b}.

For data that must be positive, like time, smoothing could produce
negative values. A remedy is to transform the data (e.g.~$\log(\mbox{time})$),
smooth on that scale, then transform back.

Smoothing is not common, because does not generalize well to
multivariate and factor data, and it is rarely needed to make
bootstrap distributions continuous.
For statistics like the sample mean, if the original distribution
was continuous then the number of distinct values in the theoretical
bootstrap distribution is ${2n-1 \choose n}$, so the distribution
is practically continuous except for small $n$ \citeP{hall86}.

\subsection{Avoiding Narrowness Bias}
\label{section:narrownessBias}

The smoothed bootstrap is one way to correct for narrowness bias,
but cannot be used in all situations, e.g.~for discrete data or factor data.
Two other procedures are more general.

One is to draw samples of size $n-1$ instead of $n$; for
two-sample or stratified applications, reduce by one in each group.

The other is {\em bootknife sampling} \citeP{hest04b};
to draw a bootstrap sample, omit one observation (randomly,
or systematically across all $r$ resamples), then draw a sample of
size $n$ with replacement from the remaining $n-1$ observations.

Both procedures add the right amount of extra variability in the
case of the sample mean; this is a good exercise for mathematical
statistics students.

\subsection{Finite Population}
\label{section:finitePopulation}

When the original sample is from a finite population and sampling
was done without replacement, we can use finite population bootstrap
sampling. This effectively incorporates a finite population correction
factor into bootstrap standard error estimates.

When $N$ is a multiple of $n$, we create a finite population with
$N/n$ copies of each observation, then draw bootstrap samples
without replacement from that finite population.

When $N$ is not a multiple of $n$, the natural approach is to
create a finite population using $\lfloor N/n \rfloor$ copies
of each observation, and selecting the remaining
$N - n \lfloor N/n \rfloor$ copies randomly without replacement.
For example, with $n=100$ and $N=425$, to use 4 copies of each
point, plus another 25 selected randomly.
However, that random selection adds extra variability (like the bootknife),
exactly the opposite of what we want to accomplish by paying
attention to the finite population.
The simplest alternative is to round up, using
$\lceil N/n \rceil$ copies of each observation.
Another is to round up or down, with the fraction of times to round each way
chosen to match the usual finite population correction factor for the mean.

\boxx{Other Bootstrap Sampling Methods}
{There are alternatives to the usual nonparametric bootstrap,
including parametric, smoothed, and finite-population methods.
There are ways to tweak bootstrap sampling to avoid narrowness bias.%
}

\section{Permutation Tests}
\label{section:permutationTests}

After a long detour we return to permutation testing.

We do four things in this section: give some details for
permutation tests, discuss two situations where permutation tests
easily apply, discuss situations where they do not,
and discuss bootstrap testing as an alternative.

\subsection{Details}
\label{section:permutationDetails}

First, about the name ``permutation test''---in the permutation
tests above, we picked $n_1$ observations without replacement
to label as the first sample, and labeled the others as the second
sample. This is equivalent to randomly permuting all labels,
hence the name.

The permutation test can be implemented deterministically or by
random sampling. In general there are
${{n \choose {n_1}}}$ possible partitions into two groups;
computing all is typically infeasible unless $n_1$ or $n_2$ is small, so
we usually use random sampling.
The special case of comparing two binomial proportions or testing
independence in a 2x2 table is
Fisher's exact test, and can be implemented deterministically using
the hypergeometric distribution.

When sampling, to compute a one-sided $P$-value we use
\begin{equation}
  \frac{x+1}{r+1}
\end{equation}
where $r$ is the number of resamples, and $x$ is the number
with $\thetahat^* \ge \thetahat$ (for upper $P$-values) or
$\thetahat^* \le \thetahat$ (for lower $P$-values).
In effect we include the original statistic as one of the resamples.
After all, the original partition is one of the
${{n \choose {n_1}}}$ possible partitions; by including it we
prevent reporting a $P$-value of zero, which is impossible.
% There is also a more technical explanation---pretend the permutation
% distribution is continuous and let $\thetahat^*_{(k)}$ be the
% $k$th order statistic, then $E(G(\thetahat^*_{(k)})) = $
% (that gets messy and not convincing).

To compute a two-sided $P$-value we calculate both one-sided $P$-values,
and use 2 times the smaller one. For example, to create
Figure~\ref{figure:permTV} we used $r=9999$, of which 44 of the
$\thetahat^*$ were greater than the original, 5 were equal, and 9950
were less.
The one-sided $P$-values are $(44+5+1)/1000$ and $(9950+5+1)/1000$,
and the two-sided $P$-value is $2\cdot50/10000$.

Permutation distributions need not
be symmetric or even centered at zero, so we measure the strength of the
evidence for each side separately. For example, in the Verizon
example in Figure~\ref{figure:permVerizon} values as large as $8.1$
are common on the right but not on the left.
That asymmetry is why we do not compute a two-sided $P$-value by
counting the fraction of resamples with $|\thetahat^*| > |\thetahat|$.

The larger $r$ is, the better. For quick demonstrations
$r=999$ is enough; $r=9999$ is better for professional use.
In the Verizon project we discuss next, we routinely used $499,999$.
The Monte Carlo standard deviation for a one-sided $P$-value is approximately
$\sqrt{p(1-p)/r}$.
% Then the SE for the P-value is < 0.00071

Different test statistics may yield equivalent results. For example,
when comparing two means, we obtain the same $P$ value using the
difference in means, either mean alone, or the $t$~statistic with pooled
variance. Conditional on the combined data, there are strictly monotone
transformations between these statistics, so
$\thetahat^* > \thetahat \Leftrightarrow h(\thetahat^*) > h(\thetahat)$,
and the count $x$ is the same.

\subsection{Test of Relationships}

There are two situations where it is relatively easy to do a permutation
test---comparing two groups, as in the examples above, and testing
the relationship between two variables, where the null hypothesis
is independence between the variables. Next we'll look at an example of the
latter.

Table~\ref{table:skating} contains scores from the 2014 Olympics
Women's Singles Figure Skating, and
Figure~\ref{figure:skatingScatterPerm} shows a scatterplot of short program
and free skate scores, together with a least-squares regression line.
The correlation is 0.86, and regression slope is 2.36.

\begin{table}[\tableplace]
 \centering
\begin{tabular}{rllrrrrr}\hline
Rank&Name              &Nation        &Total &Short&Free\\ \hline
 1&   Adelina Sotnikova&        Russia&224.59&74.64&149.95\\
 2&            Kim Yuna&   South Korea&219.11&74.92&144.19\\
 3&    Carolina Kostner&         Italy&216.73&74.12&142.61\\
 4&         Gracie Gold& United States&205.53&68.63&136.90\\
 5&   Yulia Lipnitskaya&        Russia&200.57&65.23&135.34\\
 6&           Mao Asada&         Japan&198.22&55.51&142.71\\
 7&       Ashley Wagner& United States&193.20&65.21&127.99\\
 8&        Akiko Suzuki&         Japan&186.32&60.97&125.35\\
 9&      Polina Edmunds& United States&183.25&61.04&122.21\\
10&  Ma\'e-B\'er\'enice M\'eit\'e&        France&174.53&58.63&115.90\\
11&   Valentina Marchei&         Italy&173.33&57.02&116.31\\
12&     Kanako Murakami&         Japan&170.98&55.60&115.38\\
13&      Kaetlyn Osmond&        Canada&168.98&56.18&112.80\\
14&            Li Zijun&         China&168.30&57.55&110.75\\
15&         Zhang Kexin&         China&154.21&55.80& 98.41\\
16&          Kim Haejin&   South Korea&149.48&54.37& 95.11\\
17&   Gabrielle Daleman&        Canada&148.44&52.61& 95.83\\
18&  Nathalie Weinzierl&       Germany&147.36&57.63& 89.73\\
19&Elene Gedevanishvili&       Georgia&147.15&54.70& 92.45\\
20&        Brooklee Han&     Australia&143.84&49.32& 94.52\\
21&        Park So-Youn&   South Korea&142.97&49.14& 93.83\\
22&   Elizaveta Ukolova&Czech Republic&136.42&51.87& 84.55\\
23&   Anne Line Gjersem&        Norway&134.54&48.56& 85.98\\
24&     Nicole Raji\u{c}ov\'a&      Slovakia&125.00&49.80& 75.20\\
\end{tabular}
\renewcommand{\mycaption}{2014 Olympics Women's Singles Skating.}
\caption[\mycaption]{\label{table:skating}
{\em \mycaption}
}   \end{table}

\begin{figure}[\figureplace]
% Scripts/skating.R
\centerline{\includegraphics[width=\figurewidth]{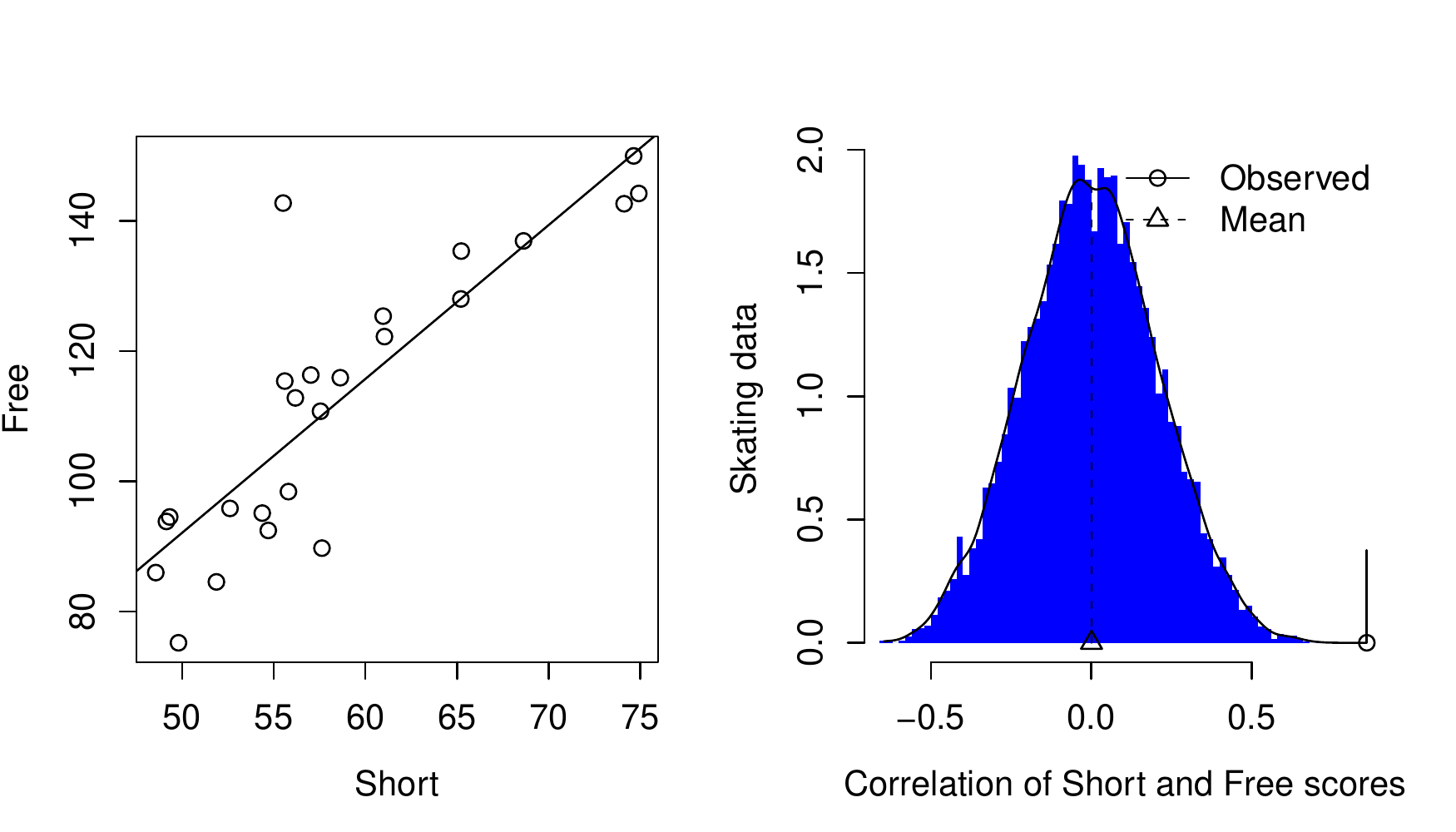}}
\renewcommand{\mycaption}{Short Program and Free Skate scores,
2014 Olympics Women's Figure Skating.}
\caption[\mycaption]{\label{figure:skatingScatterPerm}
{\em \mycaption}
The correlation is $0.86$, and regression slope is $2.36$.
The right panel shows the permutation distribution;
the two-sided $P$-value is $0.0002$.
}    \end{figure}

To test the null hypothesis of independence between the short program
and free skate scores, we create permutation resamples by randomly
permuting either the Short or Free scores, but not both.
(If we permute both columns, using the same permutation, we end up
with the original data in a different order, and the same correlation.)
Using the
partially permuted dataset, we compute the correlation, regression slope,
or another statistic that measures association between the variables.
As before, we compute a $P$-value by comparing the statistic for the
original data with the permutation distribution.

Figure~\ref{figure:skatingScatterPerm} shows the permutation distribution
for the correlation of Short and Free scores.
The correlation of 0.86 is highly significant; the two-sided $P$-value
is $0.0002$, the smallest possible with $r=9999$ resamples
(we add 1 to numerator and denominator to calculate one-sided $P$-values,
then multiply by two for two-sided).

Correlation and least-squares regression slope are equivalent
statistics for testing independence,
because conditional on the data one can be obtained from a monotone
transformation of the other.

\subsection{Limitations}

We have seen two cases where permutation testing is
straightforward---the difference between two groups, and independence
between two variables. Some extensions are easy---to use different
test statistics, or in multiple regression to test the null hypothesis
that $Y$ is independent of all the $X$'s, by permuting the $Y$
variable.

Other situations are not amenable to simple permutation testing,
without making assumptions and using procedures that are
beyond the scope of this article. For example:
\begin{itemize}
\item
We cannot test a hypothesis about a single-sample mean.
\item
We cannot test the equality of means when the variances may differ,
because then we can't pool the data.

This is not as big a hurdle as you may think.
We don't need the {\em sample} variances to be the same; if the population
is positively skewed then the sample with the larger mean naturally has
a larger sample variance. What matters is whether the {\em population}
variances differ when the null hypothesis holds.

For example,
I taught a course for a large Swiss pharmaceutical company,
who were investigating a cheaper alternative to an expensive measurement
procedure. They expected that the cheaper alternative would have somewhat
larger variance, but were willing to live with that if the means
of the two procedures matched. Permutation testing would not be appropriate
here, because we should not pool data with different variances.

\item
We cannot test ``non-zero'' hypotheses, e.g.
$H_0:\mu_1 - \mu_2 = c$ with $c \neq 0$ when comparing two samples,
or $H_0:\beta = c$ with $c \neq 0$ in linear regression.

If we were willing to assume a shift hypothesis, $H_0: F_1(x) = F_2(x_c)$,
we could subtract $c$ from each observation in sample 1, then
perform a standard two-sample permutation test.
However, that would be wrong if a shift hypothesis
were incorrect.

\item
In regression, we cannot test the null hypothesis of zero slope,
without also assuming independence.
\item
In multiple regression we can test the hypothesis that $Y$ is independent
of all $X$'s by permuting $Y$, but we cannot test whether
a single regression coefficient is non-zero by permutating that $X$.
Each regression coefficient measures the incremental impact of one
variable on $Y$, given all other $X$'s.
By permuting a single $X$,
we make that variable independent of all other $X$'s.
This tends to give one-sided $P$-values near zero or 1 when there
is collinearity between the $X$'s;
the permutation distribution for the $\beta$ of interest
is quite narrow in the absence of collinearity, so the $\beta$
for the original sample tends to fall on either side of
the narrow permutation distribution.
\item
We cannot use permutation testing to obtain confidence intervals.
\end{itemize}

\boxx{Where Permutation Tests Apply}
{It is straightforward to apply permutation tests for the
difference of two samples, or for testing independence between
two sets of variables.
There are other situations where permutation tests do not apply.
Bootstrap tests or confidence intervals might be used instead.%
}

\subsection{Bootstrap Hypothesis Testing}
\label{section:bootstrapHypothesisTesting}

Bootstrap hypothesis testing is relative undeveloped,
and is generally not as accurate as permutation testing.
For example, we noted earlier that it is better to do a permutation
test to compare two samples, than to pool the two samples
and draw bootstrap samples.
Still, there are some ways to do bootstrap tests.

The bootstrap~$t$ provides a straightforward test.
$t = (\thetahat - \theta_0)/\Shat$
where $\Shat$ is a standard error for $\thetahat$,
to the bootstrap
distribution of $t^*$; the lower $P$-value is $\Ghat(t)$
and upper $P$-value is $1-\Ghat(t)$.
This corresponds to rejecting if the confidence interval
excludes $\theta_0$, and is second-order accurate.

One general approximate approach is to base a test on
a bootstrap confidence interval---to reject the null hypothesis
if the interval fails to include $\theta_0$.

Another general approach is to sample in a way that is consistent with
the null hypothesis, then calculate a $P$-value as a tail probability
like we do in permutation tests.
For a parametric bootstrap, we sample using values of the parameters
that satisfy the null hypothesis.
For a nonparametric bootstrap, we could modify the observations,
e.g.\ to subtract $\xbar-\mu_0$ from each observation; I do not recommend
this, it is not accurate for skewed populations, can give impossible
data (e.g.\ negative time measurements), and does not generalize well
to other applications like relative risk, correlation, or regression,
or categorical data.
Better is to keep the same observations, but place unequal probabilities
on those observations;
in {\em bootstrap tilting} \citeP{efro81,davi97},
we created a weighted version
of the empirical distribution that satisfies the null hypothesis.
For example, let $\Fhat_w(x) =  \sum_{i=1}^n w_i I(x_i \le x)$,
where $w_i > 0$, $\sum w_i = 1$, and $\theta(\Fhat_w) = \theta_0$.
The $w_i$ may be chosen to maximize the empirical likelihood $\prod_i w_i$
subject to the constraints. For the sample mean,
a convenient approximation
is $w_i = c \exp(\tau x)$ where $c$ is a normalizing constant and
$\tau$ is chosen to satisfy $\sum w_i x_i = \mu_0$.

For a broader discussion of bootstrap testing see \citeP{davi97}.

\boxx{Bootstrap Tests}
{Bootstrap testing is relatively undeveloped.
One general procedure is to test based on bootstrap confidence intervals.
A special case is the bootstrap~$t$ test.%
}

\section{Summary}
\label{section:summary}

I have three goals in this article---to show the value of bootstrapping
and permutation tests for teaching statistics, to dive somewhat deeper
into how these methods work and what to watch out for,
and to compare the methods to classical methods, to show just how
inaccurate classical methods are, and in doing so to provide
impetus for the broader adoption of resampling both in teaching and practice.

Here are some of the key points in this article.
We begin with pedagogical advantages:
\begin{itemize}
\item the bootstrap and permutation testing offer strong pedagogical
benefits. They provide concrete analogs to abstract concepts.
Students
can use tools they know, like histograms and normal probability plots,
to visualize null distributions and sampling distributions.
Standard errors and bias arise naturally. $P$-values are
visible on the histogram of a permutation distribution.

\item Students can work directly with the estimate of interest,
e.g.~a difference in means, rather than working with $t$~statistics.

\item Students can use the same procedure for many different statistics,
without learning new formulas. Faculty can finally use the median
throughout the course (though with larger samples and/or $n$ even,
to avoid small-sample issues with the bootstrap and small odd $n$).

\item Students learning formulas can use resampling to check their work.

\item The process of bootstrapping mimics the central role that sampling
plays in statistics.

\item Graphical bootstrapping for regression provides pictures that
demonstrate increased variability when extrapolating, and the
difference between confidence and prediction intervals.
\end{itemize}

Many key points relate to confidence intervals:
\begin{itemize}
\item
Classical $t$~intervals and tests are terrible for skewed data.
The Central Limit Theorem operates on glacial time scales.
We need $n \ge 5000$ before the 95\% $t$~interval is reasonably accurate
(off by no more than 10\% on each side)
for an exponential population.

These procedures are {\em first-order accurate}---the
errors in one-sided coverage and rejection probabilities are $O(n^{-1/2})$.

A {\em second-order accurate} procedure has errors $O(n^{-1})$

\item Our intuition about confidence intervals for skewed data is wrong.
Given data with a long right tail, we may think
(1) we should downweight outliers, and give less weight to the extreme
observations on the right, and (2) that a good confidence
interval would be asymmetrical with a longer left side.
In fact, it needs a longer right side,
to allow for the possibility that the sample has
{\em too few} observations from the long right tail of the population.

\item The bootstrap percentile interval
(defined in Section~\ref{section:oneSampleBootstrap})
is asymmetrical with a longer
right tail---but has only one-third the asymmetry it needs
to be second-order accurate.

\item The bootstrap percentile interval is terrible in small
samples---it is too narrow. It is like a $t$~interval computed
using $z$ instead of $t$ and estimating $s$ with a divisor of $n$
instead of $n-1$, plus a skewness correction.
% And, for some kinds of bias it does exactly the wrong thing.
There is an ``expanded percentile interval'' that is better.

\item The reverse percentile interval (Section~\ref{section:reversePercentile})
has some pedagogical value,
but is does exactly the wrong thing for skewness and transformations.

\item People think of the bootstrap (and bootstrap percentile interval)
for small samples, and classical methods for large samples.
That is backward, because the percentile interval is so poor
for small samples.

\item There are better intervals, including
an expanded percentile interval,
a skewness-adjustment to the usual $t$ formulas for the mean,
and the bootstrap~$t$ for general problems;
the latter two are second-order accurate.

\item The sample sizes needed for different intervals to satisfy the
``reasonably accurate'' (off by no more than 10\% on each side)
criterion are:
are $n \ge 101$ for the bootstrap~$t$,
220 for the skewness-adjusted $t$~statistic,
2235 for expanded percentile, 2383 for percentile
4815 for ordinary $t$ (which I have rounded up to 5000 above),
5063 for $t$ with bootstrap standard errors
and something over 8000 for the reverse percentile method.

\end{itemize}

Other points include:
\begin{itemize}
\item When bootstrapping, we normally sample the same way the
original data were sampled, but there are exceptions.

\item One general exception is to condition on the observed information;
to fix sample sizes, and to fix the $x$ values in regression---to
bootstrap residuals rather than observations.
(This conditioning is less important with large $n$ in low-dimensions.)

\item The bootstrap may give no clue there is bias, when the cause
is lack of model fit.

\item Bootstrapping statistics that are not functional, like $s^2$,
can give odd results when estimating bias.

\item Permutation tests are easy and very accurate for comparing
two samples, or for testing independence.
But there are applications where they can't be used.

\item For both the bootstrap and permutation tests, the number of
resamples needs to be $15000$ or more, for 95\% probability
that simulation-based one-sided levels fall within 10\% of the
true values, for 95\% intervals and 5\% tests.
I recommend $r=10000$ for routine use, and more when accuracy matters.
\end{itemize}

\paragraph{Research needed for better intervals and tests.}
I would like to see more research into good, practical, general-purpose
confidence intervals and tests.
These should have good asymptotic properties, including
second-order accuracy,
but also handle the ``little things'' well to give good small-sample
coverage, including narrowness bias, variability in SE estimates,
and variability in skewness estimates.

For small samples it would be reasonable to
make less than a full correction for skewness,
because skewness is hard to estimate in small samples.
A shrinkage/regularization/Bayes approach could work well, something that
smoothly transitions from
\begin{squeezeitemize}
\item assuming symmetry for small $n$
\item estimating skewness from the data, for larger $n$
\end{squeezeitemize}
For comparison, the classical $z$ and $t$~intervals
are like Bayesian procedures with priors that place zero probability on
nonzero skewness.

These new better inferences should be made available in statistical
software, such as R packages, and eventually be standard in place
of current $t$~tests and intervals.
In the meantime, people can use resampling diagnostics to estimate
the properties of available methods.

\paragraph{Acknowledgments}

I thank
David Diez, Jo Hardin, Beth Chance,
Fabian Gallusser, Laura Chihara, Nicholas Horton,
Hal Varian, and Brad Efron
for helpful comments.

% \cite{Raftery1992} = (Raftery 1992)
% \citeasnoun{Raftery1992} = Raftery (1992)

%\bibliographystyle{jasa/ECA_jasa}
%\bibliographystyle{ECA_jasa}

%\bibliographystyle{plain}	% (uses file "plain.bst") Does e.g. [6].
%\bibliographystyle{asa} % asa.bst Ugly name/year combo
%\bibliographystyle{chicago} % chicago.bst % fails
\bibliographystyle{rss} % rss.bst % best so far

\bibliography{Bibliography}

\end{document}